\documentclass[a4paper,11pt]{article}
\usepackage[skins]{tcolorbox}
\usepackage{enumerate}
\usepackage{mathtools,slashed}
\usepackage{braket}
\usepackage{tikz}
\usetikzlibrary{shapes, arrows.meta, positioning}

\tikzstyle{block} = [rectangle, draw, text width=12cm, text centered, rounded corners, minimum height=4em]
\tikzstyle{arrow} = [thick, ->, >=Stealth]
\usepackage[T1]{fontenc}
\usepackage{slashed,verbatim,subfigure}
\usepackage[numbers,sort&compress]{natbib}
\usepackage{amsmath}
\usepackage{slashed}
\usepackage{mathrsfs}
\usepackage{amsbsy}
\usepackage{amssymb}
\usepackage{appendix}
\usepackage{graphicx}
\usepackage[final]{pdfpages}
\usepackage{dcolumn}
\usepackage{bm}
\usepackage[colorlinks=true,linktocpage=true,
linkcolor=blue,citecolor=blue]{hyperref}
\usepackage[a4paper]{geometry}
\usepackage{url}
\definecolor{db5}{cmyk}{0.5,0.5,0,0.5}
\definecolor{mauve}{cmyk}{0.3,0.7,0.1,0.3}
\definecolor{palemauve}{cmyk}{0.3,0.7,0.1,0.0}
\definecolor{pb}{cmyk}{0.4,0.1,0,0.1}
\definecolor{pgreen}{cmyk}{0.4,0.0,0.3,0.0}
\definecolor{pink}{cmyk}{0.0,0.5,0.3,0.0}
\newcommand{\bs}{\begin{slide}}
\newcommand{\es}{\end{slide}}
\newcommand{\tcb}{\textcolor {blue}}
\newcommand{\tcbr}{\textcolor {brown}}
\catcode`@11
\def\seceqaa{\@addtoreset{equation}{section}
	\def\theequation{A\arabic{equation}}}
\def\seceqbb{\@addtoreset{equation}{section}
	\def\theequation{B\arabic{equation}}}
\def\seceqcc{\@addtoreset{equation}{section}
	\def\theequation{C\arabic{equation}}}
\def\seceqdd{\@addtoreset{equation}{section}
	\def\theequation{D\arabic{equation}}}
\def\seceqee{\@addtoreset{equation}{section}
	\def\theequation{E\arabic{equation}}}
\def\seceqff{\@addtoreset{equation}{section}
	\def\theequation{F\arabic{equation}}}	
\catcode`@11
\newcommand{\be}{\begin{eqnarray}}
\newcommand{\ee}{\end{eqnarray}}
\usepackage{authblk}
\begin{document}
\large
\title{On the Ubiquity of (Almost) Contact  Structures  in Hot MQCD at Intermediate Coupling}
\author[,1]{Aalok Misra\footnote{email- aalok.misra@ph.iitr.ac.in}}
\author[2]{Gopal Yadav\footnote{email- gopalyadav@cmi.ac.in}\vspace{0.1in}}
\affil[1]{Department of Physics, Indian Institute of Technology Roorkee, Roorkee 247667, Uttarakhand, India}
\affil[2]{Chennai Mathematical Institute, SIPCOT IT Park, Siruseri 603103, India}

\date{}

\maketitle

\begin{center}
{\bf Abstract}
\end{center}
{\footnotesize  In the context of fluxed $G_2$-structure ${\cal M}$-theory duals of intermediate-$N$(the number of color $D3$-branes in the parent type IIB dual of \cite{metrics}) thermal QCD (Quantum Chromodynamics)-like theories, inclusive of quartic-in-curvature corrections, we show the explicit existence of (Almost) Contact 3-Structures [(A)C3S]. {\it A\ novel\ result\ in\ this\ context\ is\ that\ in\ the\ Infra-Red,\ within\ the\ parameter\ space\ of\ AC3S\ arising\ from\ $G_2$\ structures\, the\ subspaces\ of\ C3S\ and\  AC3S\, are\ not\ mutually\ ``$N$-path connected''} because the former exists only for intermediate $N$ while the latter exists only for very large $N$. Using a proposition of \cite{Ossa et al[2013]}, we also discuss the reduction of the $G_2$ structure to "transverse" $SU(3)$ structure induced by  (A)C3S in the IR.  Apart from providing a novel insight into the differential geometry of the relevant ${\cal M}$-theory non-supersymmetric manifolds, our results lead us to {\it conjecture\ that\ the\ intermediate-$N$\ limit\ which\ supports\ C3S\, is\ farther\ away\ from\ the\ non-supersupersymmetric\ AdS\ "swampland"\ than\ the\ $N\gg1$-limit\ which\ supports\ AC3S}. We also point out that a rich variety of high-temperature phenomena such as post Page-time ``Islands'' black hole entanglement entropy, the "Lyapunov exponent" and "butterfly velocity" pertaining to an "unusually" chaotic behavior discussed earlier by one of the co-authors [AM] and his former group members in \cite{MQGP-Page} and \cite{MQGP-chaos} respectively, use precisely the aforementioned intermediate-$N$ limit supporting C3S. Further, the conjectural correspondence: $\bigl(${\it non-analytic [branch-point singularities]\ gauge-coupling\ dependence\ of\ the\ ${\cal O}$(curvature$^4$)-corrected\ complexified\ "bulk\ viscosity"\ in\ terms\ of\ speed\ of\ sound\ \cite{zeta-intermediate}\ and\ strong\ magnetic\ field\ photoproduction\ spectral\ density \cite{photoprod+EoS+AC3S}}, analytic gauge-coupling depedence of the ${\cal O}$(curvature$^4$) non-renormalized complexified pressure/energy of a paramagnetic plasma $\bigr) \leftrightarrow \bigl(${\it C3S\ in\ the\ aforementioned\ intermediate-$N$\ MQGP\ limit}, AC3S in $N\gg1$ MQGP limit $\bigr)$ mimicking the aforementioned lack of $N$-path connectedness,  is reiterated.

\begin{center}
{\it One of us (AM) dedicates this paper to the memory of Professor Asoke Nath Mitra - an unparalleled academic and sentient being}
\end{center}
}

\tableofcontents

\section{Introduction}
\label{introduction}

The AdS/CFT correspondence/gauge/gravity duality \cite{AdS-CFT-Originals} is a very useful tool in understanding the properties of super Yang-Mills theory at large t'Hooft coupling. According to this correspondence, $\mathcal{N}=4$ $SU(N)$ SYM theory in the large $N$ limit is dual to type $IIB$ superstring theory on $AdS_5\times S^5$ geometry, where $AdS_5$ is the five dimensional anti-de Sitter space and $S^5$ is the five sphere. The $\mathcal{N}=4$ $SU(N)$ SYM theory is a conformal field theory, i.e., its gauge coupling does not run with the energy scale. On the other hand  QCD is non-conformal. Now, $SU(N_c)$ QCD, $N_c$ being the number of  quark colors, is an asymptotically free theory so that  the gauge coupling is scale dependent and vanishes logarithmically with large momentum or with short distance. So to deal  with QCD-like theories using Gauge/Gravity duality one needs to generalize the AdS/CFT correspondence and incorporate a  running coupling in the theory (apart from considering a finite temperature version as well).  Building up on the Klebanov-Witten \cite{KW}, Klebanov-Nekrasov \cite{KN}  and  Klebanov-Tseytlin \cite{KT} models,  a logarithmic QCD-like RG flow was obtained in the non-conformal Klebanov-Strassler model \cite{KS} by considering $M$ fractional $D3$-branes ($D5$-branes wrapping the vanishing $S^2$ of a conifold) along with $N$  $D3$ branes in a conifold geometry  wherein the IR geometry was modified resulting in a deformed conifold. Now, let us go back to the Klebanov-Strassler model  where the temperature is turned on in the field theory side corresponding to introducing a black hole in the gravity dual.  The flavor $D7$-branes are embedded in KS model via the holomorphic Ouyang embedding \cite{ouyang} and finally the M-theory uplift of the whole set up keeps the background geometry as required provided we  consider some limiting values of the parameters in the theory. The details about this are spelt out in \cite{metrics}, \cite{ouyang}. 

For over two decades, gauge/gravity duality has provided a simple, classical computational tool for understanding the strongly coupled systems that overcomes the theoretical limitations in the study of non-Abelian gauge theories. The $\frac{1}{N}$-expansion in the context of the AdS/CFT correspondence has been explored in a variety of contexts some of which are as follows. In \cite{CS-top-string}, the authors provide a worldsheet proof of the equivalence between the $U(N)$ Chern-Simons gauge theory on $S^3$ and the topological closed string theory on the resolved conifold geometry in the large-$N$ limit. Computation of loop amplitudes in AdS, which are dual to non-planar correlators in holographic CFTs is explored in \cite{Aharony-et-al}. Computation in $AdS_5$ supergravity of the trace anomaly of a $D=4, {\cal N}=2$ SCFT, and agreement with the field theory result up to next-to-leading order in the $\frac{1}{N}$-expansion was done in \cite{trace-anomaly}. Large-$N$ expansions in holographic CFTs using the Polyakov-Mellin bootstrap along with computation of loop amplitudes in $AdS$ corresponding to non-planar correlators in the CFT, was studied in \cite{Polyakov-Mellin-large-N}.  The chiral anomaly of the SYM theory, inclusive of the NNLO-in-$N$ corrections accounted for by the supergravity/string effective action at one loop was studied in \cite{NNLO-N}. Borel summability of the large-$N$ expansion of giant-graviton correlators was looked at in \cite{Giant-gravitons-large-N}. The $\frac{1}{N}$-expansion of expectation values of Wilson loops in $SO(N)$ lattice gauge theory in the strong coupling regime was studied in \cite{Large-N-Wilson}. The relationship between bulk loop diagrams in $AdS$ and the large-$N$ expansion in the boundary CFT was discussed in \cite{bulk-loop-boundary-large-N}. Rather interestingly, in \cite{numerical-AdS-CFT}, the authors showed that one can study the ABJM theory, an ${\cal N} = 6$ superconformal $U(N) \times U(N)$ Chern-Simons theory, for arbitrary $N$ and at arbitrary coupling constant by applying a simple Monte Carlo method to the matrix model. 

As mentioned earlier, in its simplest form for maximally supersymmetric $SU(N_c={\rm number\ of\ colors})$ Yang-Mills theory (${\cal N} = 4$ SYM), in the $N_c\rightarrow\infty$ limit, the gauge/gravity duality provides a tool for analysing its properties in the large `t Hooft coupling limit as well as intermediate coupling as corrections to the infinite-coupling limit. These corrections appear as higher order derivative corrections  on the gravity side. The effect of these corrections to the action is incorporated in the background metric and fluxes, perturbatively by considering perturbations of the equations of motion. However, other than higher-derivative corrections quartic in the Weyl tensor, or of the Gauss-Bonnet type, in $AdS_5\times S^5$, dual to supersymmetric thermal Super Yang-Mills \cite{previous-higher-ders}, before \cite{OR4}, there was little known about top-down string theory duals at intermediate ’t Hooft coupling of realistic thermal Quantum Chromodynamics(QCD)-like theories. Understanding thermal Quantum Chromodynamics(QCD)-like theories (equivalence class of theories that are Infra Red(IR)-confining, Ultraviolet-conformal with bi-fundamental quarks, at finite temperature) {\it at intermediate coupling} from the non-perturbative completion of string theory - ${\cal M}$ theory - is naturally an interesting area in its own right \footnote{The "Quark Gluon Plasma (QGP)" is a Physics "laboratory" where such a study will be directly relevant.}. 
In \cite{OR4}, we included terms quartic in the eleven-dimensional Riemann curvature $R$ in the eleven-dimensional supergravity action that appear as ${\cal O}(l_p^6) ( l_p$ being the 11D Planckian length)-corrections in the ${\cal M}$-theory dual of large-$N$(the number of color $D3$-branes in the string theory dual \cite{metrics}) thermal QCD-like theories (equivalence class of theories which are UV-conformal, IR-confining and have fundamental quarks).

Contact geometry is ubiquitous in Physics, be it electrodynamics \cite{A.L.Kholodenko}, fluid mechanics \cite{A.L.Kholodenko}, \cite{Arnold+Khesin}, \cite{Etnyre+Ghrist}, magnetohydrodynamics \cite{Kozlov}, \cite{A.L.Kholodenko}, thermodynamics \cite{Bravetti}, liquid crystals \cite{Lubensky}, magnetic monopoles \cite{Arnold+Khesin}, etc. Now, fluxed compactification geometries in string/${\cal M}$-theory,  naturally use manifolds of $G$-structure e.g. $SU(3)$-structure heterotic string theory was initiated in \cite{Berlin-torsion-classes}; $SU(3)$ and $G_2$-structure torsions classes of respectively six-/seven-folds in respectively type II/${\cal M}$-theory flux compactifications was extensively studied in  \cite{SYZ-free-delocalization,Dasgupta+Tatar-et-al,theta0-theta,Butti et al [2004]}. $SU(3)$-/$G_2$-structure torsion classes of type IIB/A holographic dual of thermal QCD-like theories  and their ${\cal M}$-theory uplift in the intermediate/large ``$r$" (the radial coordinate in the gravity dual which corrsponds to the energy on the gauge theory side), i.e., Ultra-Violet (UV)-Infra-Red (IR) interpolating region/UV region were obtained in the second reference in \cite{EPJC-2} and \cite{NPB}. In \cite{OR4}, for the first time, $SU(3)$-structure six-folds, $G_2$-structure seven-folds and $SU(4)/Spin(7)$-structure eight-dimensional geometries at small $r$, i.e., the IR, inclusive of the aforementioned  ${\cal O}(l_p^6)$-corrections in the $D=11$ supergravity action, were studied (note these corrections vanish in the very large-$r$ limit, i.e., the deep UV, wherein $G$-structure approaches $G$-holonomy), both in the SYZ type IIA mirror of the type IIB holographic dual constructed in \cite{metrics} of thermal QCD-like theories, as well as its ${\cal M}$-theory uplift. It is the $G_2$ structure that induces Almost Contact 3-Structures (See \cite{Ossa et al[2013]} and references therein).  We will present a prelude to \ref{MQGP limit} and \ref{LOR4} in \ref{prel_2+3.1}, followed by a prelude  to \ref{ACM3S-basics} - \ref{C3S-M7x0} in \ref{prel_4-7}. The main result of the paper is summarized in the \ref{Main-Result} with the organization of the remainder of the paper given in \ref{organization-paper}.

\subsection{The Known Background}
\label{Background}

In this subsection, we provide the necessary Physics background for the remainder of the paper.

\subsubsection{${\cal M}$-Theory Dual of Real QCD-Like Theories at Intermediate Coupling - A Prelude to \ref{MQGP limit} and \ref{LOR4}}
\label{prel_2+3.1}

Let us now turn our attention to string theory in the context of gauge/gravity duality to study real QCD-like theories, i.e., lattice/Particle Data Group-compatible gravity duals. 

Higher order derivative corrections, apart from being related to corrections to the infinite coupling limit,  also serve as the leading quantum gravity corrections to the ${\cal M}$-theory action to study the compactifications of ${\cal M}$-theory on compact eight-dimensional manifolds.  The terms quartic in curvature have been used in past \cite{O(R^4)}, while terms involving quadratic in fluxes and cubic in curvature, have been recently analyzed in \cite{O(R^3G^2)}.   The construction of a top-down holographic dual of thermal Quantum Chromodynamics (QCD) at intermediate 't Hooft coupling has been missing in the literature. The paper \cite{OR4} (involving one of the authors [AM] as a co-author) took important steps in filling this gap by studying the ${\cal M}$ theory dual of large-$N$ thermal QCD-like theories at high temperatures, at intermediate gauge and 't Hooft couplings by obtaining the ${\cal O}(l_p^6)$ corrections to the ${\cal M}$-Theory uplift of \cite{metrics} as constructed in \cite{MQGP}.

As far as we know, \cite{MQGP} is the only  holographic ${\cal M}$-Theory dual of realistic thermal QCD-like theories that 
 yields a deconfinement temperature\footnote{The very high temperature  at and above which the quarks become free, and below which they are ``'confined''.} $T_c$ from a Hawking-Page phase transition at vanishing baryon chemical potential that is compatible with lattice QCD results \cite{Misra+Gale_Conformal_Anomaly},   
yields a conformal anomaly variation with temperature again compatible with lattice results, both at  high ($T>T_c$) {\it and} low ($T<T_c$) temperatures \cite{Misra+Gale_Conformal_Anomaly}. In Condensed Mattter Physics,  inclusive of  the non-conformal corrections,  (i) a lattice-compatible shear-viscosity-to-entropy-density ratio was obtained in the first reference in \cite{EPJC-2}, (ii) the temperature variation of a variety of transport coefficients including the bulk-viscosity-to-shear-viscosity ratio,  diffusion coefficient, speed of sound was obtained in the last reference in \cite{EPJC-2}), electrical and thermal conductivity and the Wiedemann-Franz law was obtained in the first reference in \cite{EPJC-2}. In Particle Phenomenology, 
 lattice-compatible glueball spectroscopy was obtained in \cite{Sil+Yadav+Misra-glueball}, Particle Data Group(PDG)-compatible meson spectroscopy was discussed in the first reference of \cite{mesons_0E++-to-mesons-decays})\footnote{As noted in the same, we obtain $0^{--}$ (\`{a} la $J^{PC}$-assignment) pseudo-scalar mesons that thus far have not been found in the Particle Data Group (PDG) data further justifying the use of QCD-like theories in our work.}, PDG-compatible glueball-to-meson decay widths were obtained in the second reference of \cite{mesons_0E++-to-mesons-decays}, QCD-compatible supermassive inert mesinos were obtained in \cite{mesinos} thereby resolving a longstanding problem with the Sakai-Sugimoto type IIA top-down holographic dual \cite{SS} of large-$N$ QCD-like theories of meson-mesino isospectrality and unsuppressed mesino-meson interaction.  

On the Mathematics side, for the first time,  $SU(3)$-structure (for type IIB (second reference of \cite{EPJC-2})/IIA \cite{NPB}, \cite{OR4} holographic dual), $G_2$-structure  torsion classes \cite{NPB}, \cite{OR4} and $SU(4)/Spin(7)$ torsion classes \cite{OR4} of the six-, seven- and eight-folds in the UV-IR interpolating region/UV, relevant to type string/${\cal M}$-Theory holographic duals of thermal QCD-like theories at high temperatures were obtained.

\subsubsection{$G_2$-Structure Manifolds and Almost Contact 3-Structures - a Prelude to \ref{ACM3S-basics} - \ref{C3S-M7x0}}
\label{prel_4-7}

Now, continuing on with the last (Math-related) bullet, $G_2$-structure manifolds are ubiquitous to fluxed ${\cal M}$-theory compactifications \cite{G2-structure-M-theory}. Manifolds with Almost Contact Structures (which induce a complex structure on the ``transverse geometry'') could be thought of as odd-dimensional analogs of even-dimesional manifolds equipped with almost complex structures. The obstruction of the existence of nowhere vanishing vector fields on closed manifolds essential to the existence of Almost Contact Structure (ACS) , is non-vanishing Euler charactersitic. As the Euler characteristic of odd-dimensional manifolds is vanishing, therefore odd-dimensional manifolds may support  ACS. Seven-dimensional manifolds equipped with $G_2$-structure are known to provide  Almost Contact 3-Structures (AC3S). The existence of nowhere-vanishing vector fields implies that the structure group of the tangent bundle is reducible which has an implication on the amount of SUSY preserved by the vacuum. Also, a seven-fold supporting a $G_2$ structure can be shown to admit an Almost Contact Metric Structure (ACMS) and this implies that the structure group reduces to $SU(3)$. In fact, there exist at least three non-zero vector fields on a manifold with $G_2$ structure which provides the manifold with  Almost Contact Metric 3-Structures (ACM3S).

The use of $G$-structure torsion classes is a very useful tool for classifying, specially non-K\"{a}hler geometries. A complete classification of the $SU(n)$ structures  relevant to non-supersymmetric string vacua, does not exist \cite{Ossa et al[2013]}. In the context of $SU(3)$-structure manifolds, non-supersymmetric type II vacua on $SU(3)$-structure manifolds were studied in \cite{P. G. Camara+M. Grana [2007]} and classified using calibrations in \cite{D. Luest-et-al[2008]}, and similar solutions in heterotic string theory were obtained in \cite{D. Luest-et-al[2010]} - see \cite{J. G. J. Held's thesis [2012]} for $G_2$ structures relevant to non-supersymmetric vacua in heterotic(${\cal M}$-) supergravity.

A classification of $SU(3)/G_2/Spin(7)/Spin(4)$ structures relevant to {\it non-supersymmetric} (UV-complete) string theoretic dual of large-$N$ thermal QCD\footnote{Quantum Chromodynamics}-like theories (equivalence class of thoeries which are IR confining, UV conformal with quarks in the fundamental representation of color and flavor, at intermediate coupling), and its ${\cal M}$-theory uplift, has been missing in the literature.  This is what we initiated in \cite{OR4} wherein we worked out the $SU(3)$-structure torsion classes in the "MQGP" limit (first discussed/coined in \cite{MQGP}; also see (\ref{MQGP_limit}) and (\ref{intermediate-N-MQGP-limit})) of the relevant non-K\"{a}hler conifold $M_6=$ relevant to the type IIA SYZ mirror, the $G_2$-structure torsion classes  of the relevant seven-fold $M_7 = S^1 \times_w M_6$, i.e. a warped (indicated by a subscript "$w$") product of the ${\cal M}$-theory circle and $M_6$, as well as the $SU(4)$-structure and $Spin(7)$-structure torsion classes of the eight-fold $M_8$ - a warped product of the thermal $S^1$ and $M_7$ -  relevant to the ${\cal M}$-Theory uplift.  Table \ref{G-Structure} summarizes the $G$-structure torsion classes' results.
\begin{table}[h]
\begin{center}
\begin{tabular}{|c|c|c|c|}\hline
S. No. & Manifold & $G$-Structure & Non-Trivial Torsion Classes \\ \hline
1. & $M_6=$\ {\rm non-K\"{a}hler\ conifold} & $SU(3)$ & $T^{\rm IIA}_{SU(3)} = W_1 \oplus W_2 \oplus W_3 \oplus W_4 \oplus W_5: W_4 \sim W_5$ \\ \hline
2. & $M_7 = S^1_{\cal M}\times_w M_6$ & $G_2$ & $T^{\cal M}_{G_2} = W_1 \oplus W_7 \oplus W_{14} \oplus W_{27}$ \\ 
&&& \hskip 0.3in $\downarrow r\sim (1+ \alpha_a)a,\ \alpha_a\sim 0.1-0.3$ \\
&&& $W_1={\cal O}\left(\frac{1}{N^{\alpha_{W_1}}}\right), \alpha_{W_1}>1,$ \\ 
&&&  {\rm therefore\ disregarded\ up to}\ ${\cal O}\left(\frac{1}{N}\right)$;\\
& & & $W_7\approx0$ \\ 
&&&\hskip -0.5in $\Rightarrow T^{\cal M}_{G_2} = W_{14} \oplus W_{27}$ \\ \hline
3. & $M_8 = S^1_t \times_w M_7$ & $SU(4)$ & $T^{\cal M}_{SU(4)} = W_2 \oplus W_3 \oplus W_5$ \\ \hline
4. & $M_8$ & $Spin(7)$ & $T^{\cal M}_{Spin(7)} = W_1 \oplus W_2$ \\ \hline
\end{tabular}
\end{center}
\caption{ $G$-Structure Torsion Classes of Six-/Seven-/Eight-Folds in the IR, in the type IIA/${\cal M}$-Theory Duals of High-Temperature QCD-Like Theories}
\label{G-Structure}
\end{table}

We continue in this work wherein we obtain explicitly AC3S/C3S and associated transverse $SU(3)$ structures corresponding to seven/six-folds relevant to the ${\cal M}$-theory inclusive of ${\cal O}(R^4)$ terms, dual of large-$N$ thermal QCD-like theories at intermediate coupling as obtained in \cite{MQGP}, \cite{OR4}. 
The seven-fold $M_7$ that figures in the ${\cal M}$-theory uplift of the type IIB dual of large-$N$ thermal QCD-like theories at intermediate coupling \cite{OR4},
is the warped product of the ${\cal M}$-theory circle $S^1_{\cal M}$ and a non-K\"ahler six-fold that itself is the warped product of the thermal circle $S^1_{\rm thermal}$ (this is the compactified time direction at finite temperature) and a non-Einsteinian generalization of $T^{1,1}$. It is shown that $M_7$ is equipped with a positive stable three-form $\Phi$ and coframes $e^{a=1,...,7}$.

\subsection{Main Results and Organization of the Paper}

In this subsection, we summarize the main results of this work and provide a summary of the organization of the remainder of the paper.

\subsubsection{Technical summary of the results}
\label{Main-Result} 

The dynamics of the ${\cal M}$-theory dual of thermal QCD-like theories at intermediate 't-Hooft coupling/intermediate-$N$ limit (\ref{intermediate-N-MQGP-limit}), is given predominantly by the $G_2$-structure torsion classes of underlying seven-folds, and hence the Contact structure arising from the same. The seven-fold  is either  a warped product of the ${\cal M}$-theory circle and a non-K\"{a}hler six-fold which  is the warped product of the thermal circle with a non-Einsteinian deformation of $T^{1,1}$ with the seven-fold denoted by $M_7$, or a warped product of the ${\cal M}$-theory circle and a non-K\"{a}hler conifold. In this paper we prove the proposition (part 3. thereof, numerically):

\noindent {\it Proposition 1}: Near the holomorphic embedding of the flavor $D7$-branes\footnote{Referred to as Ouyang embedding.}  in the parent type IIB dual of thermal QCD-like theories \cite{metrics} $\forall (N, M, N_f, g_s)$\footnote{$M, N_f, N$ respectively denote the numbers of fractional $D3$-branes, flavor $D7$-branes and color $D3$-branes, and $g_s$ the string coupling constant in the parent type IIB dual of \cite{metrics}.},
\begin{enumerate}
\item
$M_7$ supports a $G_2$ structure with the non-trivial torsion classes \footnote{A seven-fold $M_7$ possessing a $G_2$ structure is characterized by four torsion classes $\tau_{p=0, 1, 2, 3}$ (\ref{T-G2}) which are sections of $p$ copies of the cotangent bundle of $M_7$, i.e. $\tau_{p=0, 1, 2, 3}$ are differential-$p$ forms.} $\tau =  \tau_1\oplus\tau_2\oplus\tau_3,$ in both, the $\Upsilon\gg1$-MQGP limit (\ref{MQGP_limit}) [$\Upsilon\equiv\frac{N}{\left(g_s M^2\right)^{m_1}\left(g_s N_f\right)^{m_2}}, m_{1,2}\in\mathbb{Z}_+\cup\{0\}$] and the  $\Upsilon>1(\Upsilon\slashed{\gg}1)$-MQGP limit (\ref{intermediate-N-MQGP-limit});

\item in  the $N\gg1$-MQGP limit (\ref{MQGP_limit}), 
\begin{enumerate}
\item
$M_7$ supports Almost Contact  Metric 3-Structures $(\sigma^1,\sigma^2,\sigma^3) = (e^1, e^7, -e^2)$ with associated three-tuple of dual Reeb vector fields $(R^1, R^2, R^1\times_\Phi R^2)$  {\it which do not  correspond to Contact 3-Structures}; 
\item
inherits transverse $SU(3)$ 3-structures $\left(\Omega_+^{\alpha}, \Omega_-^{\alpha}\right)$,[i.e., the real and imaginary protions of the nowhere-vanishing holomorphic three-form] wherein \\
$\Omega^{\alpha}_+ = \Phi - \sigma^{\alpha}\wedge \omega_\Phi^{\alpha},
\Omega^{\alpha}_- = \sigma^{\alpha}\lrcorner\left(\frac{1}{2}\omega_\Phi^{\alpha}\wedge \omega_\Phi^{\alpha} - *_7\Phi\right), \alpha=1, 2, 3$ from the Almost Contact 3-Structures constructed in 2;
\end{enumerate}

\item {\it $M_7$ can not support Contact 3-Structures for $\Upsilon\gg1$}, but {\it only for intermediate $N$ (i.e. $\Upsilon>1, \Upsilon\slashed{\gg}1$) satisfying (\ref{intermediate-N-MQGP-limit}) - as an example, for the real QCD-motivated values of $(M, N_f, g_s) = (3, 2\ {\rm or}\ 3, 0.1)$ (Table \tcb{2}) and therefore $N=100\pm{\cal O}(1)$}, $M_7$
\begin{enumerate}
\item 
supports Contact 3-Structures: $(\sigma^1,\sigma^2,\sigma^3) =\left(\alpha_1 e^1 + \alpha_3 e^3 + \alpha_7 e^7, \beta_1 e^1 + \beta_4 e^4 + \beta_7 e^7, \sigma^3\right),$ $ \{\alpha_{i=1, 2, 3}\}, \{\beta_{j=1, 2, 3}\}\in\mathbb{R}$, with $\sigma^3: \sigma^3(R^1\times_\Phi R^2)=1$  ;

\item
inherits transverse $SU(3)$ 3-structures from the Contact 3-Structures, $\left(\Omega_+^{\alpha},\Omega_-^{\alpha}\right)$,
where $\Omega_+^{\alpha} = \Phi - \sigma^{\alpha}\wedge\omega_\Phi^{\alpha}$ and, e.g., for $\alpha=1$, a one-parameter ($\Lambda^{(1)}_{156}$ or $\Lambda_{456}$) family of:
 \begin{eqnarray*}
& & \Omega_- = \Lambda_{AMC}e^{ABC} = \Lambda^{(1)}_{1b_0c_0}e^{1b_0c_0}   + \Lambda_{3b_0c_0}^{(1)}e^{3b_0c_0}
+ \Lambda^{(1)}_{7b_0c_0}e^{7b_0c_0}
+ \Lambda_{a_0b_0c_0}e^{a_0b_0c_0},
\end{eqnarray*}
where $a_0, b_0, c_0=2, 4, 5, 6; \Lambda^{(1)}_{1/3/7\ b_0c_0}, \Lambda_{a_0b_0c_0}$ being constant parameters appearing in $\Omega_-$, with $\Lambda_{456}$ being the only linearly independent non-vanishing $\Lambda_{a_0b_0c_0}$.
\end{enumerate}
\end{enumerate} 
In 2.(b), the contraction operator $\lrcorner$ is defined as 
\begin{equation}
\label{lrcorner-def}
\lrcorner: \Lambda^k T^* \otimes\Lambda^lT^*\rightarrow\Lambda^{l-k}T^* ,
\end{equation}
 with the convention $e^{a_1}\wedge...e^{a_k}\lrcorner e^{a_1}\wedge...e^{a_k}\wedge....e^{a_l} = e^{a_{k+1}}\wedge...e^{a_l}$.

\subsubsection{Non-technical summary and significance of results }

\begin{itemize}

\item
In the $N\gg1$-limit (\ref{MQGP_limit}), the ${\cal M}$-theory dual of thermal QCD-like theories at high temperature, mimics a three-tuple of complex lamellar fields of magnetohydrodynamics, superconductivity and liquid crystals with  
$\omega_\Phi^{\alpha = 1, 2, 3}$, the fundamental two-form: $i_{R^\alpha}\omega_\Phi^\alpha=0, \alpha = 1, 2, 3$, mimicking fluid incompressibility \cite{Etnyre+Ghrist}. On the other hand, in the  intermediate-$N$-limit (\ref{intermediate-N-MQGP-limit}), the ${\cal M}$-theory dual mimics   monopoles  on $G_2$-structure seven-folds with strengths determined by ``generalized spin connections'' (\ref{Omega^a_bc-i}), (\ref{Omega^a_bc-ii})\footnote{See (\ref{contact-x0-2}).}.

\item
The parameter space \footnote{This parameter space is parametrized by $(g_s, M, N_f; N)$.}characterizing closed seven-folds supporting $G_2$ structures and relevant to the ${\cal M}$-theory uplift of thermal QCD-like theories, is not $N$-path connected with reference to Contact Structures in the IR. 

\item Diverse quantities, e.g., the post Page-time evolution (with respect to a "boundary time") of the Page curve of the ``Islands'' entanglement entropy associated with the eternal black hole relevant to the gravity dual at large temperatures as well as the ("Lyapunov exponent" and "butterfly velocity" pertaining to an) "unusually" chaotic behavior of the ${\cal M}$-theory dual of hot QCD arising from pole-skipping, all require the use of the aforementioned limit (\ref{intermediate-N-MQGP-limit}) supporting Contact 3-Structures.

\item Conjecture (first discussed in \cite{zeta-intermediate}) The fractional gauge-coupling-dependence of the bulk-to-shear viscosity ratio when expressed in terms of the speed of sound led to the conjecture in \cite{zeta-intermediate} that the consequent non-analytic-gauge-coupling dependence (involving a branch-point singularity) of the complexified bulk-to-shear viscosity ratio is the transport coefficient analog of the aforementioned lack of $N$-path connectedness between the subspace of Almost Contact 3-Structures and Contact 3-Structures.

\item Two-Part Conjecture (discussed in \cite{photoprod+EoS+AC3S} and summarized below): \\ As \\
(I) physical quantities (e.g. photoproduction spectral density, bulk viscosity in terms of speed of sound) obtained from gauge-field fluctuations (using Kubo's formulae and Son-Starinets prescription for computation of Minkowskian correlators \cite{Son+Starinets-prescription}) receive ${\cal O}(R^4)$ corrections (verified explicitly in the presence of a strong magnetic field in \cite{photoprod+EoS+AC3S}) and correspond to Contact 3-Structures (C3S), and physical quantities (e.g. pressure, energy) obtained from finite background gauge field receive no ${\cal O}(R^4)$ corrections (verified explicitly in the presence of a strong magnetic field in \cite{photoprod+EoS+AC3S}) and correspond to Almost Contact 3-Structures (AC3S), \\
and\\
(II)  $l_p\sim\frac{\sqrt{G_{x^{10}x^{10}}^{\cal M}}}{g_s^{2/3}}\sim\frac{1}{|\log r_h|}\sim\frac{\left(g_s N_f\right)^{2/3}\left(g_s M^2\right)^{1/3}}{N^{1/3}}$ is the size of the ${\cal M}$-theory circle \cite{zeta-intermediate}), 
\begin{itemize}
\item
the subspaces of AC3S and C3S not being $N$-path connected is the differential geometric version of the statement that gauge fluctuations can not be finite,
\item
C3S under an RG flow from the IR to the UV would approach AC3S (since $M$ and $N_f$ figuring in the aforementioned expression of $l_p$ are in fact, respectively, the effective number of fractional $D3$-branes and flavor $D7$-branes, both of which become vanishingly small in the UV to guarantee UV conformality).
\end{itemize} 

\item
{\it Conjecture} ({\bf novel to this work}): The non-supersymmetric  non-AdS string/${\cal M}$-theory dual of thermal QCD-like theories, is farther away from the non-supersupersymmetric AdS swampland\footnote{The set of effective field theories which possess no UV completion upon addition of gravity.} in the intermediate-$N$ MQGP limit (\ref{intermediate-N-MQGP-limit}) which supports Contact 3-Structures (Lemma 4)  than the large-$N$ MQGP limit (\ref{MQGP_limit}) which supports Almost Contact 3-Structures (Lemmas 2 and 3). 
\end{itemize}

A more detailed discussion on the aforementioned Physics-related significance of  the induced Contact Structures (from $G_2$-structure) obtained in the intermediate-$N$ MQGP limit (\ref{intermediate-N-MQGP-limit}) to very wide range of areas namely Page curves of eternal black holes and unusually chaotic behavior of the ${\cal M}$-theory duals of thermal QCD-like theories at high temperatures, is given in \ref{applications-Physics}.  

\subsection{Organization of the rest of the paper}
\label{organization-paper}

 The remainder of the paper which is based on the proof of the aforementioned Proposition 1, is organized as follows. Section \ref{MQGP limit} summarizes the type IIB/IIA dual of large-$N$ thermal QCD-like theories and the "MQGP limit", and the ${\cal M}$-theory uplift. In Section \ref{lp6-corrections}, via three subsections, the ${\cal M}$-theoretic ${\cal O}(R^4)$-corrections, their applications and when one would need to go beyond the same, are discussed. Section \ref{ACM3S-basics} summarizes the basics of Almost Contact Structures (ACS), Contact Structures (CS) and Almost Contact Metric Structures (ACMS) as well as Almost Contact 3-Structures (AC3S).  In Lemma 1 in \ref{G2-M7x0}, we provide an explicit $G_2$ structure on  $M_7 = S^1_{\cal M}\times_w M_6, M_6=S^1_{\rm thermal}\times_w M_5$ where $w$ implies a warped product and $M_5$ is a non-Einsteinian generalization of $T^{1,1}$, and evaluate the four $G_2$-structure torsion classes of $M_7$. In \ref{ACM3S-constructions}, explicit Almost Contact 3-Structures are constructed in Lemma 2 in \ref{ACM3S-M7-x0}.  In Lemma 3 in \ref{AC3S-M7x0}, it is shown that the Almost Contact Structure on $M_7$ constructed in \ref{ACM3S-M7-x0} is in fact an Almost Contact Metric Structure. 
In Lemma 4 in \ref{C3S-M7x0}, we provide explicit Contact 3-Structures on $M_7$. In \ref{Transverse-ACMS}, we provide an explicit transverse $SU(3)$-structure arising from the $G_2$ structure constructed in Lemma 5 in \ref{G2-M7x0}. In Lemma 6 in \ref{Transverse-SU3-CMS}, we provide an explicit transverse $SU(3)$-structure arising from the Contact Structure constructed in Lemma 4. Now, \ref{applications-Physics} is about understanding the implications and applications of the Math results obtained in sections \ref{G2-M7x0} - \ref{SU3-from-G2} wherein we present a conjecture pertaining to the non-supersymmetric swampland on one hand, and connect up on the other, the existence of Contact 3-Structures with entanglement entropy of relevant eternal black holes as well as chaos via pole-skipping. We also briefly review from previous works  from our group, a connection between bulk-to-shear-viscosity ratio in terms of speed of sound and Contact 3-Structures, as well as a connection respectively between strong magnetic field photoproduction/pressure(energy)-anisotropic paramagnetic plasma Equation of State, and Contact/Almost Contact 3-Structures.   The summary and concluding remarks are presented in \ref{summary}. Appendix  \ref{Omega_-}  provides details pertaining to solution of (\ref{iROmega-0}) and (\ref{Transverse-condition-iii}). 

{\bf The new results are obtained in sections \ref{G2-M7x0} - \ref{SU3-from-G2}.} {\it Connections between the results obtained therein and previous results obtained by our group in publications subsequent to the posting of arXiv:2211.13186[hep-th]v1 which was a preliminary version of this paper, are discussed in \ref{applications-Physics}. The aforementioned results are in a variety of areas in Physics (black-hole ``Islands'' quantum entanglement entropy-related Page curve, chaos and bulk viscosity), along with a novel conjecture pertaining to non-supersymmetric swampland; the connections between bulk viscosity as well as magnetic photoproduction and a paramagnetic plasma Equation of State in a top-down holographic dual of hot QCD-like theories were first discussed respectively in \cite{zeta-intermediate} and \cite{photoprod+EoS+AC3S}, and are also reviewed in \ref{applications-Physics}.}

\section{Review of the Type IIB/IIA Dual of Large-$N$ Thermal QCD-Like Theories, its ${\cal M}$-Theory Uplift and the MQGP Limit}
\label{MQGP limit}

Holographic dual of thermal QCD-like theories at finite coupling was successfully constructed in \cite{MQGP,NPB}. Since one requires finite coupling on gauge theory side therefore one is required to study the strong coupling limit of string theory i.e. ${\cal M}$-theory to be consistent with gauge - gravity duality. The same was effected via the "MQGP \footnote{Short for ${\cal M}$-theoretic Quark Gluon Plasma (QGP) - essentially implying study of QGP-like thermal QCD systems at intermediate/finite coupling, holographically, from a top-down approach.} limit" defined as \cite{MQGP,NPB}: 
\begin{equation}
\label{MQGP_limit}
g_s^{-1}\equiv {\cal O}(1)-{\cal O}(10); N_f, M \equiv {\cal O}(1), N \gg1, \frac{\left(g_s M^2\right)^{m_1}\left(g_s N_f\right)^{m_2}}{N}\ll1, m_{1,2}\in\mathbb{Z}_+\cup\left\{0\right\},
\end{equation}
where $(g_s, N, M, N_f) \equiv$(string coupling, number of color $D3$-branes, number of fractional $D3$-branes/$D5$-branes wrapping the vanishing $S^2$ of a resolved conifold, number of flavor $D7$-branes) in the type IIB dual \cite{metrics} of thermal QCD-like theories at high (i.e. above the deconfinement temperature) temperatures.

In this work, in particular in sections \ref{C3S-M7x0} and \ref{SU3-from-G2}, we use the specific values of $g_s, M, N_f$ as given in Table \tcb{2} (as also given in \cite{MQGP-chaos}) which is {\it purely motivated by the desire the make contact with real QCD} as well as to work with {\it intermediate $N$ duals of thermal QCD-like theories}. However, it should be noted that {\bf all results pertaining to the $G_2$ structure, Almost Contact 3-Structures and the resultant transverse $SU(3)$ structures of this work are valid $\forall (g_s, M, N_f)$ satisfying (\ref{MQGP_limit})}.
\begin{table}[h]
\label{Parameters-real-QCD}
\begin{center}
\begin{tabular}{|c|c|c|c|} \hline
S. No. & Parameterc & Value chosen consistent with (\ref{MQGP_limit}) & Physics reason \\ \hline
1. & $g_s$ & 0.1 & {\footnotesize QCD fine structure constant (EW scale)} \\ 
\hline
2. & $M$ & 3 & {\footnotesize Number of colors in the IR after a } \\ 
&&& {\footnotesize Seiberg-like duality cascade} \\
&&& {\footnotesize to match real QCD} \\ \hline
3. & $N_f$ & 2 or 3 & {\footnotesize u, d (and s) quarks} \\
&&& {\footnotesize - the light quarks of QCD} \\  \hline
\end{tabular}
\end{center}
\caption{Values of $g_s, M, N_f$ in the IR motivated by realistic QCD}
\end{table}
We will see that for the values of $g_s, M, N_f$ as given in Table \tcb{2} [even though this table appears in the published \cite{MQGP-chaos} with both the authors as co-authors, chronologically, it first appeared in a preliminary version of this paper that appeared as arXiv:2211.13186 [hep-th]v1],  {\bf $N=100\pm{\cal O}(1)$ is the value of $N$ picked out to obtain explicit Contact 3-Structures (and the associated transverse $SU(3)$ 3-structures)} ; a different choice of $(g_s, M = {\cal O}(1), N_f = {\cal O}(1))$ would pick out another (intermediate) $N$. 

The ${\cal M}$-theory uplift of the type IIB string dual \cite{metrics} of thermal QCD-like theories, was obtained by first constructing its  type IIA Strominger-Yau-Zaslow (SYZ) type IIA mirror via triple T duality along a delocalized  special Lagrangian (sLag) $T^3$ resolved/deformed conifold which could be identified with the $T^2$-invariant sLag of \cite{M.Ionel and M.Min-OO (2008)} with a large base ${\cal B}(r,\theta_1,\theta_2)$ \cite{NPB,EPJC-2}, and then uplifted to ${\cal M}$-theory.  As regards delocalization, (as summarized in \cite{OR4} \footnote{This is explained in \cite{delocalized-mirror-global-D5-resolved-S2}. A resolved warped {\it deformed} conifold (in the type IIB gravity dual (See Fig. \tcb{1})) does not possess an isometry along $\psi$. Therefore, to construct its type IIA SYZ mirror and its subsequent ${\cal M}$-theory uplift, for starters, one works in the delocalized limit $\psi = \langle\psi\rangle$ wherein one replaces $S^2(\theta_{1,2},\phi_{1,2})$ by $T^2(\theta_{1,2},x/y)$ via (\ref{xyz-defs}). Then, similar to \cite{delocalized-mirror-global-D5-resolved-S2} in the context of $D5$-branes wrapped around the resolved squashed $S^2$ of a resolved conifold, it can be shown that freeing the uplift of the delocalization generates a bonafide $G_2$ structure, and therefore the ${\cal M}$-theory uplift and thus its type IIA descendant, are both free of delocalization.}) the ${\cal M}$-theory uplift (excluding the higher derivative corrections) of the type IIB holographic dual of \cite{metrics} of our manuscript, was constructed in the MQGP limit in references \cite{MQGP}, \cite{NPB}, by first constructing the delocalized SYZ type IIA mirror (wherein a pair of squashed $S^2$s are replaced by a pair of $T^2$s, and the correct T-duality coordinates are identified). Analogous to \cite{delocalized-mirror-global-D5-resolved-S2}, the ${\cal M}$-theory uplift corresponds to a bona-fide $G_2$ structure satisfying the EOMs even if one removes the delocalization, i.e., take the uplift to be valid for all angles $\theta_{1,2}, \psi$.  Further, working in the aforementioned vanishing-Ouyang-embedding’s-modulus limit (essentially limiting to the first-generation quarks[+s quark]), it is evident that one will have to work near small values of $\theta_{1,2}$. {\it As an example}, we work in the neighborhood of 
\begin{eqnarray}
\label{alpha_theta_12}
& & (\theta_1, \theta_2) = \left(\frac{\alpha_{\theta_1}}{N^{1/5}}, \frac{\alpha_{\theta_2}}{N^{3/10}}\right),\nonumber\\
& & \alpha_{\theta_{1,2}}\equiv{\cal O}(1).
\end{eqnarray}
 Further, the slightly different powers of $N$ in the delocalized $\theta_{1,2}$ is also to remind that in the pair of squashed $S^2$’s, the vanishing $S^2(\theta_1,\phi_1)$ and resolved $S^2(\theta_2,\phi_2)$ are not on the “same footing”.  At the level of on-shell action, the results up to ${\cal O}(\frac{1}{N})$ are made independent of the delocalization (as explained in \cite{OR4}) by replacing the ${\cal O}(1)$ delocalization parameters $\alpha_{\theta_{1,2}}$ respectively by $N^{1/5}\sin\theta_1$ or $N^{3/10}\sin\theta_2$. One can then choose a different delocalization by then replacing $\sin\theta_{1,2}$ by
\begin{eqnarray}
\label{alpha_theta_12_prime} 
& & \left(\frac{\tilde{\alpha}_{\theta_1}}{N^{\gamma_{\theta_1}}}, \frac{\tilde{\alpha}_{\theta_2}}{N^{\gamma_{\theta_2}}}\right), \gamma_{\theta_1}\neq\frac{1}{5}, \gamma_{\theta_2}\neq\frac{3}{10};\
 \tilde{\alpha}_{\theta_{1,2}}\equiv{\cal O}(1).
\end{eqnarray}
 {\it The results pertaining to which $G_2$-structure torsion classes of the closed $M_7$ are non-vanishing, and the existence of (Almost)Contact(3)(Metric)Structures  and transverse $SU(3)$ structures of this work, remain unchanged and independent of delocalization.}

The UV-complete (unlike \cite{SS}) Type IIB string dual of \cite{metrics}, involves $N$ color $D3$-branes placed at the tip of a resolved conifold, $M$ $D5$-branes and $\overline{D5}$-branes both wrapping the vanishing $S^2$ but at antipodal points of the resolved $S^2$, and $N_f$ flavor $D7$- and $\overline{D7}$-branes ``wrapping" a non-compact four cycle $\mathbb{R}_{>0}\times S^3$ involving the vanishing $S^2$ but at antipodal points of the resolved conifold. 

SYZ mirror symmetry is triple T-duality along three isometry directions ($\phi_1$,$\phi_2$,$\psi$). By performing first T-duality along $\psi$ direction, one obtains $N$ $D4$ branes which are wrapping $\psi$ direction and $M$ $D4$-branes straddling a pair of orthogonal NS5-branes. Further, from T-dualities along $\phi_1$ and $\phi_2$, one obtains a pair of Taub-Nut spaces and $N$ $D6$ branes. Effect of triple T-dualities on the flavor $D7$ branes is that $D7$ branes are replaced by $D6$-branes. The ${\cal M}$-theory mirror of the type IIA mirror yields KK monopoles (variants of Taub-NUT spaces). Therefore, we can see that there are no branes in ${\cal M}$-theory uplift and we have ${\cal M}$-theory on a $G_2$-structure manifold with fluxes. This is summarized in Fig. \tcb{1}.

\begin{figure}
\begin{center}
\includegraphics[width=0.83\textwidth]{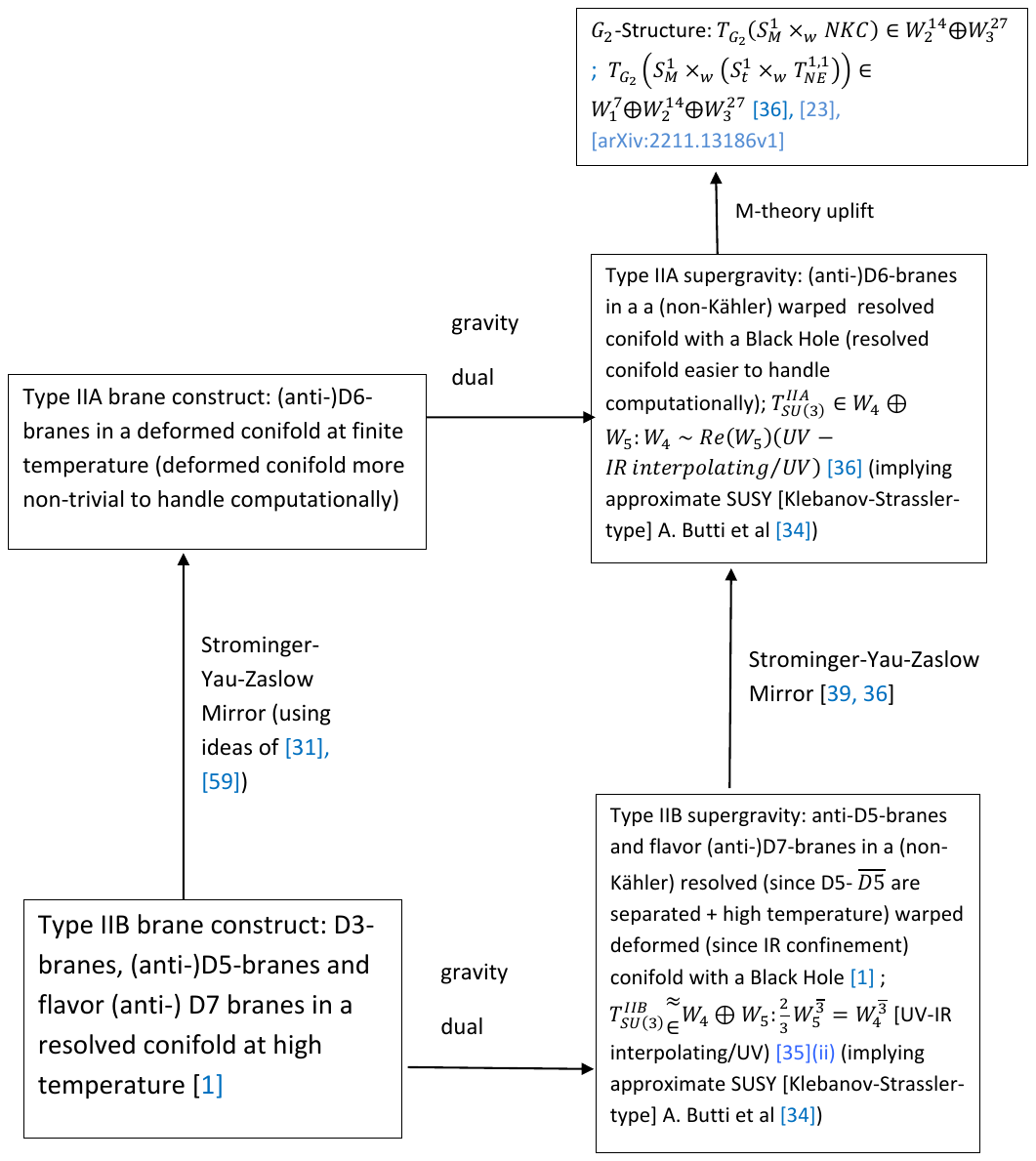}
\end{center}
\vskip -0.4in
\caption{The Status of the Type IIB/IIA/${\cal M}$-theory dual of large-N QCD at high temperature \cite{metrics}, \cite{MQGP}, \cite{NPB} inclusive of ${\cal O}(R^4)$ M-theory corrections \cite{OR4}; $T^{1,1}_{NE}$ denotes a non-Einsteinian deformation of $T^{1,1}$}
\label{Flowchart}
\end{figure}

\begin{tcolorbox}[enhanced,width=6.8in,center upper,size=fbox,
    drop shadow southwest,sharp corners]
    \begin{flushleft}
 As was explained in, e.g., \cite{Vikas+Gopal+Aalok},  after application of repeated Seiberg-like dualities at finite temperature, $N\ D3$-branes are cascaded away in the IR yielding an $SU(M)$ gauge theory that is UV-conformal, IR-confining wherein the quarks transform in the fundamental representation of flavor and color groups. As $M$ then gets identified with the number of colors in the IR , in the MQGP-limit (\ref{MQGP_limit}) $M$ can not only be taken to ${\cal O}(1)$ but in fact even the realistic-QCD-inspired value of 3. Further, the type IIB dual of \cite{metrics} is valid at {\it all} temperatures.
\end{flushleft}
\end{tcolorbox}

\section{${\cal O}(l_p^6)$ Corrections and When to Go Beyond - A Brief Review}
\label{lp6-corrections}

In this section, via two subsections, we briefly review ${\cal N}=1, D=11$ supergravity action up to terms quartic in curvature in subsection \ref{LOR4} and a competition between IR-enhancement and large-$N$ suppression thereby answering the question when one would require to go beyond the quartic-in-curvature corrections, in \ref{IR-large-N}.

\subsection{${\cal O}(l_p^6)$ terms in ${\cal N}=1, D=11$ Supergravity Action}
\label{LOR4}

The ${\cal N}=1, D=11$ supergravity action inclusive of ${\cal O}(l_p^6)$ terms is pretty well known, and has been summarized in several previous publications from our group, e.g., \cite{OR4}. Apart from the Einstein-Hilbert, the boundary Gibbons-Hawking-York and flux terms at the leading order, the higher derivative corrections start at terms quartic in the curvature (as well as terms which are cubic in curvature and quadratic in the four-form flux $G_4=dC_3$ with $C_3$ being the ${\cal M}$-theory three-form potential). The ${\cal O}(R^4)$ terms come in three varieties - the "$J_0 = t_8^2 R^4$" (see, e.g., \cite{OR4} for the definition of the $t_8$ tensor), the eleven-dimensional generalization of the eight-dimensional Eulerian density "$E_8 = \epsilon_{11}^2R^4$" ($\epsilon_{11}$ being the 11-dimensional Levi-Civita symbol) and "$X_8$" given in terms of the second and the square of the first Pontryagin classes of the 11-fold (relevant to anomaly inflow); $X_8$ was shown to vanish in \cite{MQGP} (See \cite{O(R^3G^2)} and \cite{OR4} for a discussion on a completion of the 1-loop ${\cal O}(R^4)$ in the presence of NS-NS $B$ in type IIA compatible with T duality and its ${\cal M}$-theory uplift). As in \cite{OR4}, the ${\cal O}(R^4)$ corrections are $\beta\sim l_p^6$($l_p$ being the Planckian length)-suppressed.

Now, the ${\cal M}$-theory uplift corresponding to high temperatures in QCD is given as follows \cite{MQGP}, \cite{OR4}:
\begin{eqnarray}
\label{TypeIIA-from-M-theory-Witten-prescription-T>Tc}
\hskip -0.1in ds_{11}^2 & = & e^{-\frac{2\phi^{\rm IIA}}{3}}\Biggl[\frac{1}{\sqrt{h(r,\theta_{1,2})}}\left(-g(r) dt^2 + \left(dx^1\right)^2 +  \left(dx^2\right)^2 +\left(dx^3\right)^2 \right)
\nonumber\\
& & \hskip -0.1in+ \sqrt{h(r,\theta_{1,2})}\left(\frac{dr^2}{g(r)} + ds^2_{\rm IIA}(r,\theta_{1,2},\phi_{1,2},\psi)\right)
\Biggr] + e^{\frac{4\phi^{\rm IIA}}{3}}\left(dx^{11} + A_{\rm IIA}^{F_1^{\rm IIB} + F_3^{\rm IIB} + F_5^{\rm IIB}}\right)^2,
\end{eqnarray} 
where $A_{\rm IIA}^{F^{\rm IIB}_{i=1,3,5}}$ correspond to the RR Type IIA one-form gauge field generated from the type IIB $F_{1,3,5}^{\rm IIB}$ via the SYZ mirror of the type IIB string dual \cite{metrics}. Also, $g(r) = 1 - \frac{r_h^4}{r^4}$.

The $D=11$ action  is holographically renormalizable by construction of appropriate counter terms ${\cal S}^{\rm ct}$.  It was shown in \cite{Gopal+Vikas+Aalok} that inclusive of ${\cal O}(R^4)$-corrections, the bulk on-shell $D=11$ supergravity action  is given by:
\begin{equation}
\label{on-shell-D=11-action-up-to-beta}
\hskip -0.3in S_{D=11}^{\rm on-shell} = -\frac{1}{2}\Biggl[-2S_{\rm EH}^{(0)} + 2 S_{\rm GHY}^{(0)}+ \beta\left(\frac{20}{11}S_{\rm EH}^{(1)} - 2\int_{M_{11}}\left(\sqrt{-g}\right)^{(1)}R^{(0)}
+ 2 S_{\rm GHY}^{(1)} - \frac{2}{11}\int_{M_{11}}\sqrt{-g^{(0)}}g_{(0)}^{MN}\frac{\delta J_0}{\delta g_{(0)}^{MN}}\right)\Biggr],
\end{equation}
where the superscripts "(0)" and "(1)" refer to the contributions of the relevant term at ${\cal O}(\beta^0)$ and ${\cal O}(\beta)$ respectively. The UV divergences of the on-shell action of (\ref{on-shell-D=11-action-up-to-beta}) are of the following types:
\begin{eqnarray}
\label{UV_divergences}
& &\left. \int_{M_{11}}\sqrt{-g}R\right|_{\rm UV-divergent},\ \left.\int_{\partial M_{11}}\sqrt{-h}K\right|_{\rm UV-divergent} \sim r_{\rm UV}^4 \log r_{\rm UV},\nonumber\\
& & \left.\int_{M_{11}} \sqrt{-g}g^{MN}\frac{\delta J_0}{\delta g^{MN}}\right|_{\rm UV-divergent} \sim
\frac{r_{\rm UV}^4}{\log r_{\rm UV}}.
\end{eqnarray}
It was shown in \cite{Gopal+Vikas+Aalok} that a certain linear combination of the boundary  terms: $\left.\int_{\partial M_{11}}\sqrt{-h}K\right|_{r=r_{\rm UV}}$ and $\left.\int_{\partial M_{11}}\sqrt{-h}h^{mn}\frac{\partial J_0}{\partial h^{mn}}\right|_{r=r_{\rm UV}}$ serves as the appropriate counter terms to cancel the UV divergences as given in (\ref{UV_divergences}).

Now, it was shown in \cite{OR4} that if one makes an ansatz:
\begin{eqnarray}
\label{ansaetze}
& & \hskip -0.8ing_{MN} = g_{MN}^{(0)} +\beta g_{MN}^{(1)},\nonumber\\
& & \hskip -0.8inC_{MNP} = C^{(0)}_{MNP} + \beta C_{MNP}^{(1)},
\end{eqnarray}
to be substituted into the equations of motion, one can self-consistently set $C_{MNP}^{(1)}=0$. Further, as proved in \cite{OR4} (as Lemma 1), in the neighborhood of the Ouyang embedding  of flavor $D7$-branes  (see \cite{metrics}) (that figure in the type IIB string dual of thermal QCD-like theories  at high temperatures \cite{metrics}) effected by working in the neighborhood of small $\theta_{1,2}$ (assuming a vanishingly small Ouyang embedding parameter), in the MQGP limit (\ref{MQGP_limit}), $\lim_{N\rightarrow\infty}\frac{E_8}{J_0}=0, \lim_{N\rightarrow\infty} \frac{t_8t_8G^2R^3}{E_8}=0$. Therefore, $E_8$ and $t_8^2G^2R^3$-contributions (were) are disregarded (in \cite{OR4}).

\subsection{When ${\cal O}(l_p^6)$ Is (Not) Enough}
\label{IR-large-N}

Based on the results of this paper and its applications as discussed in detail in \cite{Vikas+Gopal+Aalok}, \cite{Gopal+Vikas+Aalok}, we now address the question when it becomes necessary to go beyond ${\cal O}(R^4)$ corrections in ${\cal M}$-theory.

An extremely important lesson that we learn from the ${\cal O}(R^4)$ ${\cal M}$-theory corrections obtained in \cite{OR4}, can be abstracted from Table \ref{IR-enhancement-vs-Large-N-suppression}.
\begin{table}[h]
\begin{center}
\begin{tabular}{|c|c|c|c|} \hline
S. No. & $G^{\cal M}_{MN}$ & IR-Enhancement Factor & $N$-Suppression \\
& &  $\frac{\left(\log {\cal R}_h\right)^m}{{\cal R}_h^n}, m,n\in\mathbb{Z}^+$ & Factor \\
& & in the ${\cal O}(R^4)$ Correction & in the ${\cal O}(R^4)$ Correction   \\ \hline
1 & $G^{\cal M}_{\mathbb{R}^{1,3}}$ & $\log {\cal R}_h$ & ${N^{-\frac{9}{4}}}$ \\  \hline
2 & $G^{\cal M}_{rr, \theta_1x}$ & 1 & ${N^{-\frac{8}{15}}}$ \\  \hline
3 & $G^{\cal M}_{\theta_1z,\theta_2x}$ & ${{\cal R}_h^{-5}}$ & ${N^{-\frac{7}{6}}}$ \\  \hline
4 &  $G^{\cal M}_{\theta_2y}$ & $\log {\cal R}_h$ & ${N^{-\frac{7}{5}}}$ \\ \hline
5 &  $G^{\cal M}_{\theta_2z}$ & $\log {\cal R}_h$ & ${N^{-\frac{7}{6}}}$
\\ \hline
6 &  $G^{\cal M}_{xy}$ & $\log {\cal R}_h$ & ${N^{-\frac{21}{20}}}$ \\  \hline
7 &  $G^{\cal M}_{xz}$ &  $\left(\log {\cal R}_h\right)^3$ & ${N^{-\frac{5}{4}}}$ \\ \hline
8 &  $G^{\cal M}_{yy}$ & $\log {\cal R}_h$ & ${N^{-\frac{7}{4}}}$ \\  \hline
9 &  $G^{\cal M}_{yz}$ & $\frac{\log {\cal R}_h}{{\cal R}_h^7}$ & ${N^{-\frac{29}{12}}}$ \\ \hline
10 &  $G^{\cal M}_{zz}$ & $\log {\cal R}_h$ & ${N^{-\frac{23}{20}}}$ \\ \hline
11  &  $G^{\cal M}_{x^{10}x^{10}}$ & $\frac{\log {\cal R}_h^3}{{\cal R}_h^2}$ & ${N^{-\frac{5}{4}}}$ \\ \hline
\end{tabular}
\end{center}
\caption{IR Enhancement vs. large-$N$ Suppression in ${\cal O}(R^4)$-Corrections in the M-theory Metric in the $\psi=2n\pi, n=0,1,2$ Patches; ${\cal R}_h \equiv \frac{r_h}{{\cal R}_{D5/\overline{D5}}}<1$, ${\cal R}_{D5/\overline{D5}}$ being the $D5-\overline{D5}$ separation}
\label{IR-enhancement-vs-Large-N-suppression}
\end{table}

 One notes that in the IR: $r = \Lambda r_h, \Lambda\equiv {\cal O}(1)$, and up to ${\cal O}(\beta)$:
\begin{equation}
\label{IR-beta-N-suppressed-logrh-rh-neg-exp-enhanced}
f_{MN} \sim \beta\frac{\left(\log {\cal R}_h\right)^{m}}{{\cal R}_h^n N^{\beta_N}},\ m\in\left\{0,1,3\right\},\ n\in\left\{0,2,5,7\right\},\
\beta_N>0.
\end{equation}
Note $|{\cal R}_h|\ll1$ and as estimated in \cite{Bulk-Viscosity-McGill-IIT-Roorkee}, $|\log {\cal R}_h|\sim \kappa_{r_h}N^{\frac{1}{3}}, 0<\kappa_{r_h}<1$. This implies Planckian and large-$N$ suppression, and infra-red enhancement arising from $m,n\neq0$ in (\ref{IR-beta-N-suppressed-logrh-rh-neg-exp-enhanced}), are mutually competing effects. As shown in \cite{OR4}, choosing a hierarchy: $\beta\sim e^{-\gamma_\beta N^{\gamma_N}}$ \cite{MQGP-Page}, $\gamma_\beta,\gamma_N>0: \gamma_\beta N^{\gamma_N}>7\kappa_{r_h}N^{\frac{1}{3}} + \left(\frac{m}{3} - \beta_N\right)\log N$, ensures that the IR-enhancement does not dominate over Planckian suppression. Hence, if $\gamma_\beta N^{\gamma_N}\sim7\kappa_{r_h}N^{\frac{1}{3}}$, one would have to go to a higher order in $\beta$.


\section{Almost Contact Metric Structures - The Basics}
\label{ACM3S-basics}

Let $M_7$ be an odd-dimensional ((non-)closed) Riemannian manifold with metric $g$, then if it admits the existence of an endomorphism $J$ of the tangent bundle $T M_7$, a unit vector field $R$ (with respect to the metric $g$) called the Reeb vector field, and a one form $\sigma$ which satisfy
    $$J^2=-\mathbf{1}+R\otimes\sigma,\quad\sigma(R)=1$$
    $M_7$ is said to admit an almost contact structure $(J,R,\sigma)$ (ACS) and $\sigma$ is called the contact potential \cite{Sasaki}. The structure group of the tangent space reduces to $U(n)\times{\bf 1}$ where $2n+1$ is the dimension of $M_7$.

 A Riemannian manifold $M_7$ with an ACS $(J,R,\sigma)$ has an almost contact metric structure $(J,R,\sigma,g)$ (ACMS) if,
    $$g(Ju,Jv)=g(u,v)-\sigma(u)\sigma(v),\quad\forall u,v\in\Gamma(T M_7).$$ The fundamental two-form $\omega$ of the almost contact manifold is then defined as,
    $$\omega(u,v)=g(Ju,v),\quad\forall u,v\in\Gamma(T M_7)$$. The ACS is said to be a contact structure if $$\sigma\wedge d{\sigma}....\wedge d{\sigma}\neq 0\quad \forall{\rm points }\in M_7.$$

The ACMS determines a foliation ${\cal F}_R$ of $M_7$ by one-dimensional integral curves of $R$
w.r.t. $G_2$ metric: $ds^2 = \sigma^2 + ds_\perp^2$, where $ds_\perp^2$ is the metric on the transverse geometry of ${\cal F}_R$ induced by the ACMS on $M_7$.

 Almost contact 3-structures (AC3S) on a manifold $M_7$ are defined by three distinct ACSs $(J^{\alpha},R^{\alpha},\sigma^{\alpha}),\alpha=1,2,3$ on $M_7$ which satisfy \cite{Kuo-AC3S},
    \begin{eqnarray*}
    \label{AC3S-conditions}
        J^{\gamma}&=&J^{\alpha}J^{\beta}-R^{\alpha}\otimes\sigma^{\beta}=J^{\beta}J^{\alpha}-R^{\beta}\otimes\sigma^{\alpha}\nonumber\\
        R^{\gamma}&=&J^{\alpha}(R^{\beta})=-J^{\beta}(R^{\alpha})\nonumber\\
        \sigma^{\gamma}&=&\sigma^{\alpha}\circ J^{\beta}=-\sigma^{\beta}\circ J^{\alpha}\\
        \sigma^{\alpha}(R^{\beta})&=&\sigma^{\beta}(R^{\alpha})=0
    \end{eqnarray*}
where $\{\alpha,\beta,\gamma\}$ are cyclic permutation of $\{1,2,3\}$. An AC3S consisting of three contact structures satisfying (\ref{AC3S-conditions}), defines a 3-Sasakian geometry \cite{3-Sasakian-geometry}.
    
$M_7$ admitting  AC3S must have dimensionality $4n+3, n\in\mathbb{Z}^+$, and the structure group reduces to $Sp(n)\times\mathbf{1}_3$. An almost contact metric 3-structure (ACM3S) on a Riemannian manifold $M_7$ with metric $g$ possesses AC3S satisfying,
    $$g(J^{\alpha}u,J^{\alpha}v)=g(u,v)-\sigma^{\alpha}(u)\sigma^{\alpha}(u),\quad\forall u,v\in\Gamma(T M_7)$$ for $\alpha\in{1,2,3}$. With $J^\alpha(u) = R^\alpha\times_\Phi u$, one can see that with $R^\alpha.R^\beta=\delta^{\alpha\beta}$,  $R^1, R^2, \left(R^1\times R^2\right)$ provide  AC3S.

\section{$G_2$-Structure of  $M_7=S^1_{\cal M}\times_w\left(S^1_{\rm thermal}\times_w M_5\right)$}
\label{G2-M7x0}

In this section we now proceed to determine the $G_2$-structure torsion classes $\tau_p$'s of the seven-fold $M_7=S^1_{\cal M}\times_w\left(S^1_{\rm thermal}\times_w M_5\right)$, $M_5$ being a non-Einsteinian generalization of $T^{1,1}$, and close to the Ouyang embedding  of the flavor $D7$-branes in the parent type IIB dual in the limit of very-small-Ouyang-embedding parameter limit ($|\mu_{\rm Ouyang}|\ll1$).

Given that the adjoint of $SO(7)$ decomposes under $G_2$ as ${\bf 21}\rightarrow{\bf 7}\oplus{\bf 14}$ where ${\bf 14}$ is the adjoint representation of $G_2$, one obtains four $G_2$-structure torsion classes:
\begin{equation}
\label{T-G2}
\tau \in \Lambda^1 \otimes g_2^\perp = W_1 \oplus W_7 \oplus W_{14} \oplus W_{27} = \tau_0 \oplus \tau_1 \oplus \tau_2 \oplus \tau_3,
\end{equation}
$g_2^\perp$ being the orthogonal complement of $g_2$, the subscript $a$ in $W_a$ denoting the dimensionality of the torsion class $W_a$, and $p$ in $\tau_p$ denoting the rank of the associated differential form. 

\noindent {\it Lemma 1}: Near the Ouyang embedding locus of the flavor $D7$-branes and the $\psi=2n\pi, n=0, 1, 2$-coordinate patches, assuming $|\mu_{\rm Ouyang}|\ll1$, in the MQGP limit (\ref{MQGP limit}), 
\begin{equation}
\label{G2-torsion}
\tau\left(M_7\right) = \tau_1\oplus\tau_2\oplus\tau_3.
\end{equation}

\noindent {\it Proof}: We set
\begin{equation}
\label{e1-const-r}
e^1 = \sqrt{G^{\cal M}_{x^0x^0}}dx^0,
\end{equation}
and use the following definitions from \cite{OR4},
\begin{eqnarray}
\label{Theta_ia+X_a+Y_a+Z_a}
& & \hskip -0.3 in d\theta_{i=1/2} = \sum_{a=2}^6\Theta_{ia}e^a,\  dx = \sum_{a=2}^6{\cal X}_ae^a,\  dy = \sum_{a=2}^6{\cal Y}_ae^a,\
dz = \sum_{a=2}^6{\cal Z}_ae^a.
\end{eqnarray} 
 The delocalized $T^3(x, y, z)$ coordinates are defined near $r=\langle  r \rangle\in$IR and $\langle \theta_{1,2} \rangle$ close to the Ouyang embedding of the flavor $D7$-branes in the parent type IIB dual \cite{metrics}, as \cite{MQGP} \footnote{As explained in \cite{Knauf-thesis}, the $T^3$-valued $(x, y, z)$ are defined via:
 {\footnotesize
\begin{eqnarray*}
\label{xyz-definitions}
& & \phi_1 = \langle \phi_1 \rangle + \frac{x}{\sqrt{h_2}\left[h(\langle r \rangle,\langle \theta_{1,2} \rangle)\right]^{\frac{1}{4}} \sin\langle \theta_{1} \rangle\ \langle r \rangle},\nonumber\\
& & \phi_2 = \langle \phi_2 \rangle + \frac{y}{ \sqrt{h_4}
\left[h( \langle r \rangle,\langle \theta_{1,2} \rangle)\right]^{\frac{1}{4}}\sin\langle \theta_{2} \rangle\ \langle r \rangle}\nonumber\\
& & \psi = \langle \psi \rangle + \frac{z}{\sqrt{h_1} \left[h( \langle r \rangle,\langle \theta_{1,2} \rangle)
\right]^{\frac{1}{4}}\ \langle r \rangle}.
\end{eqnarray*}
}
In the IR, it was shown \cite{theta0-theta} that the delocalized $\langle \theta_{1,2} \rangle$ can be promoted to global $\theta_{1,2}$; we do so in all the results in the paper.}:
\begin{eqnarray}
\label{xyz-defs}
& &   dx = \sqrt{h_2}\biggl[h\biggl(\langle r \rangle,\langle \theta_{1,2} \rangle\biggr)\biggr]^{\frac{1}{4}} \sin\langle \theta_{1} \rangle\ \langle r \rangle  d\phi_1,\nonumber\\
& & dy = \sqrt{h_4}
\biggl[h\biggl( \langle r \rangle,\langle \theta_{1,2} \rangle\biggr)\biggr]^{\frac{1}{4}}\sin\langle \theta_{2} \rangle\ \langle r \rangle d\phi_2,\nonumber\\
& &  dz = \sqrt{h_1} \biggl[h\biggl( \langle r \rangle,\langle \theta_{1,2} \rangle\biggr)\biggr]^{\frac{1}{4}}\ \langle r \rangle d\psi,\nonumber\\
& &
\end{eqnarray}
$h(\langle r \rangle,\langle \theta_{1,2} \rangle)$ being the delocalized warp factor \cite{metrics}:
{\footnotesize
\begin{eqnarray}
\label{eq:h}
&& h(\langle r \rangle,\langle \theta_{1,2} \rangle) =\frac{L^4}{\langle r \rangle^4}\Bigg[1+\frac{3g_sM_{\rm eff}^2}{2\pi N}{\rm log}\langle r \rangle\left\{1+\frac{3g_sN^{\rm eff}_f}{2\pi}\left({\log} \langle r \rangle+\frac{1}{2}\right)+\frac{g_sN^{\rm eff}_f}{4\pi}{\rm log}\left({\sin}\frac{\langle \theta_{1} \rangle}{2}
{\rm sin}\frac{\langle \theta_{2} \rangle}{2}\right)\right\}\Bigg],\nonumber\\
\end{eqnarray}
}
wherein $L\equiv 4\pi g_s N \alpha'^2$, with the effective number of fractional $D3$-branes, $M_{\rm eff}$,  and the effective number of flavor $D7$-branes, $N_f^{\rm eff}$, defined, e.g., in \cite{Vikas+Gopal+Aalok}.  The squashing factors are defined below \cite{metrics}:
\begin{equation}
\label{h_{1,2,4}-defs}
h_1 = \frac{1}{9} + {\cal O}\left(\frac{g_sM^2}{N}\right),\ h_2 = \frac{1}{6} + {\cal O}\left(\frac{g_sM^2}{N}\right),\
h_4 = h_2 + \frac{4a^2}{\langle r\rangle^2},
\end{equation}
($a$ being the radius of the blown-up $S^2$).

One can show that
\begin{itemize}
\item
\begin{eqnarray}
\label{Omega^a_bc-i}
& & de^a = \Omega^a_{bc}e^b\wedge e^c,
\end{eqnarray}
where $a, b, c = 2,...,6$ and, the ``generalized spin connection'' $\Omega^a_{bc}$ are defined as:
{\footnotesize
\begin{eqnarray}
\label{Omega^a_bc-ii}
& & \hskip -0.8in \Omega^a_{bc}(r={\rm constant}\in{\rm IR}) = \partial_{[\theta_2}e^a_{\ \theta_1]}\Theta_{2b}\Theta_{1c}
+ \partial_{[x}e^a_{\ \theta_1]}{\cal X}_b\Theta_{1c} + 
\partial_{[y}e^a_{\ \theta_1]}{\cal Y}_b\Theta_{1c} + \partial_{[z}e^a_{\ \theta_1]}{\cal Z}_{b}\Theta_{1c}
+ \partial_{[x}e^a_{\ \theta_2]}{\cal X}_b\Theta_{2c}\nonumber\\
& &  \hskip -0.8in  + \partial_{[y}e^a_{\ \theta_2]}{\cal Y}_b\Theta_{2c} + \partial_{[z}e^a_{\ \theta_2]}{\cal Z}_{2b}\Theta_{2c}
+ \partial_{[y}e^a_{\ x]}{\cal Y}_b{\cal X}_{c} +\partial_{[z}e^a_{\ x]}{\cal Z}_b{\cal X}_{c} +\partial_{[z}e^a_{\ y]}{\cal Z}_b{\cal Y}_{c}.
\end{eqnarray}
}
\item
Similarly,
\begin{eqnarray}
\label{de1-i}
& & de^1 (r={\rm constant}\in{\rm IR}) = \Omega^1_ae^{a1},
\end{eqnarray}
where,
{
\begin{eqnarray}
\label{de1-ii}
& & \Omega^1_a(r={\rm constant}\in{\rm IR}) = \frac{1}{2G^{\cal M}_{x^{0}x^{0}}}\left(\partial_{\theta_1}G^{\cal M}_{x^{0}x^{0}}\Theta_{1a} + \partial_{\theta_2}G^{\cal M}_{x^{0}x^{0}}\Theta_{2a}+\partial_xG^{\cal M}_{x^0x^0}{\cal X}_a\right.\nonumber\\
& & \left. + \partial_yG^{\cal M}_{x^{0}x^{0}}{\cal Y}_a + \partial_zG^{\cal M}_{x^{0}x^{0}}{\cal Z}_a\right).
\end{eqnarray}
}
\item
Also,  
\begin{equation}
\label{e7}
e^7 = \sqrt{G^{\cal M}_{x^{10}x^{10}}}dx^{10},\ de^7(r={\rm constant} = \langle r\rangle \in{\rm IR}) = \Omega^7_ae^{a7},
\end{equation}
where,
{
\begin{eqnarray}
\label{Omega^7_a}
& & \Omega^7_a(r={\rm constant} = \langle r \rangle \in{\rm IR}) = \frac{1}{2G^{\cal M}_{x^{10}x^{0}}}\left(\partial_{\theta_1}G^{\cal M}_{x^{10}x^{10}}\Theta_{1a} + \partial_{\theta_2}G^{\cal M}_{x^{10}x^{10}}\Theta_{2a}+\partial_xG^{\cal M}_{x^{10}x^{10}}{\cal X}_a\right.\nonumber\\
& & \left. + \partial_yG^{\cal M}_{x^{10}x^{10}}{\cal Y}_a + \partial_zG^{\cal M}_{x^{10}x^{10}}{\cal Z}_a\right).
\end{eqnarray}
}
\end{itemize}

After a detailed but fairly straightforward computation, one realizes (in the following, $\sim$ represents equality up to ${\cal O}(1)$ multiplicative factors):
\begin{eqnarray}
\label{similar-Omega^1_a-beta0}
& & \left(\Omega^1_2\right)\sim \left(\Omega^1_3\right)\sim
-\left(\Omega^1_5\right)\sim - \left(\Omega^7_2\right),\nonumber\\
& & \left(\Omega^7_3\right)\sim - \left(\Omega^7_4\right)\sim
- \left(\Omega^7_5\right);\nonumber\\
& & \left(\Omega^2_{23}\right)\sim -\left(\Omega^2_{25}\right)
\sim - \left(\Omega^2_{26}\right)\sim \left(\Omega^2_{56}\right)\sim \left(\Omega^2_{35}\right)\sim -\left(\Omega^2_{36}\right),\nonumber\\
& & \left(\Omega^2_{34}\right)\sim \left(\Omega^2_{45}\right)\sim
\left(\Omega^2_{46}\right);\nonumber\\
& & \left(\Omega^3_{23}\right)\sim \left(\Omega^3_{24}\right)
\sim \left(\Omega^3_{25}\right)\sim \left(\Omega^3_{26}\right)
\sim \left(\Omega^3_{35}\right)\sim \left(\Omega^3_{36}\right)
\sim \left(\Omega^3_{56}\right),\nonumber\\
& & \left(\Omega^3_{34}\right)\sim \left(\Omega^3_{45}\right)\sim \left(\Omega^3_{46}\right);
\nonumber\\
& & \left(\Omega^4_{23}\right)\sim \left(\Omega^4_{25}\right)
\sim \left(\Omega^4_{26}\right)\sim \left(\Omega^3_{35}\right)
\sim \left(\Omega^4_{36}\right)\sim \left(\Omega^4_{56}\right),\nonumber\\
& & \left(\Omega^4_{34}\right)\sim \left(\Omega^4_{36}\right)
\sim \left(\Omega^4_{45}\right)\sim \left(\Omega^4_{46}\right);\nonumber\\
& & \left(\Omega^5_{23}\right)\sim \left(\Omega^5_{25}\right)
\sim  \left(\Omega^5_{26}\right)\sim  \left(\Omega^5_{35}\right)
\sim -  \left(\Omega^5_{36}\right)\sim  \left(\Omega^5_{56}\right),
\nonumber\\
& &  \left(\Omega^5_{24}\right)\sim  \left(\Omega^5_{34}\right)
\sim  \left(\Omega^5_{45}\right)\sim  \left(\Omega^5_{46}\right)\sim  \left(\Omega^6_{24}\right)\sim \left(\Omega^6_{34}\right)
\sim \left(\Omega^6_{45}\right)\sim \left(\Omega^6_{46}\right),\nonumber\\
& &  \left(\Omega^6_{23}\right)\sim -\left(\Omega^6_{25}\right)
\sim -\left(\Omega^6_{26}\right)\sim \left(\Omega^6_{35}\right)
\sim -\left(\Omega^6_{36}\right)\sim - \left(\Omega^6_{56}\right).
\end{eqnarray}
One consequently obtains:
{
\begin{eqnarray}
& & 
\Phi \wedge *_7 d\Phi=e^{127456}\left(2 \Omega^7_2 - 2 \Omega^2_{23} - 2 \Omega^5_{23}\right)+e^{127356}\left(-2 \Omega^7_2 +\Omega^3_{23} - \Omega^5_{23}\right)+e^{127346}\left(2 \Omega^7_2 + \Omega^2_{23} + 2 \Omega^5_{23}\right)\nonumber\\
& & +e^{127345}\left(- 2 \Omega^2_{23} - \Omega^5_{23}\right)+e^{234567}\left( \Omega^2_{23}+\Omega^3_{24} + \Omega^5_{23}\right)+e^{123456}\left(-\Omega^2_{23} - \Omega^3_{23}+\Omega^5_{23}\right)+e^{134567} \Omega^5_{23}\nonumber\\
& & \approx e^{234567}\Omega^3_{24}-e^{123456}\Omega^3_{23}\ {\rm in\ the\ MQGP\ limit}\ (\ref{MQGP_limit}),
\end{eqnarray}
}
implying,
\begin{equation}
\label{tau1}
\tau_1 = \frac{1}{12}*_7\left(\Phi \wedge *_7 d\Phi\right) \sim e^1\Omega^3_{24}-e^7\Omega^3_{23}.
\end{equation}

Also \footnote{Thanks to S. Sarkar for pointing out an error in a previous version of this result.}, in the MQGP limit (\ref{MQGP_limit})
\begin{equation}
\label{tau0}
\tau_0 = d\Phi\lrcorner*_7\Phi =0.
\end{equation}

To obtain $\tau_2$, one notes that:
\begin{eqnarray}
\label{tau1lrcornerPhi}
& & \tau_1\lrcorner\Phi \sim \left(e^{27} + e^{35} - e^{46}\right)\Omega^3_{24} - J \Omega^3_{23},
\end{eqnarray}
where $J = e^{12} + e^{34} + e^{56}$. 

Similarly, to obtain $\tau_3$, one notes that:
\begin{eqnarray}
\label{tau1lrcornerPsi}
& & \tau_1\lrcorner*_7\Phi \sim \left(e^2\wedge J + e^{457} + e^{367}\right)\Omega^3_{24}
- \left(e^{246} - e^{235} - e^{456}\right)\Omega^3_{23}.
\end{eqnarray}
Therefore, this proves (\ref{G2-torsion}) and Lemma 1.

It should be noted that Lemma 1 is applicable for $\forall (g_s, M, N_f, N)$ satisfying the MQGP limit (\ref{MQGP_limit}).

\section{Almost Contact  3-Structures}
\label{ACM3S-constructions}

In this section, in \ref{ACM3S-M7-x0}, we provide explicit construction of Almost Contact 3-Structures at fixed IR-valued $r$ supported on $M_7$ (which is a warped product of the ${\cal M}$-theory circle and $M_6$ where $M_6$ is a warped product of the thermal circle and a non-Einsteinian generalization of $T^{1,1}$). As a small aside, we also provide a similar construction of AC3S for $\tilde{M}_7$ which is a warped product of the ${\cal M}$-theory circle and a non-K\"{a}hler conifold $\tilde{M}_6$. In \ref{AC3S-M7x0}, we show that the ACS of \ref{ACM3S-M7-x0} is in fact an Almost Contact Metric Structure (ACMS).

\subsection{Almost Contact 3-Structures arising from $\tau_{i=1, 2, 3}(M_7)$}
\label{ACM3S-M7-x0}

From (\ref{e1-const-r}), (\ref{e7}) and the co-frames of $\mathbb{R}_{\geq0}\times M_5, M_5$ being the non-Einsteinian deformation of $T^{1,1}$, we arrive at the following lemma:\\
\noindent {\it Lemma 2}: Near the Ouyang embedding of the flavor $D7$-branes in the parent type IIB dual of \cite{metrics} near the $\psi=2n\pi, n=0, 1, 2$-coordinate patches in the MQGP limit, one obtains Almost Contact 3-Structures $(J^\alpha, R^\alpha, \sigma^\alpha), \alpha = 1, 2, 3$:
\begin{eqnarray}
\label{ACM3S-x0}
& & \hskip -0.3in  R^1= \sqrt{G^{{\cal M}\ 00}}\partial_{x^0},\nonumber\\
& & \hskip -0.3in  R^2 = \sqrt{G^{{\cal M}\ x^{10}x^{10}}}\partial_{x^{10}},\nonumber\\
& & \hskip -0.3in  R^3 = \Phi^\mu_{x^0x^{10}}\sqrt{G^{{\cal M}\ 00}G^{{\cal M}\ x^{10}x^{10}}}\partial_\mu  = -G^{\mu\nu}_{\cal M} \sqrt{G^{{\cal M}\ 00}G^{{\cal M}\ x^{10}x^{10}}}\left(e^{217}\right)_{\nu x^0 x^{10}}\partial_\mu = -G^{\mu\nu}_{\cal M}e^2_\nu\partial_\mu,
\end{eqnarray}
where $\mu = m, x^0, x^{10}; m = \theta_1, \theta_2, x, y, z$.
Further, $\sigma^\alpha = G^{\cal M}_{mn}R_\alpha^n dx^m$ implying
\begin{equation}
\label{ACM3S-x0-sigmas}
(\sigma^1, \sigma^2, \sigma^3) = (e^1, e^7, -e^2);
\end{equation}
the fundamental two-forms $\omega^\alpha_\Phi = d\sigma^\alpha = i_{R^\alpha}\Phi$, which using (\ref{Omega^a_bc-i}), (\ref{de1-i}) and (\ref{e7}), can be evaluated. One hence also sees that $i_{R^\alpha}\omega^\alpha=0$, which as mentioned in footnote 2, mimics the condition of incompressibility of fluids. We demonstrate the existence of Contact 3-Structures in \ref{C3S-M7x0}.

One can similarly see that near the Ouyang embedding of the flavor $D7$-branes in the parent type IIB dual of \cite{metrics} near the $\psi=2n\pi, n=0, 1, 2$-coordinate patches, the non-compact $\tilde{M}_7$ - a warped product of the ${\cal M}$-theory circle and a non-K\"{a}hler conifold $\tilde{M}_6$ -supports Almost Contact 3-Structures:
{\footnotesize
\begin{eqnarray}
\label{ACM3S-r}
& & \hskip -0.3in  R^1= \sqrt{G^{{\cal M}\ rr}}\partial_r,\nonumber\\
& &\hskip -0.3in  R^2 = \sqrt{G^{{\cal M}\ x^{10}x^{10}}}\partial_{x^{10}},\nonumber\\
& &\hskip -0.3in  R^3 = \Phi^\mu_{\ rx^{10}}\sqrt{G^{{\cal M}\ rr}G^{{\cal M}\ x^{10}x^{10}}}\partial_\mu  = -G^{\mu\nu}_{\cal M} e^{-\frac{2}{3}\Phi^{\rm IIA}}\sqrt{G^{{\cal M}\ rr}}G^{{\cal M}\ x^{10}x^{10}}\left(e^{217}\right)_{\nu r x^{10}}\partial_\mu = -e^{-\frac{2}{3}\Phi^{\rm IIA}}\sqrt{G^{{\cal M}\ x^{10}x^{10}}}e^{2,\mu}\partial_{\mu}.
\end{eqnarray}
}
Further, $\sigma^\alpha = G^{\cal M}_{mn}R_\alpha^n dx^m$; fundamental two-forms $\omega^\alpha = d\sigma^\alpha = i_{R^\alpha}\Phi$:
\begin{equation}
\label{sigma123-M7r}
(\sigma^1, \sigma^2, \sigma^3) = (e^1, e^7, -e^{-\frac{2}{3}\Phi^{\rm IIA}}\sqrt{G^{{\cal M}\ x^{10}x^{10}}}e^2).
\end{equation}

\subsection{Almost Contact Metric 3-Structures on $M_7$}
\label{AC3S-M7x0}

In this section, we prove the lemma: \\
\noindent {\it Lemma 3}: The Almost Contact 3-Structures of \ref{ACM3S-M7-x0} in fact corresponds to Almost Contact Metric 3-Structures $(J^\alpha, R^\alpha, \sigma^\alpha, g)$. 

\noindent {\it Proof}: We will explicitly show the existence of Almost Contact Metric Structure (ACMS) for $(R^1,\sigma^1)$ in (\ref{ACM3S-x0-sigmas}). 

An we know, ACS is ACMS provided $g(Ju, Jv) = g(u,v) - \sigma(u)\sigma(v)$. We now show that the same is satisifed by $(R^1,\sigma^1)$. In components, the aforementioned requirement is written as:
\begin{eqnarray}
\label{ACM3S-i}
& & g_{mn}J^{m\ (a)}_{\ \ m_1}J^{n\ (a)}_{\ \ n_1} = g_{m_1n_1} - \sigma^{(a)}_{m_1}\sigma^{(a)}_{n_1}\nonumber\\
& & {\rm or}\ g_{mn}\Phi^m_{\ m_1l}\Phi^n_{\ n_1\tilde{l}}R^{l\ (a)}R^{\tilde{l}\ (a)} = g_{m_1n_1} - \sigma^{(a)}_{m_1}\sigma^{(a)}_{n_1}.
\end{eqnarray}

\begin{itemize}

\item $m_1 = n_1 = x^0$: The RHS of (\ref{ACM3S-i}) in this case is null. The LHS of (\ref{ACM3S-i}) will be proportional to $g_{mn}\Phi^m_{\ x^0x^0}\Phi^n_{\ x^0x^0}=0$ and therefore (\ref{ACM3S-i}) checks out.

\item
$m_1=n_1=x^{10}$: The RHS of  (\ref{ACM3S-i}) in this case is $g_{x^{10}x^{10}}$. The LHS of 
 (\ref{ACM3S-i}) is:
\begin{eqnarray}
\label{ACM3S-ii}
& & g^{mn}\Phi_{mx^{10}x^0}\Phi_{nx^{10}x^0}\left(R^{x^0}\right)^2
= g^{mn}\left(e^{217}\right)_{mx^{10}x^0}\left(e^{217}\right)_{nx^{10}x^0}\left(R^{x^0}\right)^2 = g^{mn}e^2_me^2_n g_{x^{10}x^{10}} = g_{x^{10}x^{10}}.\nonumber\\
& & 
\end{eqnarray}
 Therefore, (\ref{ACM3S-i}) checks out.

\item $m_1, n_1 \neq x^0, x^{10}$ The LHS of (\ref{ACM3S-i}) yields $g_{m_1n_1}$. The LHS of (\ref{ACM3S-i}) obtains:
\begin{eqnarray}
\label{ACM3S-iii}
& & g_{mn}\Phi^m_{\ m_1x^0}\Phi^n_{\ n_1x^0}\left(R^{x^0}\right)^2 = \left(e^{27} + e^{35} - e^{46}\right)_{nm_1}\left(e^{27} + e^{35} - e^{46}\right)^n_{\ n_1}\left(e^1_{x^0}\right)^2\left(R^{x^0}\right)^2.\nonumber\\
& & 
\end{eqnarray}
Now (\ref{ACM3S-iii}) will involve:
\begin{eqnarray}
\label{ACM3S-iv}
& (a) & \left(e^{27}\right)_{m_1n}\left(e^{27}\right)^n_{n_1} = e^2_{m_1}e^7_nE^{2n}e^7_{n_1} - e^2_ne^7_{m_1}E^{2n}e^7_{n_1} + e^7_ne^2_{m_1}E^{7n}e^2_{m_1}
= 0 - 0 + e^2_{m_1}e^2_{n_1};\nonumber\\
& (b) & \left(e^{35}\right)_{m_1n}\left(e^{35}\right)^n_{n_1} = e^3_nE^{3n}e^5_{m_1}e^5_{n_1} + e^5_nE^{5n}e^3_{m_1}e^3_{n_1} - e^3_nE^{5n}e^5_{m_1}e^3_{n_1} = e^5_{m_1}e^5_{n_1} + e^3_{m_1}e^3_{n_1} - 0;\nonumber\\
& (c) &  \left(e^{46}\right)_{m_1n}\left(e^{46}\right)^n_{n_1} =e^4_nE^{4n}e^6_{m_1}e^6_{n_1} + e^6_nE^{6n}e^4_{m_1}e^4_{n_1} - e^4_nE^{6n}e^6_{m_1}e^4_{n_1} = e^6_{m_1}e^6_{n_1} + e^4_{m_1}e^4_{n_1} - 0.\nonumber\\
& &  
\end{eqnarray}
One hence obtains: 
\begin{eqnarray}
\label{ACM3S-v}
& & LHS = \left(\sum_{a=2}^6 e^a_{m_1}e^a_{n_1}\right)\left(e^1_{x^0}\right)^2\left(R^{x^0}\right)^2 = g_{m_1n_1},
\end{eqnarray}
and therefore, (\ref{ACM3S-i}) checks out.

\item
$m_1=x^0, n_1=x^{10}$ The LHS of (\ref{ACM3S-i}) then becomes proportional to $\Phi_{mx^0x^0}\Phi^m_{\ x^{10}x^0}=0$ and the RHS of (\ref{ACM3S-i}) is the difference of $g_{x^0x^{10}}=0$ and  $\sigma^{(1)}_{x^0}\sigma^{(1)}_{x^{10}}=0$. Therefore,  (\ref{ACM3S-i}) checks out. Similarly,  for $m_1 = x^{10}, n_1=x^0$.

\item
One can similarly argue that the LHS and RHS of (\ref{ACM3S-i}) vanish identically for $m_1/n_1=x^0, n_1/m_1\neq x^{0, 10}$.

\item
$m_1 = x^{10}, n_1\neq x^{0,10}$ The LHS of (\ref{ACM3S-i}) becomes 
\begin{eqnarray}
\label{x10n1-i}
& & g^{n\tilde{n}}\Phi_{nx^{10}x^0}\Phi_{\tilde{n}n_1x^0}\left(R^{x^0}\right)^2 = g^{n\tilde{n}} \left(e^{27}\right)_{nx^{10}}\left(e^{27} + e^{35} - e^{46}\right)_{\tilde{n}m} 
\end{eqnarray}
Now,
\begin{eqnarray}
\label{x10n1-ii}
& (a) & g^{n\tilde{n}}\left(e^{27}\right)_{nx^{10}}\left(e^{27}\right)_{\tilde{n}n_1} = g^{n\tilde{n}} \left(e^2_ne^2_{\tilde{n}}e^7_{x^{10}}e^7_{n_1} - e^2_ne^7_{\tilde{n}}e^2_{n_1}e^7_{x^{10}} - e^7_ne^2_{\tilde{n}}e^2_{x^{10}}x^7_{n_1} + e^2_{x^{10}}e^2_{n_1}e^7_n e^7_{\tilde{n}}\right) \nonumber\\
& & = \left(e^7_{x^{10}}e^7_{n_1} + e^2_{n_1} e^2_{x^{10}}\right) = 0 ({\rm as}\ e^2_{x^{10}} = e^7_{n_1}=0);\nonumber\\
& (b) &  g^{n\tilde{n}} e^{27}_{nx^{10}}e^{35}_{\tilde{n}n_1} =g^{n\tilde{n}} e^{27}_{nx^{10}}e^{46}_{\tilde{n}n_1} = 0 ({\rm as}\ g^{mn}e^a_me^b_n = \delta^{ab}).
\end{eqnarray}
Also, as $g_{x^{10}n_1} = \sigma^{(1)}_{x^{10}} = 0$,  (\ref{ACM3S-i}) checks out.
\end{itemize}

One can similarly argue that the Almost Contact Structure corresponding to $\sigma^{2,3}$ of (\ref{ACM3S-x0-sigmas}), corresponds to Almost Contact Metric Structure.

\section{Contact 3-Structures on  $M_7$}
\label{C3S-M7x0}

As $\sigma\wedge\omega^3=0$ in \ref{ACM3S-M7-x0} in the $\psi=2n\pi, r=$constant-coordinate patches, the same does not provide a contact structure. However, remarkably, by choosing, e.g., the QCD-motivated values $M=3, N_f=3$ or $2$ and $g_s\sim g_{\rm QCD}^2\sim 0.1$ (Table \tcb{2}),  in this section we numerically show the emergence of a {\it finite-$N$ (i.e., at intermediate coupling)} Contact Structure on the seven-fold $M_7$, for $N=100\pm{\cal O}(1)$. Motivated by real QCD, this Contact Structure can be shown to exist  not only for Table \tcb{2}, but $\forall (g_s, M, N_f, N)$ in the intermediate-$N$ MQGP limit:
\begin{eqnarray}
\label{intermediate-N-MQGP-limit}
& &  g_s^{-1}\equiv{\cal O}(1)-{\cal O}(10), M\sim{\cal O}(1), N_f\sim{\cal O}(1);
\nonumber\\
& & {\rm intermediate\ N}:\ \frac{\left(g_s M^2\right)^{m_1}\left(g_s N_f\right)^{m_2}}{N}<1, m_{1,2}\in\mathbb{Z}_+\cup\left\{0\right\}.
\end{eqnarray}   

We now prove the lemma: \\
\noindent {\it Lemma 4}: Near the $\psi=2n\pi$-coordinate patch, in the IR, \\ $(\sigma^1,\sigma^2,\sigma^3) = \left(\alpha_1 e^1 + \alpha_3 e^3 + \alpha_7 e^7, \beta_1 e^1 + \beta_4 e^4 + \beta_7 e^7,\sigma^3\right), $ $\{\alpha_{i=1, 2, 3}\}, \{\beta_{j=1, 2, 3}\}\in\mathbb{R}$, with $\sigma^3(R^1\times_\Phi R^2)=1$ with\\
(a)
\begin{eqnarray}
\label{contact-x0-5}
& & R^x_1=\frac{\alpha_3 e^3_{\theta_1}}{G^{\cal M}_{x\theta_1}};\nonumber\\
& & R^y_1=\frac{\alpha_3 (e^3_{\theta_2} G^{\cal M}_{x\theta_1}-e^3_{\theta_1} G^{\cal M}_{x\theta_2})}{G^{\cal M}_{x\theta_1}
   G^{\cal M}_{y\theta_2}};\nonumber\\
& & R^z_1=\frac{\alpha_3 (e^3_y G^{\cal M}_{x\theta_1}-e^3_{\theta_1} G^{\cal M}_{xy})}{G^{\cal M}_{x\theta_1} G^{\cal M}_{yz}};\nonumber\\
& & R^{\theta_1}_1=\frac{\alpha_3 (-e^3_{\theta_1} G^{\cal M}_{x\theta_2} G^{\cal M}_{yz} G^{\cal M}_{xz}-e^3_{\theta_1}
   G^{\cal M}_{xz} G^{\cal M}_{xy} G^{\cal M}_{z\theta_2}-e^3_x G^{\cal M}_{x\theta_1} G^{\cal M}_{yz} G^{\cal M}_{z\theta_2}+e^3_y
   G^{\cal M}_{x\theta_1} G^{\cal M}_{xz} G^{\cal M}_{z\theta_2}+e^3_z G^{\cal M}_{x\theta_1} G^{\cal M}_{x\theta_2}
   G^{\cal M}_{yz})}{G^{\cal M}_{x\theta_1} G^{\cal M}_{x\theta_2} G^{\cal M}_{yz} G^{\cal M}_{z\theta_1}};\nonumber\\
& & R^{\theta_2}_1=\frac{\alpha_3 (e^3_{\theta_1} G^{\cal M}_{xz} G^{\cal M}_{xy}+e^3_x G^{\cal M}_{x\theta_1}
   G^{\cal M}_{yz}-e^3_y G^{\cal M}_{x\theta_1} G^{\cal M}_{xz})}{G^{\cal M}_{x\theta_1} G^{\cal M}_{x\theta_2} G^{\cal M}_{yz}},
\end{eqnarray}

(b) $\alpha_1\sim\alpha_3\sim\alpha_7; \beta_1\sim\beta_4\sim\beta_7$ with $\sim$ implying equality up to ${\cal O}(1)$ terms, and $R_2=R_1(e^3\rightarrow e^4,\ \alpha_3\rightarrow\beta_4)$, \\
provide Contact 3-Structures. As mentioned in Sec. \ref{introduction}, this mimics a three-tuple of monopole-like solutions \footnote{For $\sigma: \sigma\wedge d\sigma\neq0$ on a three-dimensional contact manifold $M_3$, "helicity" is defined as ${\cal H}(\xi) \equiv \int_{M_3}\sigma\wedge d\sigma,$ where $\xi$ is a divergence-less null-homologous vector field on $M_3$ (i.e., the two-form $\omega_\xi$ associated with $\xi$ is
$\omega_\xi = d\sigma$). This is very helpful in studying magnetic monopoles (with a charge strength $g$ defined via the divergence of the magnetic intensity - for a monopole at the origin in $\mathbb{R}^3$, ${\bf\nabla}.{\bf H} = 4\pi g \delta^{(3)}({\bf r})$) on $S^3$, the contact 1-form $\sigma = -( d\psi + \cos\theta d\phi)$. Writing $H^{(2)}_{ij}\equiv\epsilon_{ijk}H_k$, one sees that the two-form $H^{(2)} = g d\sigma = g d\psi\wedge\Omega_2 = g\mu_{S^3}\neq0, \Omega\equiv\sin\theta d\theta\wedge d\phi, \mu_{S^3}$ being the volume form on $S^3$ with the resulting helicity hence just given by the monopole strength $g$. In fact it can be shown that the Helicity, a generalization of the degree of the map
$f: S^3\rightarrow S^2$, is integral (for $g=1$) \cite{Arnold+Khesin}.}   with strengths determined by ``generalized spin connections'' (\ref{Omega^a_bc-i}), (\ref{Omega^a_bc-ii}). 

\noindent {\it Proof}:

\noindent (a) To find a contact structure, one makes the ansatz:
\begin{equation}
\label{contact-x0-1}
\sigma^1 = \alpha_1 e^1 + \alpha_3 e^3 + \alpha_7 e^7,
\end{equation}
which, using (\ref{Theta_ia+X_a+Y_a+Z_a})-(\ref{Omega^7_a}),  implies:
\begin{eqnarray}
\label{contact-x0-2}
& & \omega^1 = d\sigma^1 = \alpha_1\Omega_{a1}e^{a1} + \alpha_3 \Omega^3_{bc}e^{bc} + \alpha_7\Omega_{a7}e^{a7},\ a, b, c=2,...,6;\nonumber\\
& & \sigma^1\wedge\left(\omega^1\right)^3 \sim \alpha_1\alpha_3^2\alpha_7\Omega^3_{23}\Omega^3_{45}\Omega_{61}e^{1234567}\neq0.
\end{eqnarray}
Now, (\ref{contact-x0-1}) implies (in the $\psi=2n\pi$-coordinate patch):
\begin{eqnarray}
\label{contact-x0-3}
& & \sigma^1_{x^0} = G^{\cal M}_{x^0x^0}R^{x^0}_1 = \alpha_1 \sqrt{G^{\cal M}_{x^0x^0}},\ {\rm implying}\ 
R^{x^0}_1 = \frac{\alpha_1}{\sqrt{G^{\cal M}_{x^0x^0}}};\nonumber\\
& & \sigma^1_{x^{10}} = G^{\cal M}_{x^{10}x^{10}}R^{x^{10}}_1 = \alpha_7 \sqrt{G^{\cal M}_{x^{10}x^{10}}},\ {\rm implying}\ 
R^{x^{10}}_1 = \frac{\alpha_7}{\sqrt{G^{\cal M}_{x^{10}x^{10}}}};\nonumber\\
& & \sigma^1_{\theta_i} = G^{\cal M}_{\theta_i ...}R^{...}_1 = \alpha_3 e^3_{...},\ ...\equiv x, y, z
(G^{\cal M}_{\theta_i\theta_j}(\psi=2n\pi)=0);\nonumber\\
& & \sigma^1_{...} = G^{\cal M}_{...\beta}R^\beta_1 = \alpha_3 e^3_{...},\ ...\equiv x, y, z; \beta = x, y, z, \theta_1, \theta_2.
\end{eqnarray}
Also,
\begin{equation}
\label{contact-x0-4}
G^{\cal M}_{mn}R^m_1R^n_1 = 1,\ m,n=r, x^{10}, \theta_{1,2}, x, y, z.
\end{equation}
The last equation in (\ref{contact-x0-3}) and (\ref{contact-x0-4}) are solved to yield (\ref{contact-x0-5}).

$\sigma^1(R_1)=1$ implies:
\begin{equation}
\label{contact-x0-7}
\alpha_1^2 + \alpha_3 R_1.e^3 + \alpha_7^2 = 1.
\end{equation}   
where, $r=\langle r \rangle\in$ IR either as valued in \cite{Bulk-Viscosity-McGill-IIT-Roorkee} or
$R.e^2(\langle r \rangle) =0$ arising from $i_{R^\alpha}\Phi = \omega^\alpha, \alpha=1, 2, 3$ - \ref{R1.e2 ii}, (\ref{R.e^2_0}) in particular.
 
Similarly, making an ansatz:
\begin{equation}
\label{contact-x0-9}
\sigma^2 = \beta_1 e^1 + \beta_4 e^4 + \beta_7 e^7,
\end{equation}
implying
\begin{eqnarray}
\label{contact-x0-10}
& & \omega^2 = d\sigma^2 = \beta_1\Omega_{a1} e^{a1} + \beta_4\Omega^4_{bc}e^{bc} + \beta_7 \Omega_{a7} e^{a7};\nonumber\\
& & \sigma^2\wedge\left(\omega^2\right)^3\sim \beta_1\beta_4^2\beta_7\Omega^4_{23}\Omega^4_{56}\Omega_{61}e^{1234567}.
\end{eqnarray}
Now, (\ref{contact-x0-9}) implies:
\begin{eqnarray}
\label{contact-x0-11}
& & \sigma^2_{x^0} = G^{\cal M}_{x^0x^0}R^r_2 = \beta_1 \sqrt{G^{\cal M}_{x^0x^0}},\ {\rm implying}\ 
R^{x^0}_2 = \frac{\alpha_1}{\sqrt{G^{\cal M}_{x^0x^0}}};\nonumber\\
& & \sigma^2_{x^{10}} = G^{\cal M}_{x^{10}x^{10}}R^{x^{10}}_2 = \beta_7 \sqrt{G^{\cal M}_{x^{10}x^{10}}},\ {\rm implying}\ 
R^{x^{10}}_2 = \frac{\beta_7}{\sqrt{G^{\cal M}_{x^{10}x^{10}}}};\nonumber\\ 
& & \sigma^2_{\theta_i} = G^{\cal M}_{\theta_i ...}R^{...}_2 = \beta_4 e^4_{...},\ ...\equiv x, y, z
(G^{\cal M}_{\theta_i\theta_j}(\psi=2n\pi)=0);\nonumber\\
& & \sigma^2_{...} = G^{\cal M}_{...\beta}R^\beta_2 = \beta_4 e^3_{...},\ ...\equiv x, y, z; \beta = x, y, z, \theta_1, \theta_2.
\end{eqnarray}
Also,
\begin{eqnarray}
\label{contact-x0-12}
 & & G^{\cal M}_{mn}R^m_2R^n_2 = 1,\ m,n=x^0, x^{10}, \theta_{1,2}, x, y, z;\nonumber\\
 & & \sigma^1(R_2) = \sigma^2(R_1) = 0: G^{\cal M}_{mn}R^m_1 R^n_2 = 0,
\end{eqnarray}
along with
\begin{equation}
\label{contact-x0-13}
R_2=R_1(e^3\rightarrow e^4,\ \alpha_3\rightarrow\beta_4),
\end{equation}
implies: $\beta_1^2 + \beta_4 R_2.e^4 + \beta_7^2 = 1$.
The second constraint in (\ref{contact-x0-12}), implies:
\begin{eqnarray}
\label{contact-x0-15}
& & \alpha_1 \beta_1 + \alpha_7 \beta_7 + \Sigma_{\sigma^1(R_2)}(N, M, N_f; r_h) = 0,
\end{eqnarray}
where,
\begin{eqnarray}
\label{contact-x0-16}
& & \Sigma_{\sigma^1(R_2)}(N, M, N_f; r_h) \equiv G^{\cal M}_{x\theta_1} (R^x_2 R^{\theta_1}_1 + R^{\theta_1}_2 R^x_2) + 
 G^{\cal M}_{x\theta_2} (R^x_2 R^{\theta_2}_1 + R^{\theta_2}_2 R^x_1) \nonumber\\
& & + 
 G^{\cal M}_{y\theta_1}(R^y_2 R^{\theta_1}_1 + R^{\theta_1}_2 R^y_1) + 
 G^{\cal M}_{z\theta_1}(R^z_2 R^{\theta_1}_1 + R^{\theta_1}_2 R^z_1)  + 
 G^{\cal M}_{z\theta_2}(R^z_2 R^{\theta_2}_1\nonumber\\
& &   + R^{\theta_2}_2 R^z_1) + G^{\cal M}_(R^x_2 R^y_1 + R^y_2 R^x_1)  
 G^{\cal M}_{yz}(R^y_2 R^z_1 + R^z_2 R^y_1) + G^{\cal M}_{xz}(R^x_2 R^z_1 + R^z_2 R^x_1)\nonumber\\
& &\sim -\beta_4^2\left(\kappa^4_{\theta_1;1;\beta^0}\right)^2\kappa^4_{y;1;\beta^0}\frac{r^4}{\left(r^2-3a^2\right)^2{\cal A}^{2/3}\left(\log r\right)^4\left(\log r_h\right)^2}\frac{\left(1+\alpha_{\theta_2}^3\right)N^{9/5}}{\alpha_{\theta_1}^4},
\end{eqnarray}
where ${\cal A}(r)\equiv \frac{3}{8\pi} \left[2 \log N N_f - N_f \log (r^6 + 9 a^2 r^4)\right]$ and
$\kappa^4_{\theta_1;1;\beta^0}, \kappa^4_{y;1;\beta^0}$ are constants that figure in the co-frame components $e^4_{\theta_1}, e^4_y$ at ${\cal O}(\beta^0)$. One expects that the $\sigma^3$ dual to \\ $R^3=R^1\times_\Phi R^2$ also provides a Contact Structure.

\noindent (b) Requiring $i_{R^\alpha}\Phi = \omega^\alpha$, we work out constraints on $\alpha_{1,3,7}$ that figure in (\ref{contact-x0-1}) corresponding to $R^1$ of (\ref{contact-x0-3}) and (\ref{contact-x0-5}). Similar constraints can be worked out for $R^2$ of (\ref{contact-x0-13}) and $R^3=R^1\times_\Phi R^2$. In the following ``$\sim$'' implies equality up to pre-factors of ${\cal O}(1)$.

Using the first equation in (\ref{contact-x0-2}),
\begin{eqnarray}
\label{constraints-i}
& &\omega^1 = d\sigma^1 = i_{R^1}\Phi = \alpha_7 J + \alpha_1 e^{27} - R^{1}.e^2 e^{17} + R^1.e^3 e^{47} - R^1.e^4 e^{57}
+ R^1.e^5 e^{67} - R^1.e^6 e^{57}\nonumber\\
& & + \alpha_1 e^{35} - R^1.e^3 e^{15} + R^1.e^5 e^{13} - \alpha_1 e^{46} + R^1.e^4 e^{16}\nonumber\\
& & - R^1.e^6 e^{14} - R^1.e^2 e^{36} + R^1.e^3 e^{26} - R^1.e^6 e^{23} - R^1.e^2 e^{45} + R^1.e^4 e^{25} - R^1.e^5 e^{24}\nonumber\\
& &  = \alpha_1\Omega_{a1}e^{a1} + \alpha_3 \Omega^3_{bc}e^{bc} + \alpha_7\Omega_{a7}e^{a7},\ a, b, c=2,..., 6,
\end{eqnarray}
one obtains the following (wherein $m=2,..., 6$) :
\begin{equation}
\label{R.e^2_0}
R^m_1e^2_m = 0.
\end{equation}

Now, to estimate the $r\in$IR around which $R_1.e^2\sim0$, we will expand all $r$-dependent terms except $\log r$ in $R_1.e^2$ about $r=r_h$ to yield:
\begin{eqnarray}
\label{R1.e2 i}
& & R_1. e^2 \sim 10^{-4} \frac{{\alpha_3} \left(\lambda_5^2 {\log N}^2+\lambda_5
   \log N (31.32 \log r-0.25 \log N)+68.73
   \log r\right)}{ \log r^2 r_h^2 (N_f
   (\log N-3 \log r))^{2/3}}\nonumber\\
& &  + 10^{-6}\times{\cal O}(r-r_h),
\end{eqnarray}
$\lambda_5$ being the parameter parametrizing the one-parameter family of eigenvectors relevant to the diagonalization of the $M_5(\theta_{1,2},x,y,z)$-metric in \cite{OR4}. Assuming $\lambda_5\gg1$ and in the large-$N$ limit, one sees from (\ref{R1.e2 i}) vanishes in the IR-valued $r$ in the neighborhood of $r = \langle r \rangle$:
\begin{eqnarray}
\label{R1.e2 ii}
& & \log \langle r\rangle=-{\cal O}(10^{-2})\lambda_5\log N\nonumber\\
& & {\rm or}\nonumber\\
& & \langle r\rangle\sim N^{-{\cal O}(10^{-2})\lambda_5}{\cal R}_{D5/\overline{D5}}.
\end{eqnarray}
Note in \cite{OR4}, \cite{MQGP} and in this work, we work in ${\cal R}_{D5/\overline{D5}}=1$-units; ${\cal R}_{D5/\overline{D5}}$ is the $D5-\overline{D5}$ separation in the parent type IIB dual of \cite{metrics} which we take to be $\sqrt{3}a,\ a$ being the radius of the blown-up $S^2$ of the warped resolved deformed conifold in the parent type IIB gravity dual of  \cite{metrics}, and given by in the type IIA dual gravity dual for high temperatures (i.e. $T>T_c$ on the gauge theory side) by $a =r_h\left[\frac{1}{\sqrt{3}} - \kappa_b r_h\left(\log r_h\right)^{9/2}N^{-\frac{9}{10}-\alpha_b} + \frac{g_sM^2}{N}\left(c_1 + c_2\log r_h\right)\right],\ \kappa_b\equiv{\cal O}(1),\ \alpha_b>0,\ c_{1,2}<0$ \cite{EPJC-2}, \cite{Bulk-Viscosity-McGill-IIT-Roorkee}, \cite{OR4}. Further,  $r\in[r_h, {\cal R}_{D5/\overline{D5}}]$ is taken to be the IR (strictly speaking the IR and the IR-UV interpolating region) and $r>{\cal R}_{D5/\overline{D5}}$ the UV. Given the SYZ type IIA mirror and its ${\cal M}$-theory uplift constructed in \cite{MQGP}, \cite{NPB}, \cite{OR4} involve a triple-T duality along the toroidal analogs of the angles $\phi_1, \phi_2, \psi$, the IR-UV boundaries along the radial $r$ are taken to be the same in the type IIA SYZ mirror and its ${\cal M}$-theory dual.

In the proof of {\it Lemma 4} and later, {\it Lemma 6} (See \ref{Omega_-} in particular), we will be dealing with expressions of the type $f\left(g_s, M, N_f, N; \alpha_{\theta_1}, \alpha_{\theta_2}; r\right)$. As an example, to show explicitly the existence of  Contact (3) Structure(s) for intermediate $N$ satisfying (\ref{intermediate-N-MQGP-limit}), we work with the QCD-motivated (Table \tcb{2}) $f\left(g_s=0.1, M=3, N_f=3, N; \alpha_{\theta_1}, \alpha_{\theta_2}; r\right)$. We then work out $r=\langle r\rangle$ valued in the IR, whereat $\omega^\alpha = d\sigma^\alpha = i_{R^\alpha}\Phi, \alpha=1, 2, 3$. Given that we work with a large (integral) $N = N_0$ with $\frac{N - N_0}{N_0}\ll1$ and hence assuming $N$ to be varying continuously \footnote{Alternatively, one could promote discretely varying (very) large $N$ to a continuously varying (very) large ${\cal N}$ with the understanding that one sets $N = [{\cal N}]$ towards the end.},  $f\left(g_s=0.1, M=3, N_f=3, N; \alpha_{\theta_1}, \alpha_{\theta_2}; \langle r\rangle, \log \langle r\rangle\right)$ with $\langle r\rangle$ and $\lambda_5$ given by $(\ref{R1.e2 ii})\cap(\ref{omega21-omega27})\cap(\ref{R1.e2 iii})$ and $\alpha_{\theta_1}, \alpha_{\theta_2}\sim{\cal O}(1): \frac{\sqrt{\alpha_{\theta_2}}}{\alpha_{\theta_1}}\sim{\cal O}(1)$, is expanded as:
{\footnotesize
\begin{eqnarray}
\label{Taylor-around-N_0}
& & \hskip -0.8in  f\left(g_s=0.1, M=3, N_f=3, N; \alpha_{\theta_1}, \alpha_{\theta_2}; \langle r\rangle\right)
 =  \sum_{n=0}^\infty f^{(n)}\left(g_s=0.1, M=3, N_f=3, N_0; \alpha_{\theta_1}, \alpha_{\theta_2}; \langle r\rangle\right)\left( \frac{N - N_0}{N_0}\right)^n.
\end{eqnarray}
}
For the aforementioned QCD-motivated $g_s=0.1, M=3, N_f=3$ (Table \tcb{2}), $N_0=100\pm{\cal O}(1)$ was found numerically to ensure the existence of a Contact (3) Structure. {\it It can be verified that $\exists$ a large-but-finite-$N_0$ $\forall (g_s, M, N_f)$ satisfying (\ref{intermediate-N-MQGP-limit}), and not just the ones in Table \tcb{2}; the computations mirror the following specific example that use the QCD-motivated values as given in Table \tcb{2}}. Computations based on (\ref{Taylor-around-N_0}) and the consequent derivation of the constraints (\ref{alphas-same-order}) on $\alpha_{1, 3, 7}$ (similarly $\beta_{1, 4, 7}$), is outlined below.

One can argue that  (\ref{constraints-i}) is equivalent to the following system of equalities:
{\footnotesize
 \begin{eqnarray}
\label{omega_i_RPhi}
&  & \hskip -0.4in \alpha_7 = - \alpha_1\Omega^1_2 = \alpha_3\Omega^3_{34} = \alpha_3\Omega^3_{56};\nonumber\\
&  & \hskip -0.4in \alpha_1\Omega^1_5 = - R^1.e^3,\ \alpha_1 \Omega^1_6 = R^1.e^4,\ \alpha_1\Omega^1_4 = R^1.e^6,\ \alpha_1\Omega^1_3 = - R^1.e^5;\nonumber\\
& & \hskip -0.4in \alpha_7\Omega^7_4 = R^1.e^3,\ \alpha_7\Omega^7_3 = - R^1e^4,\  \alpha_7\Omega^7_5 = - R^1e^6,\ \alpha_7\Omega^7_6 =  R^1.e^5;\nonumber\\
&  & \hskip -0.4in \alpha_3\Omega^3_{25} = - R^1e^4,\ \alpha_3\Omega^3_{26} = - R^1.e^3,\ \alpha_3\Omega^3_{24} = - R^1.e^5,\ \alpha_3\Omega^3_{23} = - R^1.e^6,\ \alpha_3\Omega^3_{36} =  R^1.e^2,\ \alpha_3\Omega^3_{45} = - R^1.e^2.
\end{eqnarray}
}
One can show that near (\ref{R1.e2 ii}), the system (\ref{omega_i_RPhi}) is satisfied if:
\begin{equation}
\label{omega21-omega27}
N^{7/20}\alpha_\theta^2 = \left(\log \langle r\rangle\right)^2 = \kappa_{\log r_h}^2N^{7/20}\left(\log N\right)^2,
\end{equation}
\begin{equation}
\label{R1.e2 iii}
\lambda_5\approx10^2 \frac{N^{7/40}}{\log N}\alpha_\theta,
\end{equation}
(where near, e.g., (\ref{alpha_theta_12}), $\alpha_\theta\equiv\frac{\sqrt{\alpha_{\theta_2}}}{\alpha_{\theta_1}}\equiv{\cal O}(1)$) as well as:
\begin{equation}
\label{alphas-same-order}
\alpha_1\sim\alpha_3\sim\alpha_7.
\end{equation}
It is easily seen that
\begin{equation}
\label{betas}
\left.\omega^2 (= d\sigma^2) = i_{R^2}\Phi\right|_{(\ref{contact-x0-15})\cap(\ref{alphas-same-order})}
\end{equation}
is easily implemented by following steps identical to (\ref{constraints-i}) - (\ref{omega_i_RPhi}), arriving at
\begin{equation}
\label{betas-same-order}
\beta_1\sim\beta_4\sim\beta_7.
\end{equation}
The ``$\sim$'' in (\ref{alphas-same-order}) and (\ref{betas-same-order}) again imply equality up to ${\cal O}(1)$ pre-factors in the IR. We hence prove part (b) of {\it Lemma 4}.

\begin{tcolorbox}[enhanced,width=6.8in,center upper,size=fbox,
    drop shadow southwest,sharp corners]
    \begin{flushleft}
Part (b) of Lemma 4, (\ref{alphas-same-order}) and (\ref{betas-same-order}), have an important implication. Not only do Contact 3-Structures not exist for $N({\rm or} \Upsilon)\gg1$ for closed seven-folds containing the ${\cal M}$-theory and thermal circles that are relevant to ${\cal M}$ theory uplift of UV-complete string dual of thermal QCD-like theories at high temperatures (above the deconfinement temperature),  the Contact  (3-) Structures obtained in Lemma 4 for intermediate $N({\rm or} \Upsilon)$ ($N>1, N\slashed{\gg}1$) in the MQGP limit (\ref{MQGP_limit}), can not be connected to the ACM(3)S obtained in \ref{ACM3S-M7-x0} via an $N$-flow \footnote{Of course, $N$ varies in steps of one, but the assumption is that $\frac{1}{N_{\rm final}-N_{\rm initial}}\ll1, N_{\rm initial,\ final}$ respectively being the initial and final values of $N$ in the $N$-trajectory in the aforementioned space of closed seven-folds with $G_2$ structure.} as the latter sets $\alpha_3 = \beta_4 = 0$; Contact Structure, as per Lemma 4 requires $\alpha_3\sim\alpha_1\sim\alpha_7; \beta_4\sim\beta_1\sim\beta_7$. {\it This is what we refer to as lack of $N$-connectedness.}
\end{flushleft}
\end{tcolorbox}

The significance of the results of \ref{ACM3S-constructions} - \ref{C3S-M7x0} is that {\it the four-parameter space ${\cal X}_{G_2}(g_s, M, N_f; N)$ characterizing the closed seven-folds supporting $G_2$ structures and relevant to the aforementioned ${\cal M}$-theory uplift of thermal QCD-like theories, is not $N$-path connected with reference to Contact Structures in the IR, i.e., the $N\gg1$ Almost Contact 3- Structures arising from the $G_2$ structure, do not connect to Contact 3-Structures (in the IR) which is shown to exist only for an appropriate intermediate $N$.}

\section{$SU(3)$ Structure on a Manifold with $G_2$ Structure $\Leftrightarrow$ Reduction of $G_2$ Structure to  $SU(3)$ Structure}
\label{SU3-from-G2}
 
A seven-fold $M_7$ admitting a $G_2$ structure $\Phi$ admits an almost contact metric structure (ACMS) and thereby reduces the structure group to $SU(3)$ \cite{G2-to-SU3}.
The reason stems from the fact that any such manifold admits a nowhere vanishing vector field $R$ which can be normalized with respect to the metric $g_\Phi$. 
(In the context of supersymmetric theories) A nowhere-vanishing vector field $R$ on $(M_7,\Phi)$ and nowhere-vanishing spinor $\eta$ (implicit in the choice of $G_2$ structure) induce a second spinor $R\eta$ which together can be used to construct an $SU(3)$ structure \cite{Friedrich+Kath [1997]}.

Alternatively, the reduction of the structure group from $G_2$ to $SU(3)$ can be effected by construction of the latter from $\Phi_{G_2}$ and ACS:\\
\noindent {\it Proposition} \cite{Ossa et al[2013]}:
The ACMS induces a reduction of the $G_2$ structure to an $SU(3)$ structure $(\omega_\Phi, \Omega)$ on the transverse geometry of the foliation with $\omega_{\Phi}$ being the fundamental 2-form on $M_7$ ($\omega_\Phi = i_R\Phi$) and $\Omega$  the transverse $(3,0)$-form w.r.t. $J (J(u) = R\times_\Phi u,\ \forall u\in\Gamma(T M_7))$.
$(\omega_\Phi, \Omega)$ are determined by the ACS decomposition of the $G_2$ structure $\Phi$ on $M_7$:
\begin{eqnarray}
\label{Transverse-conditions}
& & \Phi = \sigma\wedge\omega_\Phi + \Omega_+,\nonumber\\
& & \psi = *_7\Phi = - \sigma\wedge\Omega_- + \frac{1}{2}\omega_\Phi\wedge\omega_\Phi.
\end{eqnarray}

In this section, via two lemmas, we will construct an explicit transverse $SU(3)$ structure in \ref{Transverse-ACMS} from the ACMS of \ref{AC3S-M7x0}, and then from a CMS of \ref{C3S-M7x0} in \ref{Transverse-SU3-CMS}, on the closed $M_7$ restricted to the Ouyang embedding  of the parent type IIB dual with a vanishingly small $|\mu_{\rm Ouyang}|$.

\subsection{Transverse $SU(3)$ Structure from Almost Contact  Structure}
\label{Transverse-ACMS}

\noindent {\it Lemma 5}: $M_7$, near the Ouyang embedding in the limit of very small limit of the Ouyang embedding parameter and near the $\psi=2n\pi, n=0, 1, 2$-coordinate patch, inherits transverse $SU(3)$ 3-structures $\left(\Omega_+^{\alpha}, \Omega_-^{\alpha}\right)$ wherein
$\Omega^{\alpha}_+ = \Phi - \sigma^{\alpha}\wedge \omega_\Phi^{\alpha},
\Omega^{\alpha}_- = \sigma^{\alpha}\lrcorner\left(\frac{1}{2}\omega_\Phi^{\alpha}\wedge \omega_\Phi^{\alpha} - *_7\Phi\right), \alpha=1, 2, 3$ from the AC(3)S constructed above.
 
\noindent {\it Proof}: Using (\ref{ACM3S-x0-sigmas}), (\ref{Omega^a_bc-i}) and (\ref{de1-i}), one hence obtains,
{\footnotesize
\begin{eqnarray}
\label{Omega+-12_i}
& & \Omega_+^{(1)} = \Phi - e^1\wedge\omega^{(1)}_\Phi,\ \Omega_-^{(1)} = e^1\lrcorner\left(\frac{1}{2}\omega^{(1)}_\Phi\wedge\omega^{(1)}_\Phi - *_7\Phi\right);\nonumber\\
& & \Omega_+^{(2)} = \Phi - e^7\wedge\omega^{(2)}_\Phi,\ \Omega_-^{(2)} = e^7\lrcorner\left(\frac{1}{2}\omega^{(2)}_\Phi\wedge\omega^{(2)}_\Phi - *_7\Phi\right);\nonumber\\
& & \Omega_+^{(3)} = \Phi + e^2\wedge\omega^{(3)}_\Phi,\ \Omega_-^{(3)} = -e^2\lrcorner\left(\frac{1}{2}\omega^{(3)}_\Phi\wedge\omega^{(3)}_\Phi - *_7\Phi\right).
\end{eqnarray}
}

\subsection{Transverse $SU(3)$ Structure from Contact  Structure}
\label{Transverse-SU3-CMS}

\noindent {\it Lemma 6}: $M_7$, near the Ouyang embedding in the limit of very small limit of the Ouyang embedding parameter and near the $\psi=2n\pi, n=0, 1, 2$-coordinate patch, inherits a transverse $SU(3)$ structure from the Contact 3-Structures constructed in 3, $\left(\Omega_+^{\alpha},\Omega_-^{\alpha}\right)$,
where $\Omega_+^{\alpha} = \Phi - \sigma^{\alpha}\wedge\omega_\Phi^{\alpha}$ and, e.g., for $\alpha=1$, a one-parameter ($\Lambda^{(1)}_{156}$ or $\Lambda_{456}$) family of:
 \begin{eqnarray*}
& & \Omega_- = \Lambda_{AMC}e^{ABC} = \Lambda^{(1)}_{1b_0c_0}e^{1b_0c_0}   + \Lambda_{3b_0c_0}^{(1)}e^{3b_0c_0}
+ \Lambda^{(1)}_{7b_0c_0}e^{7b_0c_0}
+ \Lambda_{a_0b_0c_0}e^{a_0b_0c_0},
\end{eqnarray*}
where $a_0, b_0, c_0=2, 4, 5, 6$ and $\Lambda_{456}$ is the only linearly independent non-vanishing $\Lambda_{a_0b_0c_0}$.

\noindent {\it Proof}: Making the following ansatz for $\Omega_-$:
\begin{eqnarray}
\label{Omega-}
& & \Omega_- = \Lambda_{AMC}e^{ABC} = \Lambda^{(1)}_{1b_0c_0}e^{1b_0c_0} +
\Lambda^{(2)}_{13c_0}e^{13c_0} + \Lambda^{(3)}_{17c_0}e^{17c_0}\nonumber\\
& & + \Lambda^{(4)}_{137}e^{137} + \Lambda_{3b_0c_0}^{(1)}e^{3b_0c_0}
+ \Lambda^{(3)}_{37c_0}e^{37c_0} + \Lambda^{(1)}_{7b_0c_0}e^{7b_0c_0}
+ \Lambda_{a_0b_0c_0}e^{a_0b_0c_0},
\end{eqnarray}
where $a_0, b_0, c_0=2, 4, 5, 6$, we solve:
\begin{eqnarray}
\label {CS-equations}
& &  i_{R^{(1)}}\Omega_-=0,\nonumber\\
& &  *\Phi = -\sigma^{(1)}\wedge\Omega_- + \frac{1}{2}\omega_\Phi^2,
\end{eqnarray}
for $\Omega_-$ with the understanding that $\Omega_+ = \Phi - \sigma\wedge\omega_\Phi$ (the holomorphic three-form $\Omega = \Omega_+ + \iota \Omega_-$). This yields (\ref{iROmega-0-ii}) and (\ref{Transverse-condition-iii}). It turns to out to most expeditious to solve all of the equations (\ref{iROmega-0-ii}) and (\ref{Transverse-condition-iii}) save 
(\ref{Transverse-condition-iii})(iii)[$b_0 = 2, c_0 = 5$] and (\ref{Transverse-condition-iii})(i)[$b_0 = 2, c_0 = 4$]; the two excluded are solved for towards the end.

The solution using results of Appendix \ref{Omega_-}, along with addition of a regulator $\epsilon_{1745}$ (as) in $\epsilon_{1745} + \alpha_3\Lambda^{(1)}_{145} = \frac{\alpha_1\alpha_3}{\alpha_7}\Lambda^{(1)}_{745}$ that arises from (\ref{i})(3) corresponding to (\ref{Transverse-condition-iii})(i), and (\ref{iii})(4) corresponding to (\ref{Transverse-condition-iii})(iii) with $b_0=4, c_0 = 5$ therein,  is given by:
{\footnotesize
\begin{eqnarray*}
& & \Lambda^{(1)}_{125} = -\frac{0.02 \alpha_1 \epsilon_{1745} \kappa_{\log r_h}^2\Lambda^{(1)}_{724}  ((R.e^4)^2 + 
R.e^2 R.e^5)}{R.e^5 (-R.e^4 + R.e^5)},\nonumber\\
& & \Lambda^{(1)}_{126} = -\Lambda^{(1)}_{125} R.e^5/R.e^6,\nonumber\\
& & \Lambda^{(1)}_{146} = - R.e^5 (- \Lambda^{(1)}_{125} R.e^2 + \Lambda^{(1)}_{156} R.e^6)/(R.e^4 R.e^6);\nonumber\\
& & \Lambda^{(1)}_{324} = 0, \nonumber\\
& & \Lambda^{(1)}_{325} = 0.5 \Lambda^{(1)}_{125}/\alpha_1,\nonumber\\ 
& & \Lambda^{(1)}_{326} = -0.5\Lambda^{(1)}_{125} R.e^5/(\alpha_1 R.e^6),\nonumber\\
& & \Lambda^{(1)}_{345} = 0.5\Lambda^{(1)}_{145}/\alpha_1,\nonumber\\
& & \Lambda^{(1)}_{346} = -0.5\Lambda^{(1)}_{145} R.e^5/(\alpha_1 R.e^6),\nonumber\\
& & \Lambda^{(1)}_{356} = 0.5(\Lambda^{(1)}_{125} R.e^2 + \Lambda^{(1)}_{145} R.e^4)/(\alpha_1 R.e^6);\nonumber\\
& & \Lambda^{(1)}_{724}  = \epsilon_{1724}: |\epsilon_{1724}|\ll1, \nonumber\\
& & \Lambda^{(1)}_{725} = (-\Lambda^{(1)}_{724}  R.e^4 + 2 \alpha_7 \Lambda^{(1)}_{325} R.e^5)/R.e^5,\nonumber\\ 
& & \Lambda^{(1)}_{726} = \alpha_7 \Lambda^{(1)}_{126}/\alpha_1, \nonumber\\
& &  \Lambda^{(1)}_{745} = \alpha_7 (2\epsilon_{1745} + \Lambda^{(1)}_{145})/\alpha_1, \nonumber\\
& & \Lambda^{(1)}_{746} = \frac{1}{\Lambda^{(1)}_{125}\Lambda^{(1)}_{724} } \Biggl(2\alpha_7 \Lambda^{(1)}_{126} \Lambda^{(1)}_{325}\Lambda^{(1)}_{724}  + 2\alpha_7 \Lambda^{(1)}_{126} \Lambda^{(1)}_{345}\Lambda^{(1)}_{724}  - 
     2\alpha_7 \Lambda^{(1)}_{126} \Lambda^{(1)}_{325} \Lambda^{(1)}_{725}\nonumber\\
 & &  +      \Lambda^{(1)}_{126} \Lambda^{(1)}_{725}\ ^2 + 
     2\epsilon_{1745}\Lambda^{(1)}_{724}  \Lambda^{(1)}_{726}\Biggr)    = - \Lambda^{(1)}_{745} R.e^5/R.e^6 \Rightarrow {\rm identically\ satisfied}; \nonumber\\ 
& &    \Lambda^{(1)}_{756} = \frac{1}{\Lambda^{(1)}_{125}\Lambda^{(1)}_{724}\ ^2} \Biggl(-4\alpha_7^2 \Lambda^{(1)}_{126} \Lambda^{(1)}_{325} \Lambda^{(1)}_{345}\Lambda^{(1)}_{724}  + 
     2\alpha_7 \Lambda^{(1)}_{126} \Lambda^{(1)}_{325}\Lambda^{(1)}_{724}  \Lambda^{(1)}_{725}\nonumber\\
     & &  + 
     2\alpha_7 \Lambda^{(1)}_{126} \Lambda^{(1)}_{345}\Lambda^{(1)}_{724}  \Lambda^{1)}_{725} - 
     2\alpha_7 \Lambda^{(1)}_{126} \Lambda^{(1)}_{325} \Lambda^{1)}_{725}\ ^2 + 
     \Lambda^{(1)}_{126} \Lambda^{(1)}_{725}\ ^3   - 
     4\alpha_7 \epsilon_{1745} \Lambda^{(1)}_{325}\Lambda^{(1)}_{724}  \Lambda^{(1)}_{726}\nonumber\\
& &      +      2\epsilon_{1745}\Lambda^{(1)}_{724}  \Lambda^{(1)}_{725} \Lambda^{(1)}_{726}\Biggr)  = (\Lambda^{(1)}_{745} R.e^4)/R.e^6 \Rightarrow \Lambda^{(1)}_{145} = 
\frac{{\cal O}(10^{-6})}{g_s^{7/2}}\frac{\alpha_1}{\alpha_7}
\frac{\left(\epsilon_{1745}\kappa_{\log r_h}^2\right)^2\epsilon_{1724}N^{1/4}}{M^2N_f^2\left(\log N\right)^2};\nonumber\\
& & \Lambda^{(3)}_{324} = (-\alpha_1 \Lambda^{(1)}_{124} - \alpha_7\Lambda^{(1)}_{724} )/R.e^3,\nonumber\\
 & & \Lambda^{(3)}_{325} = \frac{1}{(R.e^3 (R.e^4 -  R.e^5) R.e^5)}\Biggl((\alpha_7\Lambda^{(1)}_{724}  - 
       0.02 \alpha_1^2 \epsilon_{1745} \kappa_{\log r_h}^2\Lambda^{(1)}_{724} ) (R.e^4)^2  \nonumber\\
       & & +    \alpha_7 (-2\alpha_7 \Lambda^{(1)}_{325} - \Lambda^{(1)}_{724} ) R.e^4 R.e^5 + 
    R.e^5 (-0.02 \alpha_1^2 \epsilon_{1745} \kappa_{\log r_h}^2\Lambda^{(1)}_{724}  R.e^2 + 2\alpha_7^2 \Lambda^{(1)}_{325} R.e^5)\Biggr),
\nonumber\\
& &  \Lambda^{(3)}_{326} = \frac{(\alpha_1^2 \Lambda^{(1)}_{125} R.e^5 - \alpha_7^2 \Lambda^{(1)}_{126} R.e^6)}{
 \alpha_1 R.e^3 R.e^6},
\end{eqnarray*}

\begin{eqnarray}
\label{solution-LambdaABC}
 & &  \Lambda^{(3)}_{345} = \frac{\alpha_7^2 (-2\epsilon_{1745} - \Lambda^{1)}_{145}) - \alpha_1^2 \Lambda^{1)}_{145} - \alpha_1 \Lambda_{456} R.e^6}{ \alpha_1 R.e^3}
\sim \frac{\alpha_7^2 (- \Lambda^{1)}_{145}) - \alpha_1^2 \Lambda^{1)}_{145} - \alpha_1 \Lambda_{456} R.e^6}{ \alpha_1 R.e^3}, \nonumber\\
& &  \Lambda^{(3)}_{346} = \frac{1}{\Lambda^{(1)}_{125}\Lambda^{(1)}_{724}  R.e^3 R.e^4 R.e^6}\Biggl(- \alpha_1 \Lambda^{(1)}_{125}\ ^2\Lambda^{(1)}_{724}  \
R.e^2 R.e^5 + ((\alpha_7 (-1\Lambda^{(1)}_{126} \Lambda^{1)}_{725}\ ^2 \nonumber\\
& & + \alpha_7 \Lambda^{(1)}_{126} (-2\Lambda^{(1)}_{325}\Lambda^{(1)}_{724}  - 
                2\Lambda^{1)}_{345}\Lambda^{(1)}_{724}  + 
                2\Lambda^{(1)}_{325} \Lambda^{1)}_{725}) - 
             2\epsilon_{1745}\Lambda^{(1)}_{724}  \Lambda^{1)}_{726})\nonumber\\
             & &  +  \Lambda^{(1)}_{125}\Lambda^{(1)}_{724}  \Lambda_{456} R.e^5) R.e^4 + 
       1\alpha_1 \Lambda^{(1)}_{125} \Lambda^{(1)}_{156}\Lambda^{(1)}_{724}  R.e^5) \
R.e^6\Biggr)\sim\frac{\Lambda_{456}R.e^5}{R.e^3}, \nonumber\\
& & \Lambda^{(3)}_{356} =\frac{1}{\Lambda^{(1)}_{125} \
\Lambda^{(1)}_{724}\ ^2 R.e^3}\Biggl(4\
\alpha_7^3 \Lambda^{(1)}_{126} \Lambda^{(1)}_{325} \Lambda^{1)}_{345}\Lambda^{(1)}_{724}  - 
     \alpha_1 \Lambda^{(1)}_{125} \Lambda^{(1)}_{156}\Lambda^{(1)}_{724}\ ^2 \nonumber\\
     & & + 
    \alpha_7^2 (\Lambda^{(1)}_{126} \Lambda^{1)}_{725} (-2\Lambda^{(1)}_{325}\Lambda^{(1)}_{724}  - 
          2 \Lambda^{1)}_{345}\Lambda^{(1)}_{724}  + 2 \Lambda^{(1)}_{325} \Lambda^{1)}_{725})\nonumber\\ 
          & & + 
       4 \epsilon_{1745} \Lambda^{(1)}_{325}\Lambda^{(1)}_{724}  \Lambda^{1)}_{726}) + 
    \alpha_7 (- \Lambda^{(1)}_{126} \Lambda^{1)}_{725}\ ^3 - 
       2 \epsilon_{1745}\Lambda^{(1)}_{724}  \Lambda^{1)}_{725} \Lambda^{1)}_{726}) \nonumber\\ 
       & & - 
   \Lambda^{(1)}_{125}\Lambda^{(1)}_{724}\ ^2 \Lambda_{456} R.e^4\Biggr) \sim \frac{-\alpha_1 \Lambda^{(1)}_{156} - \Lambda_{456} R.e^4 + \frac{(
 0.02 \alpha_7^3 \epsilon_{1745} \kappa_{\log r_h}^2 \Lambda^{(1)}_{145} ((R.e^4)^2 + 
    R.e^2 R.e^5))}{(\alpha_1 (- R.e^4 + R.e^5) R.e^6))}}{R.e^3};\nonumber\\
 & &    
\end{eqnarray}
}
$\kappa_{\log r_h}$ is defined in (\ref{omega21-omega27}). The need for introduction of regulators $\epsilon_{1745}$ and $\epsilon_{1724}$ is because in their absence one can show that the system (\ref{Transverse-condition-iii}) does not possess a solution.  But introducing a small $\epsilon_{1745}$ subject to the regularization: 
\begin{eqnarray}
\label{epsilon1745kappalogrIRsq}
& & \epsilon_{1745} \kappa_{\log r_h}^2\equiv {\cal O}(1),
\end{eqnarray}
($\epsilon_{1745}$ in fact turns out to be $\alpha_1\alpha_3\Omega^1_{[3}\Omega^3_{4]5} + \frac{\alpha_1}{\alpha_7}\alpha_3\alpha_7\Omega^3_{[34}\Omega^7_{5]}\sim\left.\frac{{\cal O}(1)}{\alpha_\theta \alpha_{\theta_1}^4} + \frac{{\cal O}(10)}{\alpha_{\theta_1}^4}\sim\frac{{\cal O}(10)}{\alpha_{\theta_1}^4}\right|_{\alpha_{\theta_1}=5/6}\ll1$ - with, e.g., $\kappa_{\log r_h}=15$, (\ref{epsilon1745kappalogrIRsq}) is realized)
and setting $\epsilon_{1724}$ to zero, one  obtains a one-parameter ($\Lambda^{(1)}_{156}$ or $\Lambda_{456}$) family of solutions:
{\footnotesize
\begin{eqnarray}
\label{Solutions-Lambda13abc}
& & \hskip -0.8in \left.\Lambda^{(i)}_{abc}(r\sim\langle r \rangle, g_s, M, N_f, N)\right|_{(\ref{MQGP_limit})} = \sum_{l=0}^2\xi^{(i), l}_{abc}\left.\left(\langle r \rangle, g_s, M, N_f, N; \Lambda^{(1)}_{145}, \Lambda^{(1)}_{156}, \Lambda_{456}\right)\right|_{(\ref{MQGP_limit})}\left(\epsilon_{1745}\kappa_{\log r_h}^2\right)^l,
\end{eqnarray}
}
where $i=1, 3$. Use has been made of:
\begin{eqnarray}
\label{Rea}
& & R.e^2 = 0,\nonumber\\
& & R.e^3 = -\left.\frac{{\cal O}(1)\alpha_3\lambda_5}{\left(|\log \langle r\rangle|N_f\right)^{2/3}}\right|_{\log\langle r \rangle\sim -N^{7/40}\log N \kappa_{\log r_h}, \lambda_5\sim
10^2N^{N^{7/40}}\frac{\alpha_\theta}{\log N}},\nonumber\\
& & R.e^4 = \left.\frac{\alpha_3\delta N^{3/5}\alpha_{\theta}}{{\cal O}(1)g_s^{7/2}\log N\left(\log\langle r \rangle\right)^{11/3}M^2N_f^{8/3}\langle r\rangle^2}\right|_{\log\langle r \rangle\sim -N^{7/40}\log N \kappa_{\log r_h}},\nonumber\\
& & R.e^5 \sim R.e^6 \sim \frac{\left( - 3 - \frac{1.2\alpha_\theta}{\kappa_{\log r_h}}\right)}{\kappa_{r_h}N^{7/60}N_f^{2/3}},
\end{eqnarray}
wherein ``$\sim$'' implies equality up to ${\cal O}(1)$ pre-factors.

Substituting (\ref{solution-LambdaABC}) into the remaining two equations:
\begin{enumerate}
\item
\begin{equation}
\label{consistency}
\alpha_7 \Lambda^{(1)}_{325} - \alpha_3 \Lambda^{(1)}_{725} - \frac{10^5 e^{-10^{7/20} \alpha_\theta}}{\alpha_{\theta_1}^4} = 0,
\end{equation}
for $\alpha_\theta\sim{\cal O}(1),\ \alpha_{\theta_1}=7/8$ ((\ref{iii}) - (\ref{vii})), one sees that the |left hand side of (\ref{consistency})|$\ll1$;

\item
\begin{equation}
\label{alphatheta}
\alpha_3 \Lambda^{(1)}_{124} - \alpha_1 \Lambda^{(1)}_{324} -  {\cal O}(10^3) \alpha_\theta^2=0,
\end{equation}
substituting $\Lambda^{(1)}_{324}=0$ from (\ref{solution-LambdaABC}), one determines
$\Lambda^{(1)}_{124}$. 
\end{enumerate}
In the end the thus-far undetermined $\Lambda^{(1)}_{156}$ and $\Lambda_{456}$ satisfy a real constraint:
$\frac{1}{6}\omega_\Phi^3 = \frac{1}{4}\Omega_+\wedge\Omega_-$ \cite{Ossa et al[2013]}. 

\section{Ubiquity of (Almost) Contact Structures in Top-Down Thermal Holographic QCD, and Physics Analogs and Implications of the Differential Geometric results of sections \ref{G2-M7x0} - \ref{SU3-from-G2}}
\label{applications-Physics} 

In this section, we present a conjecture pertaining to the non-supersymmetric non-AdS dual of thermal QCD-like theories supporting (Almost) Contact 3-Structures and being out of the non-supersymmetric swampland on one hand, and connect up on the other, existence of Contact 3-Structures with entanglement entropy of relevant eternal black holes as well as chaos via pole-skipping.  We also briefly review from previous works  from our group, a connection between {\it bulk-to-shear-viscosity ratio in terms of speed of sound} and Contact 3-Structures \cite{Bulk-Viscosity-McGill-IIT-Roorkee}, as well as a connection \cite{photoprod+EoS+AC3S} respectively between strong magnetic field photoproduction/pressure(energy)-anisotropic paramagnetic plasma Equation of State, and Contact/Almost Contact 3-Structures.

\begin{itemize}
\item {\bf Staying out of the non-supersymmetric ``swampland''}: Given that in the near-horizon limit, the 10D warp factor $h$ that includes the back-reaction due to fluxes as well as black-hole is given in (\ref{eq:h}) \cite{metrics} with the corresponding near-horizon $AdS$ warp factor $h_{\rm AdS} = \frac{L^4}{r^4}$, in the near-horizon limit. Near the Ouyang embedding: $\theta_{1,2}\sim N^{-\gamma_{1,2}},\ \gamma_{1,2}>0$, one sees that in the intermediate-$N$ MQGP limit (\ref{intermediate-N-MQGP-limit}),
\begin{equation}
\label{difference-h's}
h_{\rm MQGP}(=(\ref{eq:h})_{(\ref{intermediate-N-MQGP-limit})}) - h_{\rm AdS} \sim \frac{g_s M^2 \log r (g_s N_f  (\log r\ {\rm or}\ \log N))^{0, 1}}{N}.
\end{equation}
Using Proposition 1 and (\ref{difference-h's}), we propose the following conjecture \footnote{This arose due to a discussion with C. Vafa.}:

\begin{tcolorbox}[enhanced,width=6.5in,size=fbox,
    drop shadow southwest,sharp corners]
{\bf Conjecture}: The non-supersymmetric  non-AdS dual of thermal QCD-like theories (equivalence class of theories that are IR-confining, UV-conformal with bi-fundamental quarks) as constructed in \cite{metrics, MQGP, NPB, OR4}, is farther away from the non-supersupersymmetric AdS swampland in the intermediate-$N$ MQGP limit (\ref{intermediate-N-MQGP-limit}) which supports Contact 3-Structures (Lemma 4)  than the large-$N$ MQGP limit (\ref{MQGP_limit}) which supports
Almost Contact 3-Structures (Lemmas 2 and 3). 
\end{tcolorbox}
So, "(Non-Supersymmetric) Vacua Morghulis Daor'' \footnote{Modifying the title of \cite{Freivogel+Kleban} using the fictional ``High Valyrian" language developed by the linguist David J. Peterson for the show "Game of Thrones".} (not all (non-supersymmetric) vacua ``die'', i.e., are unstable \cite{Ooguri+Vafa}) \footnote{One of us (AM) thanks C. Vafa for bringing \cite{Ooguri+Vafa, Freivogel+Kleban} to our notice during the former's visit to Harvard in the fall of 2023.}. Needless to say, one will have to figure out a distance function in the space of non-supersymmetric ${\cal M}$-theory $G_2$-structure compactifications to prove this conjecture. We defer this to a future investigation.

\item {\bf $G_2/C3S$ Structure determining the Physics via the Einstein-Hilbert action}: One can show that:
\begin{equation}
\label{EH-more-than-flux}
R (r\in{\rm IR}) > G_4^2(r\in{\rm IR}).
\end{equation}
Now, Ricci scalar (essentially the Einstein-Hilbert term in the $D=11$ supergravity action) can be written in terms of the four  $G_2$-structure torsion classes \cite{Bryant-G2} as follows:
\begin{eqnarray}
\label{Ricci-scalar-G2}
& & \hskip -0.3in R(M_7(r,\theta_1,\theta_2,\phi_1,\phi_2,\psi,x^{10})) = 12 d^\dagger W_7 + \frac{21}{8}W_1^2 + 30 |W_7|^2 - \frac{1}{2}|W_{14}|^2 - \frac{1}{2}|W_{27}|^2.\nonumber\\
& &
\end{eqnarray}
One hence sees that the dynamics in the intermediate-$N$ limit (\ref{intermediate-N-MQGP-limit}) will be given predominantly by the $G_2$-structure torsion classes ($W_{7, 14, 27}$) and hence the Contact Structure arising from the same. 

\item  {\bf Making ``Contact'' with previous Physics results from our group} on {\it entanglement entropy involving eternal black holes} \cite{MQGP-Page} as well as {\it chaos via pole-skipping} \cite{MQGP-chaos}, and a reiteration from previous works from our group, of a connection between {\it bulk-to-shear-viscosity ratio in terms of speed of sound} and Contact 3-Structures \cite{Bulk-Viscosity-McGill-IIT-Roorkee}, as well as a connection \cite{photoprod+EoS+AC3S} respectively between strong magnetic field photoproduction/pressure(energy)-anisotropic paramagnetic plasma Equation of State, and Contact/Almost Contact 3-Structures:

\begin{itemize}

\item {\bf Entanglement entropy of eternal black-holes relevant to  ${\cal M}$-theory dual of thermal QCD-like theories at intermediate coupling} (Making "Contact" with the results of \cite{MQGP-Page}): 
In the context of entanglement entropy of the eternal black hole \footnote{An eternal black hole is two copies of the one-sided black hole. An eternal AdS-Schwarzschild black hole was constructed by extending the Penrose diagram of a one-sided AdS-Schwarzschild black hole in \cite{Maldacena:2001kr}. Eternal black holes contain two interiors (past and future) and two exteriors (left and right). Dual interpretation of enternal AdS black hole is the thermofield double state described by $\ket{\Psi}=\frac{1}{\sqrt{Z(\beta)}}\sum_n e^{-\beta E_n/2}\ket{E_n}_1 \ket{E_n}_2$, where $\ket{E_n}_1$ and $\ket{E_n}_2$ are the energy eigenstates in CFT$_1$ and CFT$_2$ defined at the left and right boundaries and $Z(\beta)$ is the partition function of one CFT with $\beta=\frac{1}{T}$ ($T$ is the temperature).} in the ${\cal M}$-theory dual of thermal QCD-like theories at high temperatures (i.e. above the deconfinement temperature), the ratio of the Island Surface \footnote{Island is the region inside the black hole, and the entanglement entropy associated with the Island surface dominates over the Hartman-Maldacena surface \cite{Hartman:2013qma} (time-dependent contribution in the ``Page curve") in the Page curve after Page time \cite{Page:1993wv}. The ``Island proposal'' \cite{AMMZ} started by considering the Island as an interior part of the black hole; however, later, it was argued that the Island lies outside the horizon \cite{Almheiri:2019yqk}. There are examples where Island has been found inside the horizon, e.g., non-holographic models \cite{Gan:2022jay,Yadav:2022jib} and the holographic model with Yang-Baxter deformation \cite{Yadav:2023sdg}.} entropy (as per the Islands proposal \cite{AMMZ} \footnote{See \cite{Ryu:2006bv}\cite{Hubeny:2007xt}\cite{Faulkner:2013ana}\cite{Engelhardt:2014gca} where gravitational formula for von Neumann entropy has been discussed and \cite{Penington:2019kki}\cite{Almheiri:2019qdq} for Page curve calculation from replica trick.}, the time-independent late-time entanglement entropy) and the Hawking black hole entropy in ${\cal R}_{D5/\overline{D5}}=1$-units, is given by \cite{MQGP-Page}:
\begin{eqnarray}
\label{SEEISoverSBH-arrh>>a3}
& &  \frac{S^{\beta^0,\ IS}_{\rm EE}}{S_{\rm BH}} \sim \frac{e^{-\kappa_{l_p}N^{1/3}}{g_s}^{49/12} \left({\log N}\right) M N^{11/4}}{ {N_f}^{5/3} {r_h}^{17/2} | \log ({r_h})| ^{8/3}}\left[1 + \sum_{n=1}^\infty {\cal A}_n\left(\frac{\rm constant}{ r_h}\right)^n\right], 
\end{eqnarray}
where, e.g., ${\cal A}_1 = \frac{7}{4}, {\cal A}_2 = \frac{39}{32}$, etc. and the constant is determined in terms of a pair of constants of integration appearing in the solution for the embedding function describing the Island Surface in the $\frac{\rm constant}{r_h}\ll1-$ limit. The constant $\kappa_{l_p}$ is defined via: $\beta\sim \left(g_s^{4/3}\alpha^{'2}e^{-\kappa_{l_p}N^{1/3}}\frac{r_h}{{\cal R}_{D5/\overline{D5}}}\right)^{3/2}$ \cite{MQGP-Page}. Now, one expects:
\begin{equation}
\label{SEEISoverSBH-D-dims}
\frac{S^{\beta^0,\ IS}_{\rm EE}}{S_{\rm BH}} = 2 + \sum_{n=1}a_n\left(\frac{G_N^D}{r_h^{D-2}}\right)^n,
\end{equation}
 for a $D$-dimensional black-hole  \cite{SEEISoverSBHapprox2}; "constant" in (\ref{SEEISoverSBH-arrh>>a3}) is the non-conformal analog of the conformal charge "$c$'' figuring in \cite{SEEISoverSBHapprox2}. It is rather non-trivial to obtain a somewhat similar expression for the non-conformal backgrounds in (\ref{SEEISoverSBH-arrh>>a3}). To ensure that (\ref{SEEISoverSBH-arrh>>a3}) $\sim$ (\ref{SEEISoverSBH-D-dims}),  the large-$N$ and IR(via small $r_h$) enhancements in the (\ref{SEEISoverSBH-arrh>>a3}) need to be tamed. Utilizing the estimate in \cite{Bulk-Viscosity-McGill-IIT-Roorkee} of the $r=r_0\sim r_h: N_{\rm eff}(r_0=0)$(the effective number of $D3$-branes $N_{\rm eff}$ defined in \cite{Vikas+Gopal+Aalok}), and  the exponential $N$-suppression therein ($\frac{r_h}{{\cal R}_{D5/\overline{D5}}}\sim e^{-\kappa_{r_h}N^{1/3}}$),
\begin{eqnarray}
\label{SEEISoverSBH-arrh>>a3-b}
& &  \frac{S^{\beta^0,\ IS}_{\rm EE}}{S_{\rm BH}} \sim \frac{e^{8.5\kappa_{r_h} - 1.5\kappa_{l_p}N^{1/3}}{g_s}^{49/12} \left({\log N}\right) M N^{1.9}}{2 {N_f}^{5/3}}\left[2 + \sum_{n=1}^\infty {\cal A}_n\left(\frac{\rm constant}{ r_h}\right)^n\right], 
\end{eqnarray}
where $\kappa_{r_h} = \frac{1}{3(6\pi)^{1/3}(g_sN_f)^{2/3}(g_sM^2)^{1/3}}$ \cite{Bulk-Viscosity-McGill-IIT-Roorkee}.
One sees that it is precisely {\bf in the intermediate-$N$ MQGP limit (\ref{intermediate-N-MQGP-limit}) that supports Contact 3-Structures that one obtains}:
\begin{equation}
\label{SIS-EE-over-SB}
\left. \frac{e^{8.5\kappa_{r_h} - 1.5\kappa_{l_p}N^{1/3}}{g_s}^{49/12} {\log N} M N^{1.9}}{2 {N_f}^{5/3}}\right|_{{\rm Table}\ \tcb{2}, N=100, \kappa_{r_h}=0.3, \kappa_{l_p}=2.2}=1,
\end{equation}
{\bf thereby obtaining the ${\cal M}$-theory dual of (\ref{SEEISoverSBH-D-dims}).} 
\vskip 0.3in
\item {\bf Chaos in ${\cal M}$-theory dual of thermal QCD-like theories at intermediate coupling} (making "Contact" with results of \cite{MQGP-chaos}): Apart from (\ref{EH-more-than-flux}), using also, e.g.,
\begin{eqnarray*}
\label{flux-beta0-subdominant}
& &  \left.G_{rMNP}G_r^{\ \ MNP}\right|_{(\ref{intermediate-N-MQGP-limit})} <\left.R_{rr}^{\beta^0}\right|_{(\ref{intermediate-N-MQGP-limit})},
\end{eqnarray*}
one can drop flux-dependent contributions at ${\cal O}(\beta^0)$, and using 
$J_0>E_8>t_8^2G^2R^3$ one can drop the flux-dependent contributions at
${\cal O}(\beta)$ in the $D=11$ supergravity action.

It was shown in \cite{MQGP-chaos} that the equation of motion of a gauge-invariant combination of scalar modes of metric perturbations, $Z_s(r)$ can be written as:
\begin{equation}
\label{Zs-EOM}
Z_s''(r) = l(r) Z_s'(r) + m(r) Z_s(r),
\end{equation}
where $m(r) = \sum_{i=-4}^\infty m_i(r_h; g_s, M< N_f, N)(r - r_h)^n$. It was further shown that $m_{-4} = m_{-3} = 0$ for:
\begin{eqnarray}
\label{vb-i}
& & \hskip -0.6in w =  \sqrt{\frac{2}{3}}q\left(1 + \frac{1}{6|\log r_h|} + {\cal O}\left(\frac{1}{(\log r_h)^2}\right)\right),\nonumber\\
& & 
\end{eqnarray}
with $w=\iota \lambda_L, \ \ \ \ q=\frac{\iota \lambda_L}{v_b}$, where $\lambda_L$ is the Lyapunov exponent\footnote{Lyapunov exponent characterizes the chaotic nature of a given system, and it has bound as $\lambda_L \leq \frac{2 \pi}{\beta}$ \cite{Maldacena:2015waa} with $\beta=\frac{1}{T}$. For a chaotic system, there is an exponential growth of the phase space trajectory with time in terms of the Lyapunov exponent, i.e., $\delta X(t) \sim \delta X(0) e^{\lambda_L t}$. }. The analog of the butterfly velocity $v_b$ \footnote{Butterfly velocity determines how the perturbation propagates in the system under study and the causal structure of the bulk geometry \cite{Qi:2017ttv}; $v_b=\sqrt{\frac{d+1}{d}}$ \cite{Mezei:2016wfz} for AdS-BH geometry with $d$ spatial dimensions.} in our setup, thereby is given by:
\begin{eqnarray}
\label{vb}
v_b = \sqrt{\frac{2}{3}}\left(1 + \frac{1}{6|\log r_h|}\right).
\end{eqnarray}
Hence, $r=r_h$ transmutes from an irregular to a regular singular point.
Now, (\ref{Zs-EOM}) can hence be written as:
\begin{eqnarray}
\label{EOM-Zs}
& & Z_s''(r) -\left(\frac{\gamma_2}{(r-r_h)} + \gamma_3 + \gamma_4(r-r_h)\right)Z_s'(r)
-\left(\frac{\gamma_m^{(2)}}{(r-r_h)} + \gamma_m^{(3)}\right)Z_s(r) = 0,\nonumber\\
& & 
\end{eqnarray}
where $\gamma_{2, 3, 4}, \gamma_m^{(2), (3)}$ are defined in \cite{MQGP-chaos}. 
With $Z_s(r) = (r - r_h)^\alpha z_s(r)$, $\alpha = 0, 1+\gamma_2$. We choose $\alpha=0$. Writing
$z_s(r) =\sum_{n=0}^\infty A_n(r - r_h)^n,$ 
the $z_s(r)$ EOM at every order in $n$, can be written in terms of the matrix equation \cite{Blake:2019otz}:
\begin{eqnarray}
\label{Matrix-eq}
& & \begin{pmatrix}
C_{11} & C_{12} & 0 & 0 & 0 & 0 & . & . & .  \\
C_{21} & C_{22} & C_{23}  & 0 & 0 & 0 & . & . & . \\
. & . & .  & . & . & . & . & . & . \\
. & . & .  & . & . & . & . & . & . \\
\end{pmatrix}
 \begin{pmatrix}
A_{0}   \\
A_{1} \\
.  \\
. \\
\end{pmatrix}=0.
\end{eqnarray}
Pole-skipping points $(w_n,q_n)$ [$n=0,1,2,...$] are those points in the complex $(w,q)$ plane at which retarded Green's function poles are skipped because at these special points, the numerator and denominator both vanish simultaneously in the retarded Green's function. The pole-skipping points are determined by solving the following equation:
\begin{eqnarray}
\label{PS-Eqs}
& & C_{n,n+1}=0, \ \ \ \ {\rm det} M^{(n)}=0,
\end{eqnarray}
where the matrix $M^{(n)}$ by keeping the first $n$ rows and the first $n$ column of matrix $C$. The Lyapunov exponent ($\lambda_L$) and butterfly velocity ($v_b$) can be obtained using \cite{Sil:2020jhr,Blake:2017ris}: the phenomena of pole-skipping was observed in \cite{Grozdanov:2017ajz} for momentum conservation system and in \cite{Blake:2017ris} for energy conservation system.  see \cite{Grozdanov:2018kkt} \cite{Natsuume:2019vcv}\cite{Wu:2019esr} for the study of pole-skipping in the presence of higher derivative terms in the gravity side and \cite{Sil:2020jhr} for a anisotropic plasma. {\bf It was shown in \cite{MQGP-chaos} that in the intermediate-$N$ MQGP limit (\ref{intermediate-N-MQGP-limit}) that supports Contact 3-Structures and induces transverse $SU(3)$ structure of \ref{Transverse-SU3-CMS}, for the same set of values of $(g_s, N, M, N_f)$ as in Table \tcb{2}, (\ref{PS-Eqs}) is satisfied.}

\item {\bf Bulk-to-shear viscosity ratio at intermediate coupling in terms of the speed of sound} (briefly reviewing the "Contact" already made in  \cite{zeta-intermediate}): In the context of non-conformal transport coefficients at intermediate coupling,  strongly coupled (very large `t-Hooft coupling), bulk viscosity $\zeta$ \footnote{The bulk viscosity and shear viscosity of a fluid are given by the variation of the stress-energy tensor $T^{ij}$ of the fluid. The variation of stress-energy tensor from equilibrium can be given as (considering only the spatial part)
\begin{equation*}
\delta T^{ij}=\eta \Biggl(\partial^{i} u^{j} +\partial^{j} u^{i}-\frac{2}{3}g^{ij} \partial\cdot u \Biggr)+\zeta g^{ij} \partial\cdot u,
\end{equation*}
where $\eta\equiv$ Shear Viscosity, $\zeta\equiv$ Bulk Viscosity and $u^i$ is the fluid velocity.} of the type IIA-theory dual of thermal QCD-like theories was found to vary like $\frac{1}{3} - c_s^2$ \cite{Bulk-Viscosity-McGill-IIT-Roorkee} ($c_s$ being the speed of sound\footnote{The trace of the stress-energy tensor carries the information about the pressure. In the rest frame, the fluid pressure, $P$, and the speed of the sound mode, $c_{s}$, are respectively given by: 
$P=-\frac{1}{3}{T^{i}_{i}}, c_s^2=\frac{\partial P}{\partial\epsilon}, \epsilon = T^{00}$ being the energy.}), and $\zeta$ at weak coupling (using kinetic theory and finite temperature field theory), was shown in \cite{Bulk-Viscosity-McGill-IIT-Roorkee} to vary like $\left(\frac{1}{3} - c_s^2\right)^2$.

\noindent (i) Using the results of \cite{EO}\footnote{Also see \cite{Buchel-EO}.} to compute the bulk-to-shear viscosity ratio, with the ${\cal M}$-theory three-form potential $C_3$ given in terms of $B_{\rm NS\ NS}^{\rm IIA}$ \cite{MQGP},\\
\noindent (ii) by considering the most dominant and sub-dominant components of $C_3/B_2^{\rm IIA}$ in the near horizon limit/IR \footnote{One needs to an appropriate angular regularization.}, \\
\noindent and \\
\noindent (iii) the horizon being an irregular singular point of the EOM of the gauge-invariant combination of scalar modes of metric perturbations yielding:
\begin{equation}
\label{oneover3minuscssquared}
\frac{1}{3} - c_s^2 = {\cal O}(1)\frac{\left(g_s M^2\right)\left(g_s N_f\right)|\log r_h|^3}{N}
- {\cal O}(1)\beta
\end{equation}
[implying (dropping terms of ${\cal O}\left(\frac{1}{N^2}\right)), {\cal O}(\beta^2)$, $\left(\frac{1}{3} - c_s^2\right)^2\sim \frac{\left(g_s M^2\right)\left(g_s N_f\right)|\log r_h|^3}{N}\beta$],
we  showed in \cite{zeta-intermediate} that one obtains the bulk-to-shear-viscosity ratio up to terms quartic in curvature in the ${\cal M}$-theory action:
\begin{eqnarray}
\label{EO-2}
& & \hskip -0.6in \frac{\zeta}{\eta} = \left.c_s^4 T^2\int_{\left(S^2_{\rm squashed}\times_w S^3_{\rm squashed}\right)\times_w S^1} \sqrt{g_6}g^{rr} g^{x^{10}x^{10}} \left[
     g^{\theta_1\theta_1} g^{\theta_2\theta_2} - (g^{\theta_1\theta_2})^2\right]\left(\frac{d C_{\theta_1\theta_2x^{10}}}{dT}\right)^2\right|_{r=r_h}\nonumber\\
& & \hskip -0.6in \forall T\in[T_c,\frac{2(1 + \sqrt{3}\varepsilon )}{\sqrt{3}}T_c],
\end{eqnarray}
One may wish to rewrite (\ref{EO-2}) in terms of the running/temperature-dependent YM coupling constant $g(T)$. Using standard KS(Klebanov-Strassler)-like RG-flow equations \cite{KW}, \cite{Bulk-Viscosity-McGill-IIT-Roorkee}, one sees  that (similar to \cite{effec_kin_model_zetaovereta}):
{\footnotesize
\begin{equation}
\label{g(T)}
g^2(T/T_c) \sim \left.\left(e^{-\phi^{\rm IIA}}\int_{S^2}B^{\rm IIA}\right)^{-1}\right|_{r=r_h}
\sim\frac{1}{(g_s M) (g_s N_f)}\left[\left|\log \left(\frac{\sqrt{3}T}{2(1 + \sqrt{3}\varepsilon )T_c}\right)\right| \left(\log N - 3 \log \left(\frac{\sqrt{3}T}{2(1 + \sqrt{3}\varepsilon )T_c}\right)\right)\right]^{-1}, 
\end{equation}
}
and consequently,  (\ref{EO-2}) yields \cite{zeta-intermediate} :
{\scriptsize
\begin{eqnarray}
\label{zetaovereta-g}
& & \hskip -1in \frac{\zeta}{\eta} =  M N^{5/4}N_f^{4/3}g_s^{21/4}\Biggl[-{\cal O}(10^{-4})c_2\left\{-\frac{1}{6} + \left(\frac{1}{3} - c_s^2\right)\right\} N_f^2(g_s^2 M N_f)^{2/3}\left( g^2(T/T_c)\right)^{2/3} \frac{\left(\left|\log \left(\frac{\sqrt{3}T}{2(1 + \sqrt{3}\varepsilon )T_c}\right)\right|\right)^{2/3}}{\left(c_1 + c_2\log  \left(\frac{\sqrt{3}T}{2(1 + \sqrt{3}\varepsilon )T_c}\right)\right)^2} \nonumber\\
& & \hskip -1in  + {\cal O}(10^{-5})c_2N_f^2\left\{-\frac{2}{3} \left(\frac{1}{3} - c_s^2\right)^{\beta}+ \left(\frac{1}{3} - c_s^2\right)^{2,\ \beta}\right\}(g_s^2 M N_f)^{1/3}\left( g^2(T/T_c)\right)^{1/3}\left(\left|\log \left(\frac{\sqrt{3}T}{2(1 + \sqrt{3}\varepsilon )T_c}\right)\right|\right)^{1/3}\nonumber\\
& & \hskip -1in \times \frac{\kappa_\beta}{(g_s N_f)^4}\left(\left|\log \left(\frac{\sqrt{3}T}{2(1 + \sqrt{3}\varepsilon )T_c}\right)\right|\right)^8  \left(g_s^2 M N_f T \frac{d g^2(T/T_c)}{d T}\right)^4\Biggr].
\nonumber\\
& & \hskip -1in \forall T\in[T_c,\frac{2(1 + \sqrt{3}\varepsilon )}{\sqrt{3}}T_c].
\nonumber\\
& & 
\end{eqnarray} 
} Note, the study of gauge-coupling dependence of physical quantities (entropy, free energy, etc.) with the inclusion of ${\cal O}(R^4)$ corrections in the the type IIB dual of the large-$N$ limit of ${\cal N}=4$ SYM, goes all the way back to \cite{Gubser-et-al-gYM-dep-thermodynamics}; in fact, the authors obtained fractional 't-Hooft-coupling dependence in their results reminiscent of the fractional-gauge-coupling results obtained in (\ref{zetaovereta-g}). Also, in the context of type IIB supergravity with a non-trivial dilaton and hence a curved four-dimensional background, curvature-dependent gauge coupling, and gauge-coupling dependent quark-anti-quark potential leading up to the possibility of having confinement, was worked out in \cite{R-dep-g-qqbar-potential}.  From (\ref{zetaovereta-g}), it is hence evident that at intermediate coupling effected via the intermediate-$N$ MQGP limit (\ref{intermediate-N-MQGP-limit}) that supports C3S, using the results of \cite{EO} (or the relevant Kubo's formula in the "hydrodynamical limit" as also shown in \cite{zeta-intermediate}), $\zeta$ (as well as $\frac{\zeta}{\eta}$) varies like a linear combination of the strong and weak coupling results. Further, as mentioned in \cite{zeta-intermediate}, {\bf the $N$-path disconnectedness of the parameter space with reference to Contact Structures in the IR arising from the underlying $G_2$ structure, is mapped to the appearance of fractional powers of the coupling constants which if complexified, would have corresponding to branch-point singularities in the complex coupling constant space, of the complexified $\frac{\zeta}{\eta}$}.
\end{itemize}

\item {\bf Paramagnetic Plasma EoS and Magnetic Photoproduction} ({\it review\ of\  \cite{photoprod+EoS+AC3S}\ on\ connection\ between\ paramagnetic\ EoS\ and\ strong\ magnetic\ field\ Photoproduction\ and\ results\ of\ a\ preliminary\ version\ arXiv:2211.13186 [hep-th]v1\ of\ this\ paper})

\begin{itemize}
\item {\bf Paramagnetic Equation of State}

In \cite{photoprod+EoS+AC3S} (involving one of the co-authors [AM]), from a strong magnetic field Equation-of-State computation, we demonstrated that the holographic dual, in principle, could correspond to several $T>T_c$ scenarios: stable wormhole, stable wormhole transitioning via a smooth crossover to dark energy as the universe cools (in terms of ``boundary time'' \footnote{The "boundary time", relevant to calculating the entanglement entropy of the relevant eternal black hole in the early stages of its evolution before the ``Page time'' and corresponding to the Hartman-Maldacena-like surface using Dong's formula \cite{Dong} for the computation of entanglement entropy in higher derivative theories. 
In ${\cal M}$-theory dual, Hartman-Maldacena(HM)-like surface is a co-dimension two surface which is located at $x^1=x_R$ and corresponds to the embedding $t = t(r)$.  The solution to the HM-like embedding equation $t=t(r)$ at $r=r_h$ with $t(r=r_h)=t_b$, i.e., the "boundary time" was shown in \cite{MQGP-Page} to yield:
\begin{equation}
\label{rh-tb-relation}
r_h=\frac{\left(\frac{2}{3}\right)^{2/3}}{\left(\frac{(c_2-{t_{b_0}}) \
N^{-\frac{3 n_{t_b}}{2}}}{c_1}\right)^{2/3}},
\end{equation}
where $c_{1,2}<0$ and $|c_2|\sim e^{|c_1|}$ (rendering a "Swiss-Cheese" structure to the HM entanglement entropy) and $n_{t_b}={\cal O}(1)$.}), and a paramagnetic pressure/energy-anisotropic plasma. Given that $T>T_c$ QGP is expected to be paramagnetic \cite{Bali et al-magnetic-chi}, the third possibility appears to be the preferred one. We also show that it is not possible that the anisotropic plasma leads to the formation of a compact star.

During heavy nuclei collisions when QGP is produced, the strong magnetic field is observed for a short time \cite{Skokov:2009qp}. QGP is the plasma of charged particles, and hence, it is highly responsive to a strong magnetic fields. Defining the magnetization, $M=-\frac{\partial F}{\partial B}=\frac{\partial P}{\partial B}$, we see that $M = \frac{P^{\rm UV}}{B} + {\cal O}\left(\frac{1}{B^2}\right)$ [$P^{\rm UV}$ being the pressure in the UV as worked out from the negative of the renormalized free energy in \cite{photoprod+EoS+AC3S}] for $B\gg \left(0.15\ {\rm GeV}\right)^2$, which turns out to be positive for strong magnetic fields, implying the anisotropic plasma is paramagnetic for high temperatures above $T_c$ (Fig. \tcb{2}). The magnetic susceptibility, $\chi_{m}=-\frac{\partial^2F}{\partial B^2} = {\cal O}\left(\frac{1}{B}\right)^3$ in our setup which in the large-$B$ limit, we drop and hence $\chi_m\approx0$ (as supported by the negligible value of $\chi_m$ obtained in "parton-hadron-string-dynamics transport approach" \cite{small-chi_m-i}, as well as lattice results \cite{Bali et al-magnetic-chi}). 

\begin{figure}
\begin{center}
\includegraphics[width=0.5\textwidth]{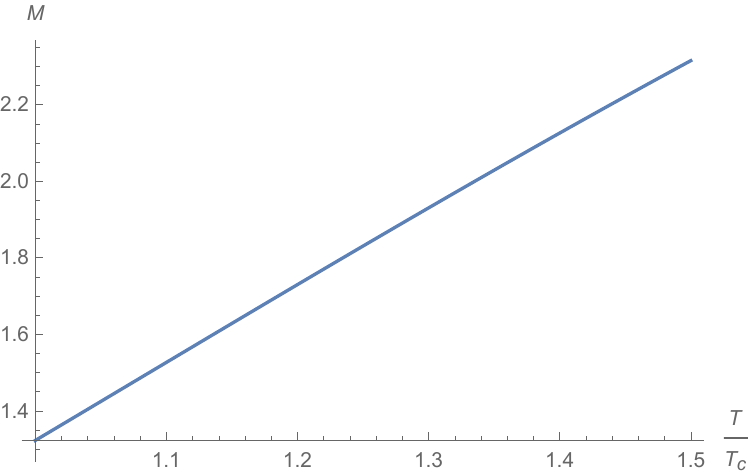}
\end{center}
\caption{Magnetization-vs-Temperature}
\end{figure}

\item
{\bf Photoproduction in the presence of strong magnetic fields}

In \cite{photoprod+EoS+AC3S} (involving one of the authors - AM), the spectral density of photon production in the UV region in a paramagnetic pressure/energy-anisotropic plasma, which is related to the differential photon production rate, was obtained. First, in the absence of an external magnetic field,  the gauge invariant gauge fluctuations, e.g., along $x^2$ axis transverse to the plane containing the external uniform and strong magnetic field along $x^3: {\bf B} = (0,0,B)$,  and the direction of propagation corresponding to $k^{\mu}=(\omega,q=\omega,0,0)$ in $\mathbb{R}^{1,3}( t,x^1,x^2,x^3)$ - $E_{\rm trans }$ [see appendix A of \cite{Holographic-Photoprod-B} for more details] - were obtained. By solving EOMs for $E_{\rm trans }$:
\begin{equation}
\label{Etrans-EOM}
E_{\rm trans}''(Z)+\biggl(\log\left({\cal L}_{DBI} G^{ZZ} G^{x^2x^2}\right)\biggr)'E_{\rm trans}'(Z)-\frac{1}{G^{ZZ}}(G^{tt}\omega^{2}+G^{x^1x^1}q^2)E_{\rm trans}(Z)=0,
\end{equation}
$G = i^*(g + B)^{\rm IIA} + F, i:\Sigma_{D6}\hookrightarrow (S^1_t\times\mathbb{R}^3)\times_w (\mathbb{R}_{\geq0}\times {\cal T}_{NE}^{1,1})$ and $F$ corresponding to $D6$-brane world-volume gauge field, the spectral density of photon production: $\chi_{\rm transverse\ polarization\ [along\ x^2]}(w) \equiv \chi_2(w) \sim \lim_{Z\rightarrow Z_{\rm UV}}\Im m \left(\frac{G^{ZZ}{\cal L}_{\rm DBI}G^{x^2x^2}E_{\rm trans}'(Z;w)}{E_{\rm trans}(Z;w)}\right)$, was obtained:
\begin{eqnarray}
\label{chioverwB0-iii}
& & \frac{\chi_2^{B=0}(\tilde{w})}{4 N^2 \tilde{w} T^2} = {\cal C} \tilde{w} e^{-\frac{16 {\cal R} \tilde{w}^2
   \left(27-\frac{58 e^5\beta}{\xi_{\rm UV}^{B=0}\ }\right)}{1323}-\frac{9}{\xi_{\rm UV}^{B=0}\ }} \left(\beta
   e^{1-\frac{114.513 {\cal R} \tilde{w}^2}{\xi_{\rm UV}^{B=0}\ }}-21\right),
\end{eqnarray}
$\tilde{w}\equiv\frac{w}{2\pi T}$. Using \cite{zeta-intermediate},
\begin{eqnarray}
\label{beta_Bulk_Viscosity_cs_paper}
& & \beta \sim\frac{\left(\frac{24 {g_s} M^2 {N_f}
   ({c_1}+{c_2} \log ({r_h})) \left(\frac{{g_s} M^2 ({c_1}+{c_2} \log
   ({r_h}))}{N}+\frac{1}{\sqrt{3}}\right)^2}{9 \left(\frac{{g_s} M^2 ({c_1}+{c_2}
   \log ({r_h}))}{N}+\frac{1}{\sqrt{3}}\right)^2+1}+\frac{3 N {N_f} (\log (N)-3 \log
   ({r_h}))}{4 \pi }\right)^3}{{g_s}^4 N^3 {N_f}^7 (\log (N)-3 \log ({r_h}))^7}\nonumber\\
& & \sim \frac{1}{|\log r_h|^4}\sim\left(\frac{\left(g_s N_f\right)^{2/3}\left(g_s M^2\right)^{1/3}}{N^{1/3}}\right)^4,
\end{eqnarray}
along with $r_h = e^{-\kappa_{r_h}(g_s, M, N_f)N^{1/3}}$ in ${\cal R}_{D5/\overline{D5}}=1$-units, which for $g_s = 0.1, M=N_f=3, N=100$ and values of $c_{1,2}$ as obtained in \cite{zeta-intermediate} to match with $SU(3)$ Gluodynamics lattice results \cite{lattice-SU3_Glue}, yields $\beta=1.1$ (essentially implying that the higher derivative corrections can not be disregarded). Numerically, one obtains a reasonable match with \cite{D5gaugedSUGRAtrunc} for
\begin{eqnarray}
\label{Const+ratio+xiUV}
{\cal C} = 0.0064,\ \xi_{\rm UV}^{B=0}\  = -3.095, {\cal R} = 0.5727.
\end{eqnarray}
In the presence of a strong magnetic field \footnote{It was shown in  \cite{photoprod+EoS+AC3S}  that $\frac{\partial T_c}{\partial B}<0$ in the presence of a strong magnetic field. The strength of the external magnetic field, in $e=1$-units, is hence determined by $T_c^2(B=0)$, i.e., $B\gg T_c^2(B=0)\sim\left(0.15\ {\rm GeV}\right)^2$ implies a strong magnetic field.}, it was shown in \cite{photoprod+EoS+AC3S},
\begin{equation}
\label{chioverNsqwtildeTsq}
\frac{\chi^B(w)}{N^2\tilde{w}T^2} = \frac{(\kappa_1  + \tilde{\kappa}_1\beta)\tilde{w}e^{-(\kappa_2 + \tilde{\kappa}_2\beta)\tilde{w}^2}}{\tilde{B}^2},
\end{equation}
where $\tilde{B} = \frac{B}{T^2}$. It is after adding a phenomenological  ${\cal O}(w^2)$ term in the numerator that one can show that one obtains a good match with, e.g., the results of \cite{D5gaugedSUGRAtrunc}.

In the consistent trunction of only $A_t^\beta$ being the only non-trivial ${\cal O}(R^4)$-correction to the background $D6$-brane world-volume gauge field, it was shown that there are no ${\cal O}(R^4)$-corrections to the $D6$-brane world-volume gauge field supporting a constant magnetic field and hence no ${\cal O}(R^4)$-corrections to the free energy/pressure and energy densities, and hence no non-analytic-in-complexified-gauge-coupling dependence in complexified pressure/energy density. For the aforementioned values of $(g_s, M, N_f)$, it turns out that $N=100$ results in complex free energy/pressure and energy densities, but, e.g., for $N=200$, one obtains real free energy/pressure and energy densities. Based on the results of  \cite{photoprod+EoS+AC3S}, {\bf this corresponds to the existence of Almost Contact 3-Structures (inducing the transverse $SU(3)$ 3-structures of \ref{Transverse-ACMS}), and not Contact 3-Structures}. We will discuss a pair of conjectures related to this in \ref{summary}.

\end{itemize}

\end{itemize}

\section{Summary and Concluding Remarks}
\label{summary}

\subsection{Summary}

Continuing with the program of studying the differential geometry of non-supersymmetric manifolds relevant to ${\cal M}$-theory dual of  large-$N$ thermal QCD-like theories (equivalence class of theories which are UV-conformal, IR-confining and have fundamental quarks) at intermediate coupling initiated in \cite{OR4}, we also obtained the $G_2$-structure torsion classes of the relevant seven-fold $M_7$ which is the ${\cal M}$-theory circle times the warped product $S^1_{\rm thermal}\times_w M_5, M_5$ being non-Einsteinian deformation of $T^{1,1}$ in the $N\gg1$ MQGP limit (\ref{MQGP_limit}) \cite{MQGP}, \cite{NPB} and the intermediate-$N$ MQGP limit (\ref{intermediate-N-MQGP-limit}):
\begin{equation}
\label{tau-G2-M7x0}
\tau\in \tau_1\oplus\tau_2\oplus\tau_3.
\end{equation}

We then explicitly showed that in the $N\gg1$-MQGP limit (\ref{MQGP_limit}), the aformentioned $M_7$ supports Almost Contact 3-Structures (AC3S). Subsequently, we verified that the AC3S are in fact  Almost Contact  Metric 3-Structures (ACM3S). It turned out that the abovementioned Almost Contact 3-Structures are not a Contact 3-Structures. We then showed that it is possible to explicitly construct a Contact Structure on the seven-fold $M_7$ in the intermediate-$N$ MQGP limit (\ref{intermediate-N-MQGP-limit}). 

Using a proposition of \cite{Ossa et al[2013]}, we not only obtained an explicit transverse $SU(3)$ structure by a reduction of the $G_2$ structure obtained earlier induced by aforementioned Almost Contact Metric Structure, but also a one-parameter family of transverse $SU(3)$ structures by a reduction of the $G_2$ structure induced by the Contact Structure obtained earlier.

Combining the Proposition of \cite{OR4} and this paper, we hence arrive at the following (combined) proposition:

\noindent {\it Proposition 2}: 
\begin{itemize}
\item
The non-K\"{a}hler warped six-fold $M_6$, obtained as a cone over a compact five-fold $M_5$ (a non-Einsteinian generalization of $T^{1,1}$), that appears in the type IIA background corresponding to the ${\cal M}$-theory uplift of thermal QCD-like theories at high temperatures, in the neighborhood of the Ouyang embedding (with vanishingly small embedding parameter) \cite{Ouyang} of the parent type IIB flavor $D7$-branes  $\biggl($that figure in the type IIB string dual of thermal QCD involving $N$ $D3$-branes, $M$ fractional $D3$-branes and $N_f$ flavor $D7$-branes ``wrapping'' $\mathbb{R}_{\geq0}\times S^3\biggr)$  effected by working in the neighborhood of small $\theta_{1,2}$ ($\theta_{1,2}\in[0,\pi]$), in the "MQGP limit"  {\it inclusive of the ${\cal O}(l_p^6),\ (l_p$} being Planck length) {\it corrections},
\begin{itemize}
\item
is a non-complex manifold (though the deviation from $W_{1,2}^{SU(3)}=0$ being $N$-suppressed) - proved in \cite{OR4},
\item
 $W_4^{SU(3)} \sim W_5^{SU(3)}$ (upon comparison with \cite{Butti et al [2004]},  interpreted as ``almost" supersymmetric [in the large-$N$ limit]) - proved in \cite{OR4}.

\end{itemize}

\item
The $G_2$-structure torsion classes of the seven-fold $M_7$ (a cone over the ${\cal M}$-theory-$S^1$-fibration over $M_5$: $p_1^2(M_{11}) = p_2(M_{11}) = 0$ ($p_a$ being the $a$-th Pontryagin's class of $M_{11}$),  are: $\kappa^0_{G_2} = W^{G_2}_{14} \oplus W^{G_2}_{27}$ - proved in \cite{OR4}.

\item
 $M_8$, a warped product of the thermal $S^1$ and $M_7$:  $W^{SU(4)/Spin(7)}_{M_8} = W_2^{SU(4)} \oplus W_3^{SU(4)} \oplus W_5^{SU(4)}/W_1^{Spin(7)} \oplus W_2^{Spin(7)}$ - proved in \cite{OR4}.

\item (Proved in this work - Proposition 1)  $\left.M_7\right|_{r={\rm const}\in{\rm IR}\cap {\rm Ouyang\ embedding}}$ 
\begin{itemize}
\item
has the following non-trivial $G_2$-structure torsion classes: $\kappa^0_{G_2} = W_7^{G_2}\oplus W_{14}^{G_2}\oplus W_{27}^{G_2}$, both, in the $N\gg1$-MQGP limit (\ref{MQGP_limit}) and the intermediate-$N$ MQGP limit (\ref{intermediate-N-MQGP-limit});
\item
supports Almost Contact 3-Metric Structures $(\sigma^1,\sigma^2,\sigma^3) = (e^1, e^7, -e^2)$ [and the dual $(R^1, R^2, R^1\times_\Phi R^2)$] in the $N\gg1$-MQGP limit (\ref{MQGP_limit}) , but the same do not  correspond to Contact 3- Structures;

\item 
supports  Contact 3-Structures: $(\sigma^1,\sigma^2,\sigma^3) =\left(\alpha_1 e^1 + \alpha_3 e^3 + \alpha_7 e^7, \beta_1 e^1 + \beta_4 e^4 + \beta_7 e^7, \sigma^3\right) $, $\{\alpha_{i=1, 2, 3}\}, \{\beta_{j=1, 2, 3}\}\in\mathbb{R}$ wherein $\alpha_1\sim\alpha_3\sim\alpha_7; \beta_1\sim\beta_4\sim\beta_7$ with $\sim$ implying equality up to ${\cal O}(1)$ terms, and $\sigma^3: \sigma^3(R^1\times_\Phi R^2)=1$ in the intermediate-$N$ MQGP limit (\ref{intermediate-N-MQGP-limit});

\item
inherits a transverse $SU(3)$ structure $\left(\Omega_+^{\alpha}, \Omega_-^{\alpha}\right)$ in the $N\gg1$-MQGP limit (\ref{MQGP_limit}), wherein
$\Omega^{\alpha}_+ = \Phi - \sigma^{\alpha}\wedge \omega_\Phi^{\alpha},
\Omega^{\alpha}_- = \sigma^{\alpha}\lrcorner\left(\frac{1}{2}\omega_\Phi^{\alpha}\wedge \omega_\Phi^{\alpha} - *_7\Phi\right), \alpha=1, 2, 3$ from the abovementioned AC3S;

\item
inherits a transverse $SU(3)$ structure in the intermediate-$N$ MQGP limit (\ref{intermediate-N-MQGP-limit})  from the aforementioned C3S, $\left(\Omega_+^{\alpha},\Omega_-^{\alpha}\right)$,
where $\Omega_+^{\alpha} = \Phi - \sigma^{\alpha}\wedge\omega_\Phi^{\alpha}$ and, e.g., for $\alpha=1$, a one-parameter ($\Lambda^{(1)}_{156}$ or $\Lambda_{456}$) family of:
 \begin{eqnarray*}
& & \Omega_- = \Lambda_{AMC}e^{ABC} = \Lambda^{(1)}_{1b_0c_0}e^{1b_0c_0}   + \Lambda_{3b_0c_0}^{(1)}e^{3b_0c_0}
+ \Lambda^{(1)}_{7b_0c_0}e^{7b_0c_0}
+ \Lambda_{a_0b_0c_0}e^{a_0b_0c_0},
\end{eqnarray*}
where $a_0, b_0, c_0=2, 4, 5, 6$ and $\Lambda_{456}$ is the only linearly independent non-vanishing $\Lambda_{a_0b_0c_0}$.
\end{itemize}
\end{itemize}


\subsection{Concluding Remarks}


Proposition 1 is significant as regards our understanding of the classification of geometries relevant to ${\cal M}$-theory uplift of thermal QCD-like theories (the equivalence class of theories that are IR(Infra Red, i.e., at low energies)-confining and UV(Ultra Violet, i.e., at high energies)-conformal with quarks in the fundamental representation of color and flavor) in the following sense. 
\begin{tcolorbox}[enhanced,width=6.5in,size=fbox,
    drop shadow southwest,sharp corners]
The four-parameter space ${\cal X}_{G_2}(g_s, M, N_f; N)$ [$g_s\in(0,1)$ and varying continuously; $M, N_f, N$ varying in steps of 1 such that $M, N_f$ are ${\cal O}(1)$ and $\frac{1}{N}\ll1$] of closed seven-folds supporting $G_2$ structures and relevant to the aforementioned ${\cal M}$-theory uplift of thermal QCD-like theories, is not $N$-path connected with reference to Contact Structures in the IR, i.e., the $N\gg1$ Almost Contact 3-Structures arising from the $G_2$ structure for $\frac{\left(g_s M^2\right)^{m_1}\left(g_s N_f\right)^{m_2}}{N}\ll1, m_{1,2}\in\mathbb{Z}_+\cup\left\{0\right\}$, do not $N$-connect to a Contact 3-Structures (in the IR) which is shown to exist only for an appropriate intermediate $N$ for  $\frac{\left(g_s M^2\right)^{m_1}\left(g_s N_f\right)^{m_2}}{N}<1, m_{1,2}\in\mathbb{Z}_+\cup\left\{0\right\}$ and the aforementioned (realistic QCD-motivated) parameters $M, N_f, g_s$.
\end{tcolorbox}
Put in terms of Physics lingo, the aforementioned implies that the space of magnetohydrodynamics/superconductivity/liquid crystals-like lamellar solutions \footnote{ Using terminology of Sec. \ref{ACM3S-basics}, the "contact 1-form" $\sigma$ of almost contact geometry, if expressible in terms of a pair of functions $f$ and $g: \sigma = f\ dg$ implies $\sigma\wedge d\sigma = 0$, known as the Frobenius theorem, mimics complex lamellar fields ${\bf V}$ in $\mathbb{R}^3$ which are defined as ones satisfying 
${\bf V}.{\bf\nabla}\times{\bf V} = 0,$
 which find applications in magnetohydrodynamics, superconductivity and liquid crystals \cite{A.L.Kholodenko} [There are unfortunately many typos though, but serves as a good reference to original works in Contact Geometry and its applications to Physics.]} and the space of monopole-like solutions in ${\cal M}$-theory duals of large-$N$ thermal QCD-like theories, are not $N$-path connected.  The results of this paper hence provide a novel insight into the differential geometry of non-supersymmetric manifolds relevant to top-down holographic QCD at intermediate coupling.

\noindent {\bf A\ pair\ of\ conjectures}:   Combined with the conjecture of \cite{zeta-intermediate}, namely the existence of Contact 3-Structures is mapped to the non-analytic-complexified-temperature-dependent-gauge-coupling dependence\footnote{The non-analyticity corresponding to branch-point singularities.} of the complexified bulk-to-shear-viscosity ratio, we hence further conjecture the following.

The failure of the space of AC3S to be $N$-path connected to the space of C3S in the parameter space of such structures induced from $G_2$-structures on closed seven-folds (warped product of ${\cal M}$-theory circle and a non-Kähler six-fold with the six-fold being a warped product of the thermal circle with a non-Einsteinian deformation of $T^{1,1}$), conjectured earlier/above in the presence of a strong external magnetic field, to be mapped to the existence of \tcb{non-analytic(corresponding to C3S)}/\tcbr{analytic(corresponding to AC3S)}-dependence of the complexified \tcb{photoproduction spectral density as well as bulk-to-shear-viscosity ratio (or equivalently the speed of sound)}/\tcbr{pressure/energy anisotropic plasma} on the complexified gauge coupling, we conjecture, is the differential geometric analog of the following pair of statements.
\begin{tcolorbox}[enhanced,width=7in,center upper,size=fbox,
    drop shadow southwest,sharp corners]
    \begin{flushleft}
 In the presence of a strong magnetic field, (i) fluctuations in world-volume gauge fields (relevant to, e.g., holographic photoproduction spectral density computation)  can not be finite, unlike the finite background world-volume gauge field (relevant to, e.g., EoS); (ii) in the zero-instanton sector, (type-IIB modular-completion-inspired) ${\cal O}(R^4)$ non-renormalized gauge fields corresponding to AC3S produce ${\cal O}(R^4)$-corrected gauge fluctuations corresponding to C3S.
\end{flushleft}
\end{tcolorbox}
Lastly, motivated by the above conjecture and a pair of observations: (i) replacing $M$ and $N_f$ by the effective number of $D5$-branes and $D7$-branes respectively in the parent type IIB dual of \cite{metrics},  (\ref{beta_Bulk_Viscosity_cs_paper}) implies that $\beta\stackrel{\rm UV}{\longrightarrow}0$ as there is no net effective $D5/D7$-brane charge in the UV validating the expected UV conformality, and (ii) the ${\cal O}(R^4)$-corrections to the ${\cal M}$-theory uplift of thermal QCD-like theories vanish in the UV \cite{Vikas+Gopal+Aalok}, we now conjecture:
\begin{tcolorbox}[enhanced,width=7in,center upper,size=fbox,
    drop shadow southwest,sharp corners]
    \begin{flushleft}
   \hskip 2.5 in C3S $\stackrel{\rm UV}{\longrightarrow}$ AC3S.
    \end{flushleft}
    \end{tcolorbox}
Note, the above is not a contradiction of the lack of $N$-path connectedness in the parameter space of (A)C3S that stemmed as a consequence of {\it Lemma\ 4(b)}. First, the latter was in the IR. What the last conjecture above is based on is given that the effective number of $D5/D7$-branes can RG-flow continuously with $r$, the ${\cal O}(R^4)$ corrections become vanishingly small in the UV.  
 
For convenience, the results along with their equation-wise occurrence in the paper, are summarized in the following table:
\begin{table}[hbt!]
\begin{center}
\begin{tabular}{|c|c|c|c|}\hline
&&&\\
S. No. & Structure Type & Description & Reference in the paper \\
&&& wherefrom Structure obtained \\
&&&\\ \hline
&&&\\
1. & $G_2$-structure torsion classes $\tau$ & $\tau\in \tau_1\oplus\tau_2\oplus\tau_3$ & Section \ref{G2-M7x0} \\
&&& \\ \hline
&&& \\
2. & Almost Contact 3-Structures & $(\sigma^1, \sigma^2, \sigma^3) = (e^1, e^7, -e^2)$ & (\ref{ACM3S-x0})-(\ref{ACM3S-x0-sigmas}) in \ref{ACM3S-M7-x0} \\
&&& \\ \hline
&&& \\
3. & Almost Contact Metric 3-Structures & $g_{mn}J^{m\ (a)}_{\ \ m_1}J^{n\ (a)}_{\ \ n_1} $ &  \ref{AC3S-M7x0} \\
&& $= g_{m_1n_1} - \sigma^{(a)}_{m_1}\sigma^{(a)}_{n_1}$ & \\
&&& \\ \hline
&&& \\
4. & Contact 3-Structures & $\sigma^1 = \alpha_1 e^1 + \alpha_3 e^3 + \alpha_7 e^7$ & \\
&& $\sigma^2 = \beta_1 e^1 + \beta_4 e^4 + \beta_7 e^7$ &  \ref{C3S-M7x0} \\ 
&& $\sigma^3:\sigma^3(R^1\times_\Phi R^2)=1$ & \\
&&& \\ \hline
&&& \\
5. & Transverse $SU(3)$  &  (\ref{Transverse-conditions}) & (\ref{Omega+-12_i})  in  \ref{Transverse-ACMS} \\
& 3-Structure from  && \\
& Almost Contact 3-Structures && \\
&&& \\ \hline
&&& \\
6. & Transverse $SU(3)$ Structure  & $(\Omega_+,\left.\Omega_-\right|_{(\ref{Omega-})})$ & (\ref{Transverse-conditions}), (\ref{solution-LambdaABC}) in \ref{Transverse-SU3-CMS}, \\
& from Contact Structure && App. \ref{Omega_-} \\
&&& \\ \hline
\end{tabular}
\end{center}
\caption{Summary of Results for $M_7:$ a warped product of the ${\cal M}$-theory circle and the warped product $S^1_{\rm thermal}\times_w M_5, M_5$ being a non-Einsteinian deformation of $T^{1,1}$ - $G_2$ Structure torsion classes, (Almost) Contact (3) (Metric) Structures and Transverse $SU(3)$ Structures in the $N\gg1$-MQGP limit (\ref{MQGP_limit}) (entries in rows marked as S. No. 1. - 3., 5. in this table), and intermediate-$N$ MQGP limit (\ref{intermediate-N-MQGP-limit}) (entries in rows marked as S. No. 4. and 6. in this table)}
\end{table}

\newpage

\section*{Acknowledgements}

 Some of the results were presented by one of us (AM) in  seminars at UC Santa Barbara, UC Berkeley, Purdue, U. British Columbia and McGill. We thank S. Sarkar for verifying some results of \ref{G2-M7x0} and \ref{ACM3S-M7-x0} as part of his Master's project.

\noindent {\it Data availability statement}: The authors declare that this paper does not make use of any publicly available data. The publicly available papers/eprints some of which also include the authors' papers/eprints whose results have either been used or compared with, have been appropriately cited.

\section{Funding and/or Conflicts of interests/Competing interests}

AM is partly supported by a Core Research Grant number SER-1829-PHY from the Science and Engineering Research Board, Govt. of India. GY was partly supported by a Senior Research Fellowship (SRF) from the Council of Scientific and Industrial Research, Govt. of India. The authors declare that there is no conflict of interest or competing interest.

\appendix

\section{$\Omega_-: i_{R^{(1)}}\Omega_-=0\cap *\Phi = -\sigma^{(1)}\wedge\Omega_- + \frac{1}{2}\omega_\Phi^2$}
\label{Omega_-}
\setcounter{equation}{0}\seceqaa

Relevant to the explicit construction of a transverse $SU(3)$ structure in \ref{SU3-from-G2} and in particular, \ref{SU3-from-G2}, arising from the contact structure constructed in \ref{C3S-M7x0},  in this appendix we discuss the details pertaining to solution of (\ref{iROmega-0}) and (\ref{Transverse-condition-iii}). 

From the ansatz (\ref{Omega-}), one obtains:
{\footnotesize
\begin{eqnarray}
\label{iROmega-0}
& & \hskip -0.3in i_{R^{(1)}}\Omega_- = \Lambda^{(1)}_{1b_0c_0}\left(R_{(1)}.e^1e^{b_0c_0}
- R_{(1)}.e^{b_0}e^{c_01} + R_{(1)}.e^{c_0}e^{b_01}\right)  + \Lambda^{(2)}_{13c_0}
\left(R_{(1)}.e^1 e^{3c_0} - R_{(1)}.e^3 e^{c_01} + R_{(1)}.e^{c_0}e^{31}\right)\nonumber\\
& & \hskip -0.3in  +  \Lambda^{(3)}_{17c_0}\left(R_{(1)}.e^1e^{7c_0}
- R_{(1)}.e^7e^{c_01} + R_{(1)}.e^{c_0}e^{71}\right) 
+  \Lambda^{(4)}_{137}\left(R_{(1)}.e^1e^{37}
- R_{(1)}.e^3e^{71} + R_{(1)}.e^7e^{31}\right)\nonumber\\
& &\hskip -0.3in + \Lambda^{(1)}_{3b_0c_0}\left(R_{(1)}.e^3e^{b_0c_0}
- R_{(1)}.e^{b_0}e^{c_01} + R_{(1)}.e^{c_0}e^{37}\right) \nonumber\\
& & \hskip -0.3in  +  \Lambda^{(3)}_{37c_0}\left(R_{(1)}.e^3e^{7c_0}
- R_{(1)}.e^7e^{3c_0} + R_{(1)}.e^{c_0}e^{37}\right)
+  \Lambda^{(1)}_{7b_0c_0}\left(R_{(1)}.e^7e^{b_0c_0}
- R_{(1)}.e^{b_0}e^{7c_0} + R_{(1)}.e^{c_0}e^{7b_0}\right)\nonumber\\
& & \hskip -0.3in + \lambda_{a_0b_0c_0}\left(R_{(1)}.e^{a_0}e^{b_0c_0} - R_{(1)}.e^{b_0}e^{a_0c_0} + R_{(1)}.e^{c_0}e^{a_0b_0}\right) = 0,
\end{eqnarray}
}
($a_0, b_0, c_0 = 2, 4, 5, 6$) that yields:
{\footnotesize
\begin{eqnarray}
\label{iROmega-0-ii}
& & \Lambda^{(1)}_{1b_0c_0}R_{(1)}.e^1 + \Lambda^{(3)}_{3b_0c_0}R_{(1)}.e^3 + \Lambda^{(1)}_{7b_0c_0}R_{(1)}.e^7 + 3 \Lambda_{a_0b_0c_0}R_{(1)}.e^{a_0} = 0,\nonumber\\
& & \Lambda^{(1)}_{1b_0c_0}R_{(1)}.e^{b_0} + \Lambda^{(2)}_{13c_0}R_{(1)}.e^3 + \Lambda^{(3)}_{17c_0}R_{(1)}.e^7 = 0,\nonumber\\
& & \Lambda^{(2)}_{13c_0}R_{(1)}.e^{(1)} - 2\Lambda^{(1)}_{3b_0c_0}R_{(1)}.e^{b_0}
- \Lambda^{(3)}_{37c_0}R_{(1)}.e^7 = 0,\nonumber\\
& & \Lambda^{(3)}_{17c_0}R_{(1)}.e^1 + \Lambda^{(3)}_{37c_0}R_{(1)}.e^3 - 2 \Lambda^{(1)}_{7b_0c_0}R_{(1)}.e^{b_0} = 0,\nonumber\\
& & \Lambda^{(2)}_{13c_0}R_{(1)}.c^{c_0} + \lambda^{(1)}_{137}R_{(1)}.e^7 = 0,\nonumber\\
& & \Lambda^{(4)}_{137}R_{(1)}.e^3 = 0,\nonumber\\
& & \Lambda^{(3)}_{17c_0}R_{(1)}.e^{c_0} = 0,\nonumber\\
& & \Lambda^{(4)}_{137}R_{(1)}.e^1 + \Lambda^{(3)}_{37c_0}R_{(1)}.e^{c_0} = 0.
\end{eqnarray}
}
From (\ref{iROmega-0-ii}), 
\begin{eqnarray}
\label{iROmega-0-iii}
& & \Lambda^{(2)}_{13c_0} = \Lambda^{(1)}_{17c_0} = \Lambda^{(3)}_{37c_0} = \Lambda^{(3)}_{17c_0} = 0,\ \Lambda^{(4)}_{137} = 0.
\end{eqnarray}
Substituting
{\footnotesize
\begin{eqnarray}
\label{Transverse-condition-i}
& & \hskip -0.5in \omega_\Phi^2 = 2 \Biggl(\alpha_1\alpha_3 \Omega^1_a\Omega^3_{bc}e^{a1bc} +
\alpha_1\alpha_7\Omega^1_a\Omega^7_be^{a1b7} + \alpha_3\alpha_7\Omega^3_{bc}\Omega^7_ae^{bca7} \Biggr)+ \alpha_3^2\Omega^3_{bc}\Omega^3_{df}e^{bcdf}\nonumber\\
& & \hskip -0.5in  = 2\Biggl(\alpha_1\alpha_3\Omega^1_3\Omega^3_{b_0c_0}e^{31b_0c_0}
+ 2\alpha_1\alpha_3 \Omega^1_{a_0}\Omega^3_{3c_0}e^{a13c_0} + \Omega^1_3\Omega^7_{b_0}e^{31b_07}\nonumber\\
& & \hskip -0.5in + \Omega^1_{a_0}\Omega^7_3e^{a_0137} + 2\alpha_3\alpha_7\Omega^3_{3c_0}\Omega^7_{a_0}e^{3c_0a_07} + \alpha_3\alpha_7\Omega^3_{b_0c_0}\Omega^7_3e^{b_0c_037}\Biggr)\nonumber\\
& & \hskip -0.5in + 2\Biggl(\alpha_1\alpha_3 \Omega^1_{a_0}\Omega^3_{b_0c_0}e^{a_01b_0c_0} +
\alpha_1\alpha_7 \Omega^1_{a_0}\Omega^7_{b_0}e^{a_01b_07} + \alpha_3\alpha_7\Omega^3_{c_0c_0}\Omega^7_{a_0}e^{b_0c_0a_07}\Biggr) + \alpha^2\Omega^3_{b_0c_0}\Omega^3_{d_0f_0}e^{b_0c_0d_0f_0};
\end{eqnarray}
\begin{eqnarray}
\label{Transverse-condition-ii}
& & \Omega^3_{b_0c_0}\Omega^3_{d_0f_0}e^{b_0c_0d_0f_0} = \Biggl(\Omega^3_{24}\Omega^3_{56} - \Omega^3_{25}\Omega^3_{46} + \Omega^3_{26}\Omega^3_{45}\Biggr)
e^{2456}\approx 0,
\end{eqnarray}
}
and
\begin{eqnarray}
\label{Transverse-condition-iiiprime}
& & \sigma^1 \wedge \omega_\Phi = \alpha_1\alpha_3\left(\Omega^3_{bc} e^{1bc} + \Omega^1_ae^{3a1}\right) + \alpha_1\alpha_7\left(\Omega^7_a e^{1a7} + \Omega^1_ae^{7a1}\right)\nonumber\\
& & + \alpha_3^2 \Omega^3_{bc}e^{3bc} + \alpha_3\alpha_7\left(\Omega^7_ae^{3a7} + \Omega^3_{bc}e^{7bc}\right),
\end{eqnarray}
into (\ref{Transverse-conditions}) one can obtain $\Omega_+$, and the following set of conditions on $\Lambda^{(i)}_{ABC}$'s:
{\footnotesize
\begin{eqnarray}
\label{Transverse-condition-iii}
& (i) & -\alpha_{[1}\Lambda^{(1)}_{3]b_0c_0} + \alpha_1\alpha_3\Omega^1_{[3}\Omega^3_{b_0]c_0} = 0, b_0, c_0\neq 2, 4;\nonumber\\
& &  -\alpha_1\Lambda^{([1)}_{3]24} + \alpha_1\alpha_3\Omega^1_{[3}\Omega^3_{2]4} = 1,\nonumber\\
& (ii)  & -\left(\alpha_1\Lambda^{(3)}_{37c_0} - \alpha_3\Lambda^{(3)}_{17c_0} + \alpha_7\Lambda^{(3)}_{13c_0}\right) + \frac{1}{2}\Omega^1_{[3}\Omega^7_{c_0]} = 0, c_0\neq6;\nonumber\\
&  &  -\left(\alpha_1\Lambda^{(3)}_{376} - \alpha_3\Lambda^{(3)}_{176} + \alpha_7\Lambda^{(3)}_{136}\right) + \frac{1}{2}\Omega^1_{[3}\Omega^7_{6]} = 1,\nonumber\\
&(iii) & -\left(\alpha_3\Lambda^{(1)}_{7b_0c_0} - \alpha_7\Lambda^{(1)}_{3b_0c_0}\right)
+ \frac{1}{2}\left(2\alpha_3\alpha_7 \Omega^3_{[3|c_0}\Omega^7_{a_0]} \right)=0, b_0, c_0\neq2,5;\nonumber\\
&  & -\left(\alpha_3\Lambda^{(1)}_{725} - \alpha_7\Lambda^{(1)}_{325}\right)
+ \frac{1}{2}\left(2\alpha_3\alpha_7 \Omega^3_{[3|5}\Omega^7_{2]} \right)=-1,\nonumber\\
& (iv)  & -\Lambda_{a_0b_0c_0} + \alpha_3^2\Omega^3_{3[c_0}\Omega^3_{a_0]b_0}=0,
a_0, b_0, c_0\neq 4, 5, 6;\nonumber\\
& & -\Lambda_{456} + \alpha_3^2\Omega^3_{3[6}\Omega^3_{4]5}=1,\nonumber\\
& (v) & \Lambda_{a_0b_0c_0} + \alpha_1\alpha_3\Omega^1_{a_0}\Omega^3_{b_0c_0} = 0, a_0, b_0, c_0\neq 2, 5, 6;\nonumber\\
& & \Lambda_{256} + \alpha_1\alpha_3\Omega^1_2\Omega^3_{56} = 0,\nonumber\\
& (vi) & -\alpha_{[1}\Lambda^{(1)}_{7]b_0c_0} + \frac{1}{2}\alpha_1\alpha_7 \Omega^1_{[c_0}
\Omega^7_{b_0]} = 0, b_0, c_0\neq 4, 5;\nonumber\\
& & -\alpha_{[1}\Lambda^{(1)}_{7]45} + \frac{1}{2}\alpha_1\alpha_7\Omega^1_{[5}\Omega^7_{4]} = 1,\nonumber\\
& (vii) & \Lambda_{a_0b_0c_0} + \frac{1}{2}\alpha_3\alpha_7\Omega^3_{b_0c_0}\Omega^7_{a_0} = 0, a_0, b_0, c_0\neq 2, 4, 6;\nonumber\\
&  & \Lambda_{246} + \frac{1}{2}\alpha_3\alpha_7\Omega^3_{46}\Omega^7_2 = 1.
\end{eqnarray}
}
We will now explicitly show that for $M=3$ (corresponding to the number of colors in the IR to be three), $N_f=3\ {\rm or}\ 2$ (corresponding to the lightest quarks in QCD consistent with the vanishing Ouyang-embedding modulus of the flavor $D7$-branes in the parent type IIB dual of \cite{metrics}) and $g_s=0.1$ (corresponding to the QCD fine structure constant at the EW scale), one obtains a transverse $SU(3)$ structure arising from a contact structure. In the following, we would be assuming $\alpha_{\theta_1}, \alpha_{\theta_2}={\cal O}(1): \alpha_{\theta}\equiv\frac{\sqrt{\alpha_{\theta_2}}}{\alpha_{\theta_1}} = {\cal O}(1)$. We show that for the aforementioned values of $g_s, M, N_f$ to be able to make direct contact with real QCD, one obtains a transverse $SU(3)$ structure arising from the contact structure of Sec. \ref{C3S-M7x0} for $N\sim100$. In the following, we work (\ref{Transverse-condition-iii}) out for the aforementioned values of $g_s, M, N_f$ and then perform an expansion about $N=100$\footnote{Such an expansion only makes sense when the discretely varying $N$ is continuous;  for large values of $N$, the same is approximately valid or one can promote $N$ to a continuous variable such that its integral values alone are relevant for making contact with QCD-like theories.}, and show that the contributions from deviations of $N$ from 100 are numerically suppressed, and in the process also obtain the upper bound on this deviation.

We will again consider the QCD-motivated values $g_s = 0.1, M=N_f=3$ (Table \tcb{2}), and 
\begin{eqnarray}
\label{deloc-constants-etc}
& & \alpha_{\theta_{1,2}},\ \alpha_\theta \equiv {\cal O}(1), \kappa_{r_h}\equiv\frac{1}{{\cal O}(1)}, \kappa_{\log r_h}\equiv\left({\cal O}(1)\right)^2
\end{eqnarray}
 $[(\ref{R1.e2 ii})\ {\rm to\ implement}:\ i^*_R\Phi = \omega = d\sigma]$ in the following. We will, once again, follow the same logic as explained in the text containing (\ref{Taylor-around-N_0}) in the proof of {\it Lemma 4} on the construction of  explicit Contact 3-Structures. In the following we show that the deviations from $N=100: \delta_N \equiv \frac{N - 100}{100}$, are extremely suppressed as compared to the contribution at $N=100$.
\begin{itemize}
\item
(\ref{Transverse-condition-iii}) (i) implies:
{\footnotesize
\begin{eqnarray*}
& (1)  & \Omega^{1,\ \beta^0}_3 \Omega^{3,\ \beta^0}_{25} - \Omega^{1, \beta^0}_5 \Omega^{3,\ \beta^0}_{32}\nonumber\\
& & \sim \frac{e^{-\alpha_\theta N^{7/40}}N^{4/5}\log N}{\alpha_{\theta_1}^4} = \frac{10^3 e^{-10 \kappa_{r_h}}}{\alpha_{\theta_1}^4}\left[1 + \delta_N + 10^{-2}\delta_N^2\right],
\end{eqnarray*}
}
which tells us that: $\Omega^{1,\ \beta^0}_3 \Omega^{3,\ \beta^0}_{25} - \Omega^{1, \beta^0}_5 \Omega^{3,\ \beta^0}_{32}\ll 1$.

Similarly,
{\footnotesize
\begin{eqnarray}
\label{i}
&(2)  &  \Omega^{1,\ \beta^0}_3 \Omega^{3,\ \beta^0}_{26} - \Omega^{1, \beta^0}_2 \Omega^{3,\ \beta^0}_{36}
= \delta_N\left[1 + \delta_N + 10^{-2}\delta_N^2\right]\ll1;\nonumber\\
& (3) &  \Omega^{1,\ \beta^0}_3 \Omega^{3,\ \beta^0}_{45} - \Omega^{1, \beta^0}_4 \Omega^{3,\ \beta^0}_{35} = \frac{1}{\left({\cal O}(1)\right)^3}\left(1 +  \delta_N + 10^{-2}\delta_N^2\right)\ll 1;\nonumber\\
& (4) & \Omega^{1,\ \beta^0}_3 \Omega^{3,\ \beta^0}_{46} - \Omega^{1, \beta^0}_4 \Omega^{3,\ \beta^0}_{36}\approx0;\nonumber\\
& (5) & \Omega^{1,\ \beta^0}_3 \Omega^{3,\ \beta^0}_{56} - \Omega^{1, \beta^0}_5 \Omega^{3,\ \beta^0}_{36} = \frac{{\cal O}(10)}{({\cal O}(1))^5}\left[1 +  \delta_N + 10^{-2}\delta_N^2\right]\ll1;\nonumber\\
& (6) & \Omega^{1,\ \beta^0}_3 \Omega^{3,\ \beta^0}_{24} - \Omega^{1, \beta^0}_4 \Omega^{3,\ \beta^0}_{32} = \alpha_3 \Lambda^{(1)}_{124} - \alpha_1 \Lambda^{(1)}_{324} \sim {\cal O}(10^3)\alpha_\theta\left[1 +  \delta_N + \delta_N^2\right].\nonumber\\
& & 
\end{eqnarray}
}
\item Relevant to  (\ref{Transverse-condition-iii}) (ii) are:
{\footnotesize
\begin{eqnarray}
\label{ii}
& (1)  & \Omega^{1, \beta^0}_{[3} \Omega^{7, \beta^0}_{2]} =
{\cal O}(10^{-2})\left[1 +\delta_N +\delta_N^2\right]\ll1;\nonumber\\
& (2) & \Omega^{1, \beta^0}_{[3} \Omega^{7, \beta^0}_{4]} =
\left.\frac{100}{\kappa_{\log r_h}^4}\left[1 + \delta_N + \delta_N^2\right]\right|_{\kappa_{\log r_h}\geq5}\ll1;\nonumber\\
& (3) &  \Omega^{1, \beta^0}_{[3} \Omega^{7, \beta^0}_{5]} = \left.
-\frac{{\cal O}(1)}{\kappa_{r_h}^4}\left[1 + \delta_N + \delta_N^2\right]\right|_{\kappa_{\log r_h}\geq5}\ll1 ;\nonumber\\
& (4)  & \Omega^{1, \beta^0}_{[3} \Omega^{7, \beta^0}_{6]} = \frac{{\cal O}(10^{-1})}{\kappa_{\log r_h}^2\alpha_{\theta_1}^2}\ll1,\nonumber\\
& & {\rm Using}\ \Lambda^{(3)}_{17a_0}=0,\nonumber\\
& & \alpha_1\Lambda^{(3)}_{376} + \alpha_7\Lambda^{(3)}_{136} =  -1.
\end{eqnarray}
}

\item Relevant to (\ref{Transverse-condition-iii}) (iii) are:
{\footnotesize
\begin{eqnarray*}
& (1) & 2\left(\Omega^{3,\ \beta^0}_{32}\Omega^{7,\ \beta^0}_4 - \Omega^{3,\ \beta^0}_{34}\Omega^{7, \beta^0}_2 + \Omega^{3,\ \beta^0}_{24}\Omega^{7, \beta^0}_3\right)\nonumber\\
& & =  - \frac{{\cal O}(10^3)\alpha_\theta^2}{\kappa_{\log r_h}} + \frac{{\cal O}(10^2)e^{-10\kappa_{r_h}} - \frac{{\cal O}(10)\kappa_{\log r_h}}{\alpha_\theta^2}}{\alpha_{\theta_1}^4}+ \xi(\kappa_{r_h}, \alpha_\theta, \kappa_{\log r_h})\delta_N + {\cal O}(\delta_N^2). \nonumber\\
\end{eqnarray*}
}
One can show that for $\alpha_{\theta_1}, \alpha_\theta \equiv {\cal O}(1), \kappa_{\log r_h}
\equiv \left({\cal O}(1)\right)^2, \kappa_{r_h}\equiv\frac{1}{{\cal O}(1)}$ (\ref{deloc-constants-etc}),\\ $- \frac{{\cal O}(10^3)\alpha_\theta^2}{\kappa_{\log r_h}} + \frac{{\cal O}(10^2)e^{-10\kappa_{r_h}} - \frac{{\cal O}(10)\kappa_{\log r_h}}{\alpha_\theta^2}}{\alpha_{\theta_1}^4}=0$. One hence obtains:
{\footnotesize
\begin{eqnarray*}
& &  2\left(\Omega^{3,\ \beta^0}_{32}\Omega^{7,\ \beta^0}_4 - \Omega^{3,\ \beta^0}_{34}\Omega^{7, \beta^0}_2 + \Omega^{3,\ \beta^0}_{24}\Omega^{7, \beta^0}_3\right)      
= \frac{\delta_N}{{\cal O}(1)}\left[1 + \delta_N^2\right]\ll1;
\end{eqnarray*}
\begin{eqnarray}
\label{iii}
&(2)   & 2\left(\Omega^{3,\ \beta^0}_{32}\Omega^{7,\ \beta^0}_5 - \Omega^{3,\ \beta^0}_{35}\Omega^{7, \beta^0}_2 + \Omega^{3,\ \beta^0}_{25}\Omega^{7, \beta^0}_3\right)\nonumber\\
& & = -\frac{{\cal O}(10^5)e^{-10\kappa_{r_h}}}{\alpha_{\theta_1}^4}\left[1 + \delta_N + {\cal O}\left(\delta_N^2\right)\right];\nonumber\\
&(3) & 2\left(\Omega^{3,\ \beta^0}_{32}\Omega^{7,\ \beta^0}_6 - \Omega^{3,\ \beta^0}_{36}\Omega^{7, \beta^0}_2 + \Omega^{3,\ \beta^0}_{26}\Omega^{7, \beta^0}_3\right) = \delta_N + \delta_N^2;\nonumber\\
&(4)  & 2\left(\Omega^{3,\ \beta^0}_{34}\Omega^{7,\ \beta^0}_5 - \Omega^{3,\ \beta^0}_{35}\Omega^{7, \beta^0}_4 + \Omega^{3,\ \beta^0}_{45}\Omega^{7, \beta^0}_3\right) = \delta_N + \delta_N^2;\nonumber\\
& (5) & 2\left(\Omega^{3,\ \beta^0}_{34}\Omega^{7,\ \beta^0}_6 - \Omega^{3,\ \beta^0}_{36}\Omega^{7, \beta^0}_4 + \Omega^{3,\ \beta^0}_{46}\Omega^{7, \beta^0}_3\right) = -\frac{{\cal O}(10^3)}{\left({\cal O}(1)\right)^6}\left[1 + \delta_N + \delta_N^2\right]\ll1;\nonumber\\
& (6) & 2\left(\Omega^{3,\ \beta^0}_{35}\Omega^{7,\ \beta^0}_6 - \Omega^{3,\ \beta^0}_{36}\Omega^{7, \beta^0}_5 + \Omega^{3,\ \beta^0}_{56}\Omega^{7, \beta^0}_3\right)\ll1\nonumber\\
& &
\end{eqnarray} 
}

\item Relevant to (\ref{Transverse-condition-iii}) (iv) are:
{\footnotesize
\begin{eqnarray}
\label{iv}
&  & \hskip -0.8in (1) -\left(\Omega^{3,\ \beta^0}_{24} \Omega^{3,\ \beta^0}_{35} + \Omega^{3,\ \beta^0}_{45} \Omega^{3,\ \beta^0}_{32} - 
   \Omega^{3,\ \beta^0}_{25} \Omega^{3,\ \beta^0}_{34}\right)=\delta_N\ll1\nonumber\\
&  & \hskip -0.8in (2) -\left(\Omega^{3,\ \beta^0}_{24} \Omega^{3,\ \beta^0}_{36} + \Omega^{3,\ \beta^0}_{46} \Omega^{3,\ \beta^0}_{32} - 
   \Omega^{3,\ \beta^0}_{26} \Omega^{3,\ \beta^0}_{34}\right)={\cal O}(10^{-1})\left[1 + 10^{-1}\delta_N + 10^{-2}\delta_N^2\right]\ll1  \nonumber\\
& & \hskip -0.8in (3)  -\left(\Omega^{3,\ \beta^0}_{25} \Omega^{3,\ \beta^0}_{36} + \Omega^{3,\ \beta^0}_{56} \Omega^{3,\ \beta^0}_{32} - 
   \Omega^{3,\ \beta^0}_{26} \Omega^{3,\ \beta^0}_{35}\right)
 ={\cal O}(10^{-1})\left[1 + \delta_N + 10^{-2}\delta_N^2\right]\ll1 ;\nonumber\\ 
& & \hskip -0.8in (4) -\left(\Omega^{3,\ \beta^0}_{45} \Omega^{3,\ \beta^0}_{36} + \Omega^{3,\ \beta^0}_{56} \Omega^{3,\ \beta^0}_{34} - 
   \Omega^{3,\ \beta^0}_{46} \Omega^{3,\ \beta^0}_{35}\right)={\cal O}(10^{-1})\left[1 + \delta_N + \delta_N^2\right]\ll1\nonumber\\
\end{eqnarray}
}

\item Relevant to (\ref{Transverse-condition-iii}) (v) are:
{\footnotesize
\begin{eqnarray}
\label{v}
&  & \hskip -1in  (1) \Omega^{1, \beta^0}_2 \Omega^{3,\ \beta^0}_{45} - \Omega^{1,\ \beta^0}_4 \Omega^{3,\ \beta^0}_{25} + 
 \Omega^{1,\ \beta^0}_5 \Omega^{3,\ \beta^0}_{24}=\frac{{\cal O}(10)}{\left({\cal O}(1)\right)^6}\left[1 + \delta_N + 10^{-1}\delta_N^2\right]\ll1\nonumber\\
&   & \hskip -1in (2) \Omega^{1, \beta^0}_2 \Omega^{3,\ \beta^0}_{46} - \Omega^{1,\ \beta^0}_4 \Omega^{3,\ \beta^0}_{26} + 
 \Omega^{1,\ \beta^0}_6 \Omega^{3,\ \beta^0}_{24}=\frac{{\cal O}(10)}{\left({\cal O}(1)\right)^6}\left[1 + \delta_N + 10^{-2}\delta_N^2\right]\ll1;\nonumber\\
&  & \hskip -1in (3) \Omega^{1, \beta^0}_4 \Omega^{3,\ \beta^0}_{56} - \Omega^{1,\ \beta^0}_5 \Omega^{3,\ \beta^0}_{46} + 
 \Omega^{1,\ \beta^0}_6 \Omega^{3,\ \beta^0}_{45}= \frac{{\cal O}(10)}{\left({\cal O}(1)\right)^6}\left[1 + \delta_N +10^{-2}\delta_N^2\right]\ll1;\nonumber\\
&  & \hskip -1in (4) \left.\Omega^{1, \beta^0}_2 \Omega^{3,\ \beta^0}_{56} - \Omega^{1,\ \beta^0}_5 \Omega^{3,\ \beta^0}_{26} + 
 \Omega^{1,\ \beta^0}_6 \Omega^{3,\ \beta^0}_{25}\right|_{\alpha_\theta,\kappa_{r_h}={\cal O}(1)} = \frac{{\cal O}(10)}{\left({\cal O}(1)\right)^6}\left[1 + \delta_N + 10^{-2}\delta_N^2\right]\ll1\nonumber\\
 & & 
\end{eqnarray}
}

\item Relevant to  (\ref{Transverse-condition-iii}) (vi) are:
{\footnotesize
\begin{eqnarray}
\label{vi}
& (1) & \Omega^{1,\ \beta^0}_{[2} \Omega^{7,\ \beta^0}_{4]} = {\cal O}(10^{-3})
\left[1 + \delta_N +\delta_N^2\right]\ll1\nonumber\\
& (2) & \Omega^{1,\ \beta^0}_{[2} \Omega^{7,\ \beta^0}_{5]}  = {\cal O}(10^{-2})
\left[1 +\delta_N +\delta_N^2\right]\ll1\nonumber\\
& (3) & \Omega^{1,\ \beta^0}_{[2} \Omega^{7,\ \beta^0}_{6]}  = {\cal O}(10^{-3})
\left[1 + 10^{-1}\delta_N + 10^{-2})\delta_N^2\right]\ll1;\nonumber\\
& (4) &  \Omega^{1,\ \beta^0}_{[4} \Omega^{7,\ \beta^0}_{5]} = \frac{100\alpha_\theta^2}{\kappa_{\log r_h}^2}\left[1 + \delta_N +\delta_N^2\right];\nonumber\\
& (5) &  \Omega^{1,\ \beta^0}_{[4} \Omega^{7,\ \beta^0}_{6]} = {\cal O}(10^{-3})\left[1 + 10^{-1}\delta_N + 10^{-1}\delta_N^2\right]\ll1;\nonumber\\
& (6) &  \Omega^{1,\ \beta^0}_{[5} \Omega^{7,\ \beta^0}_{6]} = {\cal O}(10^{-3})
\left[1 + 10^{-1}\delta_N + 10^{-1}\delta_N^2\right]\ll1;\nonumber\\.
\end{eqnarray}
}
Hence,
{\footnotesize
\begin{equation}
\label{vi-non-zero_RHS}
\alpha_7 \Lambda^{1,\ \beta^0}_{145} - \alpha_1\Lambda^{1, \beta^0}_{745} + \frac{10^2}{\kappa_{\log r^{\rm IR}}^2} = 0.
\end{equation}
}

\item Relevant to (\ref{Transverse-condition-iii}) (vii) are:
{\footnotesize
\begin{eqnarray}
\label{vii}
& & \hskip -0.8in (1)  \frac{\left(\Omega^{3,\ \beta^0}_{24} \Omega^{7,\ \beta^0}_5 + \Omega^{3,\ \beta^0}_{45} \Omega^{7,\ \beta^0}_2 - \Omega^{3,\ \beta^0}_{25} \Omega^{7, \beta^0}_4\right)}{2} = -{\cal O}(10^{-1}\left[1 + 10^{-1}\delta_N +10^{-2}\delta_N^2\right]\ll1;\nonumber\\
& & \hskip -0.8in (2)  \frac{\left(\Omega^{3,\ \beta^0}_{24} \Omega^{7,\ \beta^0}_6 + \Omega^{3,\ \beta^0}_{46} \Omega^{7,\ \beta^0}_2 - \Omega^{3,\ \beta^0}_{26} \Omega^{7, \beta^0}_4\right)}{2} = \frac{10^{2}}{\left({\cal O}(1)\right)^4}\left[1 + \delta_N +10^{-1}\delta_N^2\right]\ll1\nonumber\\
& & \hskip -0.8in (3) \frac{\left(\Omega^{3,\ \beta^0}_{45} \Omega^{7,\ \beta^0}_6 + \Omega^{3,\ \beta^0}_{56} \Omega^{7,\ \beta^0}_4 - \Omega^{3,\ \beta^0}_{46} \Omega^{7, \beta^0}_5\right)}{2} = \frac{10}{\left({\cal O}(1)\right)^6}\left[1 + \delta_N + 10^{-1}\delta_N^2\right]\ll1\nonumber\\
& & \hskip -0.8in (4)  \frac{\left(\Omega^{3,\ \beta^0}_{25} \Omega^{7,\ \beta^0}_6 + \Omega^{3,\ \beta^0}_{56} \Omega^{7,\ \beta^0}_2 - \Omega^{3,\ \beta^0}_{26} \Omega^{7, \beta^0}_5\right)}{2} = \frac{{\cal O}(1)}{\left({\cal O}(1)\right)^{10}}\left[1 + \delta_N + 10^{-1}\delta_N^2\right]\ll1\nonumber\\ 
& & 
\end{eqnarray}
}
\end{itemize}
From (\ref{iv}), (\ref{v}) and (\ref{vii}), \begin{eqnarray}
\label{Lambdaa0b0c0-v}
& & |\Lambda_{245}|,   |\Lambda_{246}|,  |\Lambda_{256}| \ll 1;\  \Lambda_{456} \approx -1.
\end{eqnarray}
Hence, only $\Lambda_{456}$ will be taken to be non-vanishing.

Further,
{\footnotesize
\begin{eqnarray}
\label{i-2}
& (i) & \alpha_3 \Lambda^{(1)}_{124} - \alpha_1 \Lambda^{(1)}_{324} = {\cal O}(10^3)\alpha_\theta,\nonumber\\
& (ii) & \Lambda^{(3)}_{27a_0} = 0\ {\rm for}\ a_0 = 2, 4, 5, 6 \nonumber\\
& (iii) & \alpha_7 \Lambda^{(1)}_{325} - \alpha_3 \Lambda^{(1)}_{725} = \frac{{\cal O}(10^5)e^{-2.2\alpha_\theta\kappa_{r_h}}}{\alpha_{\theta_1}^4};\nonumber\\
& & \alpha_1 \Lambda^{(1)}_{325} = \alpha_3 \Lambda^{(1)}_{125} ({\rm from}\ (i))\nonumber\\
& (vi) &  \alpha_7 \Lambda^{(1)}_{145} - \alpha_1 \Lambda^{(1)}_{745} + \frac{100\alpha_\theta^2}{\kappa_{\log r^{\rm IR}\ ^2}}  = 0. 
\end{eqnarray}
}
Consistency requires the following constraints:
{\footnotesize
\begin{eqnarray}
\label{Check - IV}
& & \Lambda^{(1)}_{7b_02} R.e^2
 + \Lambda^{(1)}_{7b_04} R.e^4 + \Lambda^{(1)}_{7b_05} R.e^4 + 
    \Lambda^{(1)}_{7b_06} R.e^6 = 0;\nonumber\\
& & \Lambda^{(2)}_{13a_0} R.e^{a_0} =  0\nonumber\\
& &  {\rm identically\ satisfied\ if\ \Lambda^{(2)}_{13a_0} = 0, a_0 = 2, 4, 5, 6).}  
\end{eqnarray}
}
Further, $\Lambda^{(1)}_{1a_0b_0} \neq 0$:
{\footnotesize
\begin{eqnarray}
\label{i+Check-II-3}
& & \alpha_3 \Lambda^{(1)}_{125} = \alpha_1 \Lambda^{(1)}_{325} ({\rm from}\ (\ref{i-2})(i));
\nonumber\\
& & \alpha_3 \Lambda^{(1)}_{126} = 
 \alpha_1 \Lambda^{(1)}_{326} = \frac{\alpha_1 \alpha_3}{\alpha_7} \Lambda^{(1)}_{726} ({\rm from}\ (iii));\nonumber\\
& & \alpha_3 \Lambda^{(1)}_{145} = 
 \alpha_1 \Lambda^{(1)}_{345} = \frac{\alpha_1 \alpha_3}{\alpha_7} \Lambda^{(1)}_{745};
 \nonumber\\
& & \alpha_3 \Lambda^{(1)}_{146} = 
 \alpha_1 \Lambda^{(1)}_{346} = \frac{\alpha_1 \alpha_3}{\alpha_7} \Lambda^{(1)}_{746};\nonumber\\
& & \alpha_3 \Lambda^{(1)}_{156} = 
 \alpha_1 \Lambda^{(1)}_{356} = \frac{\alpha_1 \alpha_3}{\alpha_7} \Lambda^{(1)}_{756},\nonumber\\
& & \Lambda^{(1)}_{1b_0a_0} R.e^{b_0} = 0.
\end{eqnarray}
}
Also, using (i):
 \begin{equation}
 \label{Check - III}
\Lambda^{(1)}_{3a_0b_0} R.e^{b_0} = 0.
\end{equation}

\end{document}